\documentclass[a4paper,11pt]{article}
\pdfoutput=1

\usepackage{jheppub} 
\usepackage{subfigure}
\usepackage[utf8x]{inputenc}
\usepackage{amsthm}
\usepackage{array}
\usepackage{multirow}
\usepackage{booktabs}
\usepackage{shuffle}
\usepackage{tikz}
\usepackage{graphicx}
\usepackage{amssymb}
\usepackage{mathrsfs}
\usepackage{bm}

\usepackage{mathtools}

\usetikzlibrary{calc}
\allowdisplaybreaks[4]
\usepackage[all]{xy}

\makeatletter
\protected\def\xvcenter{%
  \hbox\bgroup$\everyvbox{\everyvbox{}\aftergroup\m@th\aftergroup$\aftergroup\egroup}%
  \vcenter
}
\DeclareRobustCommand{\midscript}[1]{
  \mathchoice{\mid@script\scriptstyle{#1}}
    {\mid@script\scriptstyle{#1}}
    {\mid@script\scriptscriptstyle{#1}}
    {\mid@script\scriptscriptstyle{#1}}
}
\newcommand{\mid@script}[2]{
  \vcenter{\hbox{$\m@th#1#2$}}
}

\makeatletter

\newcommand{\bea}{\begin{eqnarray}}
\newcommand{\eea}{\end{eqnarray}}
\newcommand{\bean}{\begin{eqnarray*}}
\newcommand{\eean}{\end{eqnarray*}}
\newcommand{\nn}{\nonumber \\}

\def\O #1{\overline{#1}}

\def\W #1{\widetilde{#1}}

\def\eref#1{(\ref{#1})}

\def\a{{\alpha}}

\def\b{{\beta}}

\def\label#1{\label{#1}%
  \smash{\hbox to0pt{\raise1ex\hbox{\tiny[#1]}\hss}}}

\title{\boldmath Note on the Labelled tree graphs}

\author[a,b]{Bo Feng,}
\author[a]{Yaobo Zhang}

\affiliation[a]{Zhejiang Institute of Modern Physics, Department of Physics,
 Zhejiang University,\\
 No.38 Zheda Road, Hangzhou 310027, P.R. China.}
\affiliation[b]{Center of Mathematical Science,
  Zhejiang University,\\
  No.38 Zheda Road, Hangzhou 310027, P.R. China.}

\emailAdd{fengbo@zju.edu.cn}
\emailAdd{yaobozhang@zju.edu.cn}

\date{\today}
\abstract{In the CHY-frame for the tree-level amplitudes, the bi-adjoint scalar theory has played a fundamental role because it gives the on-shell Feynman diagrams for all other theories.
Recently, an interesting generalization of the bi-adjoint scalar theory has been given in \cite{gao2017labelled} by the "Labelled tree graphs", which carries a lot of similarity comparing to the bi-adjoint scalar theory. In this note, we have investigated the Labelled tree graphs from two different angels. In the first part of the note, we have shown that we can organize all cubic Feynman diagrams produces by the  Labelled tree graphs to the "effective Feynman diagrams". In the new picture, the pole structure of the whole theory is more manifest. In the second part, we have generalized the action of "picking pole" in the bi-adjoint scalar theory to general CHY-integrands which produce only simple poles.
 }

\begin{document}
\maketitle
\flushbottom

\section{Introduction}

In 2013, a remarkable formula  of tree-level amplitudes for massless particles has been proposed in \cite{cachazo2014scattering1,cachazo2014scattering2,cachazo2014scattering3,cachazo2015scattering,cachazo2015einstein}, which is written as
\begin{equation}
\mathcal{A}_{n}=\int \frac{\left(\prod_{i=1}^{n} d z_{i}\right)}
{\operatorname{vol}(S L(2, \mathbb{C}))} \Omega(\mathcal{E})
 \mathcal{I}(z)=\int \frac{\left(\prod_{i=1}^{n} d z_{i}\right)}{d \omega}
 \Omega(\mathcal{E}) \mathcal{I}~~~ \label{con:CHY-def}
\end{equation}
where $z_i$ are puncture locations of $n$ external particles in $\mathbb{CP}^1$ and $d \omega=\frac{d z_{r} d z_{s} d z_{t}}{z_{r s} z_{s t} z_{t r}}$ comes from the gauge fixing of the M$\ddot{o}$bius symmetry $SL(2,\mathbb{C})$ at the locations of three variables $z_{r}, z_{s}, z_{t}$ using the Faddeev-Popov method. The $\Omega$ is given by
\begin{equation}
    \Omega(\mathcal{E}) \equiv z_{i j} z_{j k} z_{k i}
    \prod_{a \neq i, j, k} \delta\left(\mathcal{E}_{a}\right)
\end{equation}
where $\mathcal{E}_a$’s are the famous scattering equations defined
as
\begin{equation}
    \mathcal{E}_{a} \equiv \sum_{b \neq a} \frac{s_{a b}}{z_{a}-z_{b}}=0, a=1,2, \ldots, n~.
\end{equation}
In the CHY-formula \eqref{con:CHY-def}, there are two parts. The first part is  the universal measure part
$\frac{\left(\prod_{i=1}^{n} d z_{i}\right)}{d \omega} \Omega(\mathcal{E})$. The second part is the special part, i.e., the so called "CHY-integrand" $\mathcal{I}$, which defines the special theory. As a function of
$z_i$'s, for \eqref{con:CHY-def} to be well defined, the CHY integrand $\mathcal{I}$ must transform covariantly under an $SL(2,\mathbb{C})$ transformation, with opposite weight as $d \omega$
\begin{equation}
\begin{aligned}
    z_{i} \rightarrow &  \frac{\alpha z_{i}+\beta}{\gamma z_{i}+\delta} \quad \quad \alpha \delta-\beta \gamma=1 \\
    \mathcal{I}_{n}  \rightarrow & \prod_{i=1}^{n}\left(\gamma z_{i}+\delta\right)^{4}
    \mathcal{I}_{n}\left(\left\{z_{i}\right\}\right)~~~\label{weight}
\end{aligned}
\end{equation}
Such a condition in \eqref{weight} is called the weight four condition for CHY-integrands. The CHY-integrand of a given theory is, in general, a sum of rational functions of $z_{i j} \equiv
z_{i}-z_{j}$. If we use a solid line (or a dashed line) to represent a factor $z_{ij}$ in the denominator (or the numerator), the weight four condition for each term can be rephrased as the 4-regular
graph, i.e., a graph such that for each node, the number of attached solid lines minus the number of attached dashed lines is four.

For theories with $n$ particles, there are  $(n-3)!$ solutions to these scattering equations.
It is well known that when $n>5$, it is impossible to have analytical solutions and one need to apply  numerical method. However, without analytic results, many theoretical studies will be difficult. There are several works, where the analytic solutions have been bypassed and final
analytic results can be obtained using computational algebraic geometry method like \cite{dolan2014polynomial,kalousios2015scattering,huang2015algebraic,sogaard2016scattering,dolan2016general,cardona2016comments,cardona2016elimination}, such as the companion matrix, the Bezoutian matrix, the elimination theorem. In \cite{baadsgaard2015integration,baadsgaard2015scattering,baadsgaard2015integration2},  integration rule has been proposed to read out Feynman diagrams for CHY-integrands containing only simple poles analytically without solving the scattering equations. For more general
CHY-integrands with higher poles, we can
use the cross ratio identities \cite{cardona2016cross} to reduce the
degree of poles one by one until all poles are simple poles, then
the integration rule can be applied.

\begin{figure}[h]
   \quad \quad \includegraphics[width=0.9\textwidth]{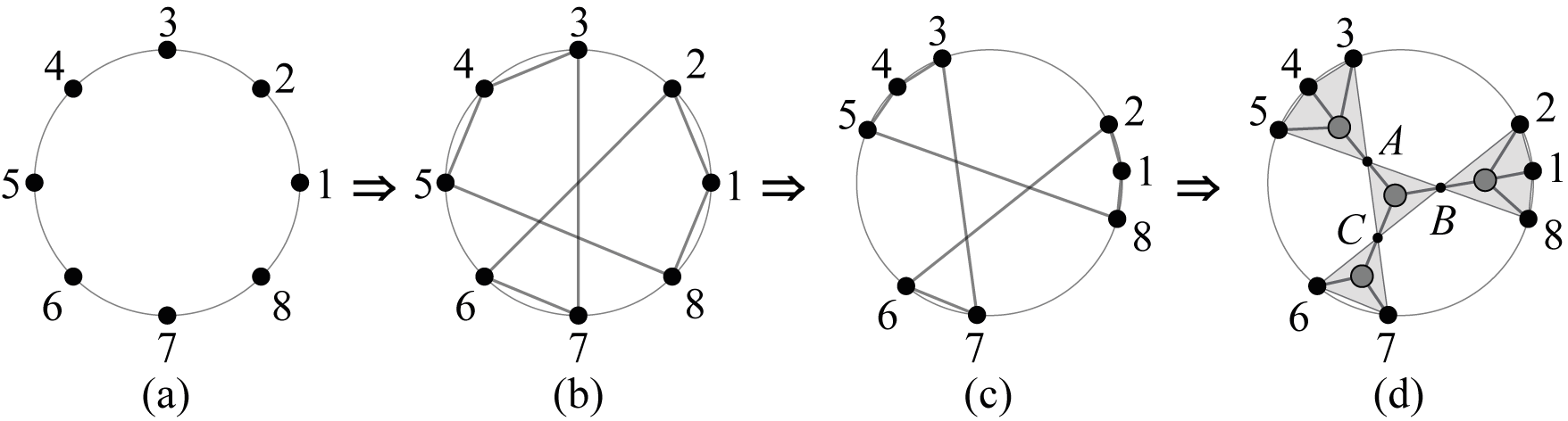}
    \caption{\label{fig:i1} The procedure to obtain the effective Feynman diagram of $I[(12345678)|(12673458)]$ and the corresponding cubic Feynman diagrams are $\frac{1}{s_{128}} \frac{1}{s_{345}} \frac{1}{s_{67}} \left(\frac{1}{s_{12}}+\frac{1}{s_{18}}\right)\left(\frac{1}{s_{34}}+\frac{1}{s_{45}}\right)$}
\end{figure}

Now we briefly review how  to read out the analytical expression for general weight four CHY-integrands. To do so, an important quantity we need to calculate is the {\bf pole
index} of each subset $A_i\subset \{1,2,...,n\}$, which is defined as\footnote{It is worth to notice that the subset $A_i$ and its complement subset $\O A_i$ have the same pole index by the weight four condition for CHY-integrands.}
\begin{equation}
    \chi (A_i)\equiv L[A_i]-2(|A_i|-1)~,~~\label{index-1}
\end{equation}
where $|A_i|$ is the number of external particles inside the subset $A_i$ and $L[A_i]$ is the linking number which is given by
\begin{equation}
     L[A_i]=\sum_{a<b;a,b\in A_i} \b_{ab}~~~\label{index-2}
\end{equation}
with the CHY-integrand given by ${\cal I}=\prod_{1\leq a<b\leq n} z_{ab}^{-\b_{ab}}$.
For a given subset $A$ with the pole index $\chi[A]\geq 0$, the amplitude could have terms with poles like ${1\over s_{A_i}^{\chi[A_i]+1}}$, where
\begin{equation}
     s_{A_i}=(\sum_{a\in A_i} k_a)^2=(\sum_{b\in \O A_i}
k_b)^2~.
\end{equation}
When $\chi[A]>0$, it is higher pole. When CHY-integrand contains higher poles,  as we have mentioned, a good way to deal with them is to use the cross ratio identity to translate the given CHY-integrands to the sum of
CHY-integrands having only simple poles, i.e., $\chi[A]\leq 0,\forall A$. Thus the basis of the analytic method is the study of CHY-integrands with only simple poles. A well studied example is the bi-adjoint scalar theory with CHY-integrands
\begin{equation}
     I[\pi|\rho]=PT(\pi)\times PT(\rho)~,~~\label{two-PT}
\end{equation}
where $\pi,\rho$ are two orderings of external particles and for the given ordering $\a$, the Park-Taylor factor is defined by
\begin{equation}
 PT(\pi) \equiv \frac{1}{(z_{\pi(1)}-z_{\pi(2)})(z_{\pi(2)}-z_{\pi(3)})\cdots(z_{\pi(n-1)}-z_{\pi(n)})
        (z_{\pi(n)}-z_{\pi(1)})}~~.~~~\label{PT}
\end{equation}
A good property of \eqref{two-PT} is that the analytic expression is the sum of certain cubic tree Feynman diagrams with the {\sl same sign}. For the bi-adjoint scalar theory, a diagrammatic method has been suggested in \cite{cachazo2014scattering2}, which can be summarized as the  {\bf effective Feynman diagrams}  \cite{huang2018permutation}. As showing in figure \ref{fig:i1}, having the effective Feynman diagram, we could easily read out the analytic expression by noticing that each vertex with $m$ legs represents the sum of  all ordered cubic  Feynman diagrams of $m$ external legs.

Bi-adjoint scalar theory is very important in the CHY-frame. Just like the Feynman diagrams in
the quantum field theory, the amplitudes of bi-PT integrands provides the on-shell cubic Feynman diagrams in the CHY-frame for all other theories, so amplitudes of other theories can be expanded
by bi-adjoint scalar amplitudes with kinematic numerators as coupling constant of these on-shell cubic Feynman diagrams. Bi-adjoint scalar theory has also the simplest CHY-integrands, i.e., as the products of two PT-factor, thus many properties of this theory can be easily abstracted. Thus it is very natural to ask how many good behaviors of bi-adjoint scalar theory can be generalized to other CHY-integrands.

Recently, in \cite{gao2017labelled}, Gao, He and Zhang have given a new class of weight two CHY-integrands (the so called "Cayley functions"), which largely generalized the Parke-Taylor factor \eqref{PT}. The Cayley
functions could be understood  better in the gauge fixed form $z_n\to \infty$. With this gauge fixing, each Cayley function is mapped to a {\bf labelled tree graph} with $(n-1)$-nodes $\{1,2,...,n-1\}$ connecting by $(n-2)$ edges\footnote{For the graph to be tree, given a pair of nodes $(i,j)$, there is at
most one edge connecting them. } (we will call them "Cayley tree" $T_{n-1}$). From a Cayley tree, we can read out the corresponding gauge fixed weight two CHY-integrand as
\begin{equation}
  C_n(T_{n-1}):= \frac{1}{\prod_{\{i,j\}\in
  Edges(T_{n-1})}(z_i-z_j)}~~\label{Cayley-1}
\end{equation}
or the $SL(2,C)$ covariant form
\begin{equation}
  C_n(T_{n-1}):= \frac{\prod_{{k}\in Vertexes(T_{n-1})}(z_k-z_n)^{v_k-2}}
  {\prod_{\{i,j\}\in Edges(T_{n-1})}(z_i-z_j)}~~\label{Cayley-2}
\end{equation}
when we put the $z_n$ back. As shown in \cite{gao2017labelled}, when applying the integration rule method, like the PT-factor \eqref{PT}, the combinations of Cayley tree's produce also the sum of certain cubic tree Feynman diagrams with the {\sl same sign}. Furthermore, they showed how
all these diagrams can be constructed by iterative algorithm. These diagrams can be
combinatorially organized to a geometric picture, i.e., the "combinatoric polytope", just like
amplitudes of bi-adjoint scalar theory having the corresponding "associahedron" picture.

As an interesting generalization of bi-adjoint scalar theories, we would like to see how many good
behaviors have been kept for these Lablled tree graphes. In this note  we will show two nice observations. First, for complicated cubic tree Feynman diagrams produced by iterative procedure,
we will show that they can be re-organized to some much simpler {\bf effective Feynman diagrams}.
As it will be seen, using these effective Feynman diagrams, it is much easier to capture the theory, since the pole structure will be much more  organized and the connection to geometric picture (i.e., the combinatoric polytope) will be more transparent. The second observation is that for CHY-integrands coming from PT-factors, as shown in \cite{feng2016chy,huang2018permutation} there is a procedure to select a subset of cubic Feynman diagrams with a particular pole structure by multiplying some cross ratio factors. This idea has been explored recently in the construction of one-loop CHY-integrands of bi-adjoint scalar theory in \cite{feng2020one}. Since the Cayley tree's are the natural generalization of the PT-factors, we would like to ask if there is a similar procedure, such that we could pick out terms with a given pole structure. In this paper, we have suggested an algorithm to achieve this goal. Although we could not give a rigorous proof for the algorithm\footnote{For the case of PT-factors, such a proof is straightforward.}, we have checked many examples. We find that our algorithm works not only for Cayley tree's, but also other weight four CHY-integrands containing only
simple poles.

We organize this paper as follows. In §\ref{The effective Feynman diagrams}, we have carefully analyzed the structure of effective Feynman diagrams, including two types of effective vertexes,
and show how to read out effective Feynman diagrams from a given  Cayley tree. From these compact
effective Feynman diagrams, we show how  to connect to the geometric picture. In §\ref{Pick up Poles}, by carefully analyzing different kinds of cross ratio factors coming from denominator and numerators, we show how to algorithmically construct their combinations to pick out particular pole from all
cubic Feynman diagrams produced by any CHY-integrand having only simple poles. We illustrate the algorithm by various non-trivial examples. In the §\ref{Conclusion}, a brief discussion of our work is presented.

\section{The effective Feynman diagrams}
\label{The effective Feynman diagrams}

In this section, we will present our first main result, i.e., the effective Feynman diagrams. We will show how to read out effective vertexes from a given Cayley tree and how to glue these vertexes together to give an effective Feynman diagram. To understand the construction, we
demonstrate with several examples. We  discuss also the enumeration problem, i.e., how
many real cubic Feynman diagrams are coded by a single effective Feynman diagram. Finally, using some
examples, we show how to use the effective Feynman diagrams to understand the geometric picture, i.e., the polytope of the amplitudes.


\subsection{Two basic types of effective Feynman vertexes}

In \cite{gao2017labelled} the iterative construction of cubic Feynman diagrams of CHY-integrands $(C_n(T_{n-1}))^2$ with  $C_n(T_{n-1})$ defined by Cayley tree's has been given\footnote{Please remember that the Cayley tree is the gauge fixed version of covariant CHY-integrand, see
\eref{Cayley-1} and \eref{Cayley-2}.  }. It is observed that with complicated Cayley tree's, there are a lot of
Feynman diagrams, thus a good way to organize these diagrams will be very useful for our further understanding
of various questions related to these theories.

\begin{figure}[h]
  \centering
  \begin{tabular}[t]{cm{0.30\linewidth}}
  \includegraphics[width=0.25\textwidth]{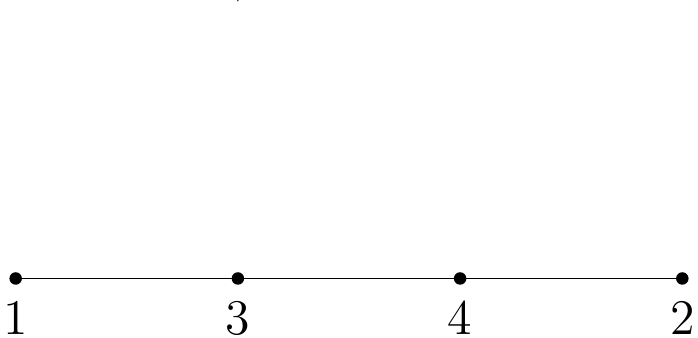} & \includegraphics[width=0.30\textwidth]{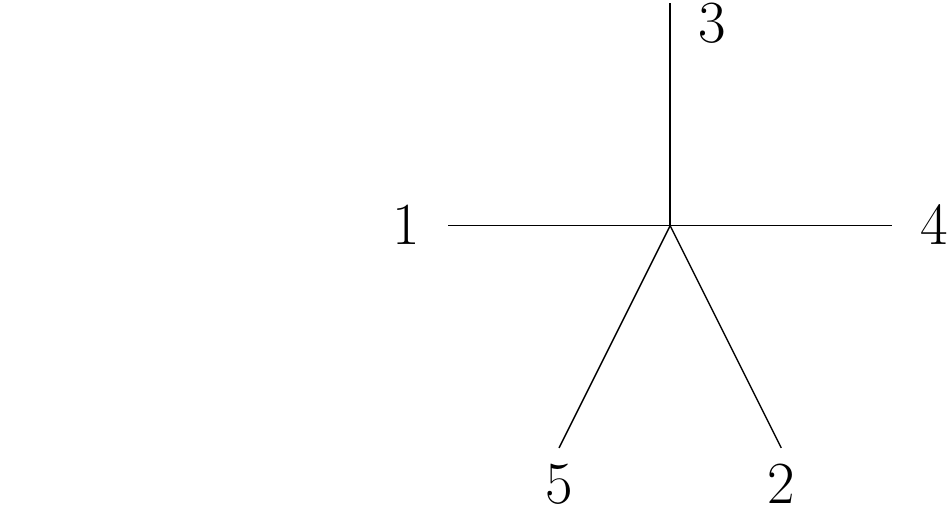} \\
  \end{tabular}
  \caption{\label{fig:1} Cayley tree of PT(1,3,4,2,5) and the corresponding effectively Feynman diagrams (vertex)}
\end{figure}

\begin{figure}[h]
  \centering
  \begin{tabular}[t]{m{0.36\linewidth}cm{0.20\linewidth}m{0.02\linewidth}m{0.20\linewidth}m{0.02\linewidth}}
    \specialrule{0em}{5pt}{15pt}
    \multirow{3}*{\includegraphics[width=0.35\textwidth]{4-PT-colours.pdf}} &\multirow{3}*{\Large{=}}  & \includegraphics[width=.20\textwidth]{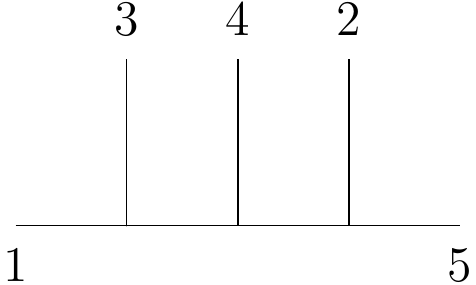} & $+$ & \includegraphics[width=.20\textwidth]{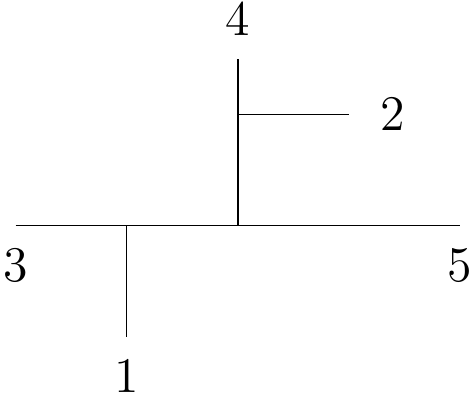}  & $+$  \\
    \specialrule{0em}{5pt}{5pt}
     & &                                                                                                             \includegraphics[width=.20\textwidth]{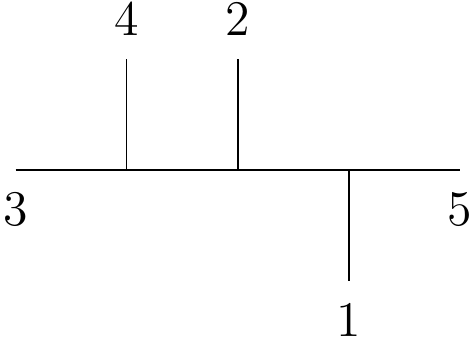} & $+$ & \includegraphics[width=.20\textwidth]{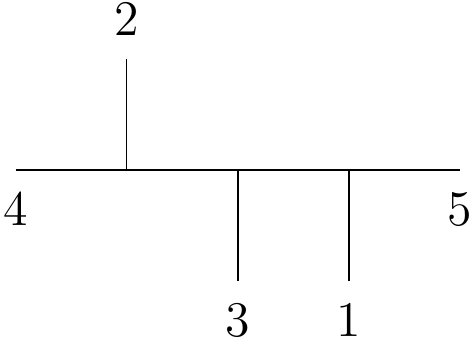}  & $+$  \\
    \specialrule{0em}{5pt}{5pt}
     & &                                                                                                             \includegraphics[width=.20\textwidth]{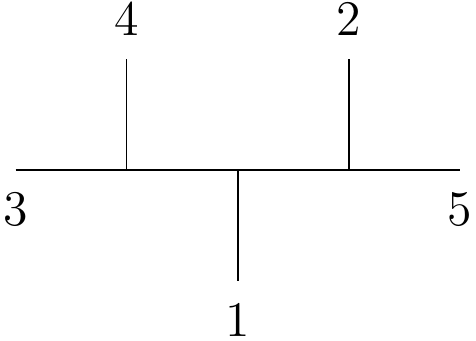}
  \end{tabular}
  \caption{\label{fig:2} the expansion of the effective vertex $V_{C}(1,3,4,2,n=5)$ and the corresponding cubic Feynman diagrams}
\end{figure}

Among all Cayley tree's, there are two special types,
for which the pole structures are clear. The first one is just a line, for example, the  Cayley tree $T_4$ with edge-list $\{\{1, 3\}, \{3, 4\}, \{4, 2\}\}$ (see the left graph in the Figure \ref{fig:1}). When rewriting into the $SL(2,C)$ covariant form, it is nothing but the familiar Parke-Taylor graph (or the "Hamiltonian graph"). Thus the left example given in the Figure \ref{fig:1} is just $PT(\{1,3,4,2,5\})$.
Using the algorithm in \cite{cachazo2014scattering2,baadsgaard2015integration,baadsgaard2015scattering}, we could easily get the related cubic Feynman diagrams: they are all five point cubic Feynman diagrams respecting the ordering $\a$ (see the Figure \ref{fig:2}). Since we know the pole structure of these Feynman diagrams, we can compactly represent them by {\bf "an effective Feynman vertex"} $V_C$, which is defined as
(see the right graph in the Figure \ref{fig:1})
\bea V_C(\a)\equiv
\{\rm{the~sum~of~all~color~ordered~cubic~Feynman~diagrams~respecting~the~ordering}~\alpha\}~~~\label{VC-def}
\eea

\begin{figure}[h]
  \centering
  \includegraphics[width=0.25\textwidth]{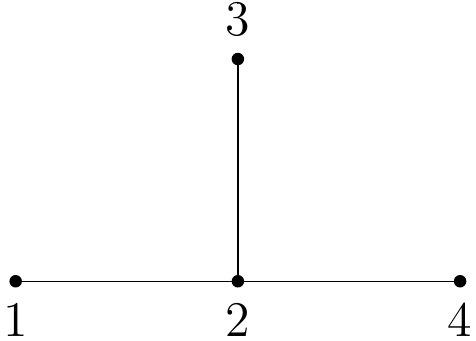}
  \caption{\label{fig:3} Star graph of $n=5$.}
\end{figure}

\begin{figure}[h]
  \centering
  \includegraphics[width=0.25\textwidth]{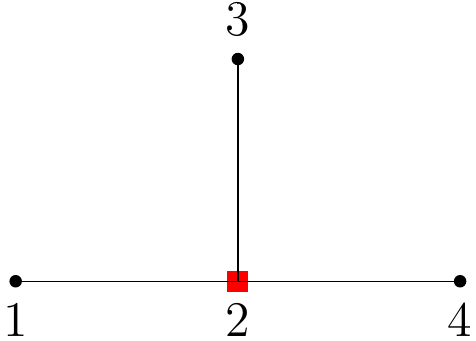} \quad\quad\quad\quad\quad \includegraphics[width=0.25\textwidth]{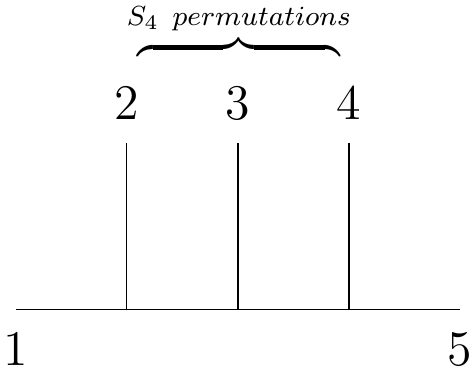}
  \caption{\label{fig:4} The 2-original Star graph and the corresponding effectively Feynman diagram (vertex).}
\end{figure}

\begin{figure}[h]
  \centering
  \begin{tabular}[t]{m{0.30\linewidth}cm{0.20\linewidth}m{0.02\linewidth}m{0.20\linewidth}m{0.02\linewidth}}
    \specialrule{0em}{5pt}{15pt}
    \multirow{3}*{\includegraphics[width=0.25\textwidth]{5-Star-Feynman.pdf}} &\multirow{3}*{\Large{=}}  & \includegraphics[width=.20\textwidth]{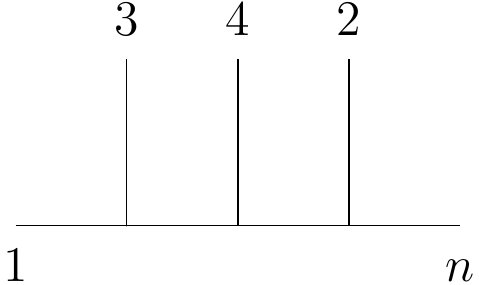} & $+$ & \includegraphics[width=.20\textwidth]{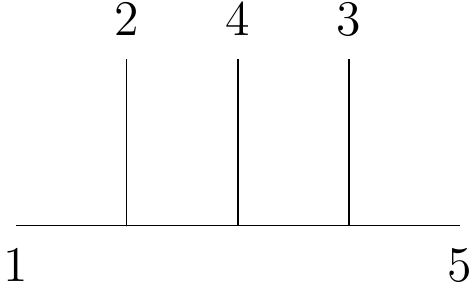}  & $+$  \\
    \specialrule{0em}{5pt}{5pt}
     & &                                                                                                             \includegraphics[width=.20\textwidth]{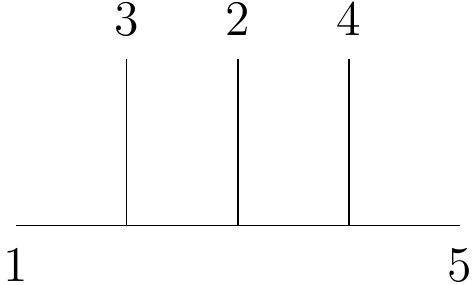} & $+$ & \includegraphics[width=.20\textwidth]{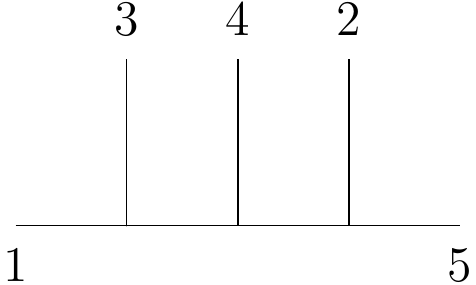}  & $+$  \\
    \specialrule{0em}{5pt}{5pt}
     & &                                                                                                             \includegraphics[width=.20\textwidth]{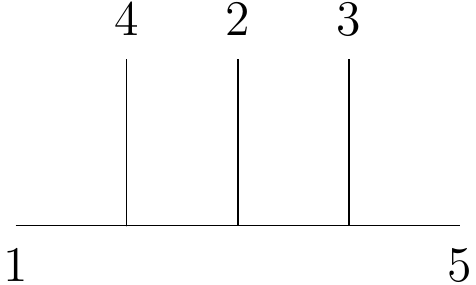} & $+$ & \includegraphics[width=.20\textwidth]{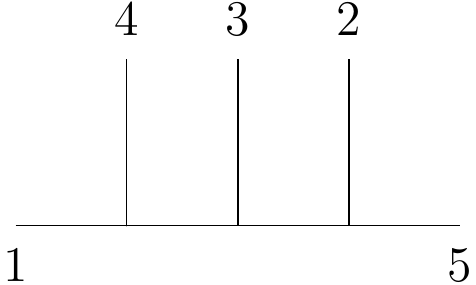}
  \end{tabular}
  \caption{\label{fig:5}  the expansion of the effective vertex $V_{P}(1 ;\{2\} \shuffle\{3\} \shuffle\{4\} ; 5)$ and the corresponding comb-like Feynman diagrams}
\end{figure}

Another special example is the star graph, where among $(n-1)$-points, $(n-2)$ of them connect to
the remaining point (see, for example, Figure \ref{fig:3}).
Using the algorithm in \cite{gao2017labelled}, we could get the related cubic Feynman diagrams with the permutation symmetry of $(n-2)$ points (see for example, Figure \ref{fig:4} and \ref{fig:5}).
To understand the pattern, first let us relate an ordered sequence with a cubic Feynman diagram:
\bea (a_1; a_2, a_3,...,a_{n-1}; a_n)\to {1\over \prod_{t=2}^{n-2}
s_{a_1 a_2...a_t}}~.~~\label{star-1}\eea
Using this notation, all Feynman diagrams coming from the  star graph in the Figure \ref{fig:3} can be
summarized as  $(1 ;\{2\} \shuffle\{3\} \shuffle\{4\} ; 5)$, which will be the sum of sequences of the form in
\eqref{star-1} (see the Figure \ref{fig:5}). In this writing, we have defined the "shuffle" algebra. For a two ordered sets, their shuffle
is defined as
\bea & & \a \shuffle \emptyset =\a,~~~\emptyset \shuffle \b=\b,~~~\nn
& & \a\shuffle\b=\{\a_1,\{\a_2,...,\a_m\}\shuffle \b\}+\{\b_1,\a\shuffle
\{\b_2,...,\b_k\}\}~~~ \label{shuffle-def} \eea
where $\a_1,\b_1$ are the first elements in the sets $\a,\b$.
Because we know very well the pattern for the star graph, we can compactly represent them by another {\bf
"effective Feynman vertex"} $V_P$, where the subscript P means the P-type vertex (the permutation type vertex).
For example, the start graph in the Figure \ref{fig:3} can be rewritten as
$V_{P}(2 ;\{1\} \shuffle\{3\} \shuffle\{4\} ; 5)$ (as shown in the Figure \ref{fig:4} and \ref{fig:5}).
The $V_{P}(2 ;\{1\} \shuffle\{3\} \shuffle\{4\} ; 5)$ vertex contains three parts:  the starting external leg $2$, the ending external leg $n=5$ and the middle sequence coming from shuffle algebra.

The shuffle algebra $\scriptstyle \shuffle $ is crucial for our compact representation of Feynman diagrams.
Later we will meet more complicated shuffle algebras, such as
\bea \{2\}\shuffle \{3\}\shuffle \{4,5,6\},~~~~ \{2\}\shuffle \{3\}\shuffle \{\{4\}\shuffle\{5\},6\}\eea
The corresponding sequences produced by them should be easily worked out using \eqref{shuffle-def}.

Up to now we see that for above two special Cayley tree's, we have complete understanding of the pole structure of corresponding Feynman diagrams. These two special Cayley tree's correspond to the line and vertex (with multiple branches) structures in the general tree graphs respectively.
Since every tree is constructed just by lines and vertexes with branches,
it is very natural to guess that we should be able to compactly represent all Feynman diagrams coming from  arbitrary Cayley tree by properly using about two special structures of Feynman diagrams. In the later subsections, we will use various examples to demonstrate the idea, especially the algorithmic way to read out all effective Feynman diagrams.

\subsection{The example of next-to-Star graph}

\begin{figure}[h]
  \centering
  \includegraphics[width=.30\textwidth]{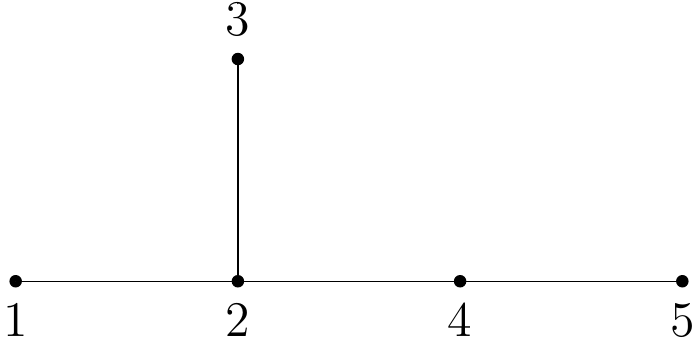}
  \qquad
  \qquad
  \qquad
  \centering
  \includegraphics[width=.30\textwidth]{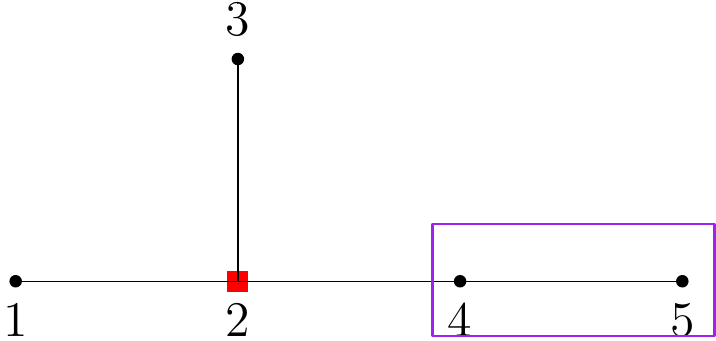}
  \caption{\label{fig:6} The Next-to-Star graph of $n=6$.}
\end{figure}

\begin{figure}[h]
  \centering
  \includegraphics[width=.30\textwidth]{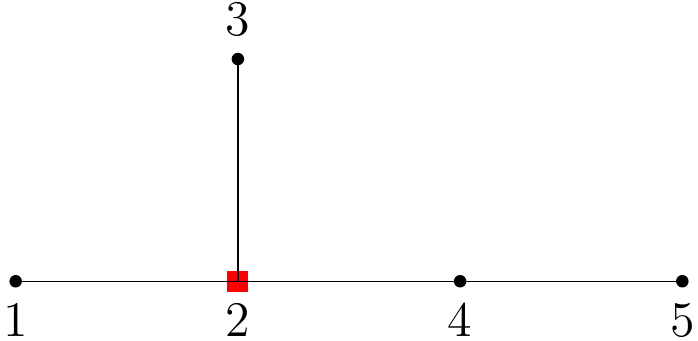}
  \qquad
  \qquad
  \qquad
  \centering
  \includegraphics[width=.30\textwidth]{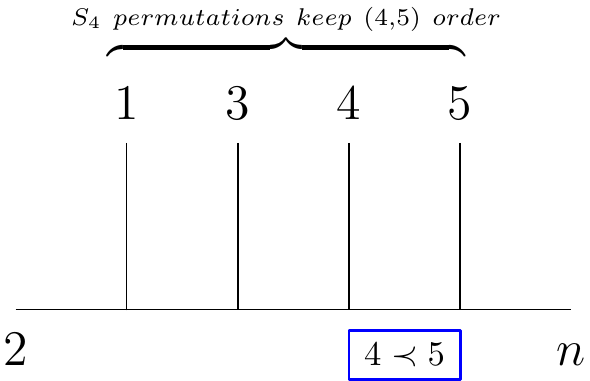}
  \caption{\label{fig:7}    $V_{P}(2 ;\{1\} \shuffle\{3\} \shuffle\{4,5\}  ; n=6)$ and the corresponding effective Feynman diagram}
\end{figure}

Starting from this subsection, we will show how to decompose an arbitrary Cayley tree to two basic structures ( i.e., the line and the vertex) and then using the corresponding effective vertexes we could glue them together to give the corresponding effective Feynman diagrams. To get the idea, let us start with the simplest nontrivial example, i.e., the $n=6$ Next-to-Star graph  as
shown in the Figure \ref{fig:6}.
 This Cayley tree gives $18$ cubic Feynman
diagrams. To organize them, let us mark the node $2$ first since it
is the vertex with three branches in the Cayley tree (we marked it with a red square in
the Figure). Now we remove the point $2$ and its adjacent edges,
i.e., the edges $\{1,2\},\{3,2\},\{4,2\}$, and get three subgraphs.
Among three subgraphs, two of them have no further structure, but
one of them, does. For the nontrivial subgraph, we construct its
substructures by considering all possible contracting of edges. The
first case is that no edge has been contracted. Using the star graph
of node $2$ and its corresponding effective Feynman vertex, we get
the effective diagram with only one effective vertex
\bea V_{P}(2 ;\{1\} \shuffle\{3\} \shuffle\{4,5\}  ;
n)~~~\label{fig-7-1}\eea
When expanding the effective diagram,  the  corresponding cubic Feynman diagrams are given in the Figure
\ref{fig:7}.
It is worth to notice that different from the pure star graph, now
the effective vertex contains the sub-structure, i.e., in the
comb-like Feynman diagrams, the leg $4$ is always near the leg $2$
than the leg $5$ by the shuffle algebra. To emphasize this important
point, we have mark the $\{4,5\}$ order on the  comb-like diagram as
$4\prec5$\footnote{In the original Cayley tree, the  node $4$ is
closer to the marked point $2$ than node $5$. In the following, we
will use  the left directed $\prec$ to represent the precursor and
omit the order mark with the 2 adjacent points.} under the legs of
$4,5$ in the effective Feynman diagram. One can check that  this
effective vertex gives $\frac{4!}{2!}=12$ comb-like cubic Feynman
diagrams.

\begin{figure}[h]
  \centering
  \includegraphics[width=.25\textwidth]{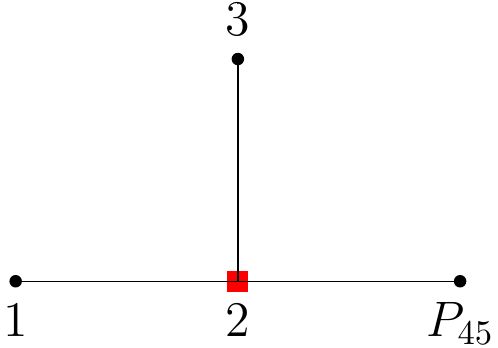}
  \qquad
  \qquad
  \qquad
  \centering
  \includegraphics[width=.13\textwidth]{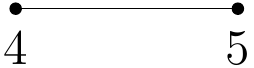}
  \caption{\label{fig:8} star graph and the line 
  }
\end{figure}

\begin{figure}[h]
  \centering
  \includegraphics[width=.25\textwidth]{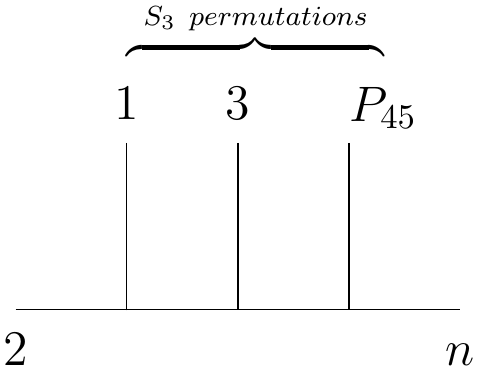}
  \qquad
  \qquad
  \qquad
  \centering
  \includegraphics[width=.16\textwidth]{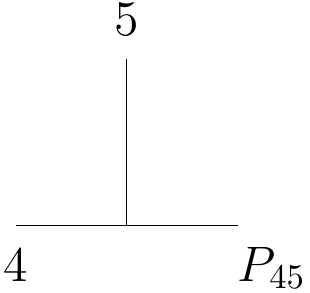}
  \caption{\label{fig:9}  $V_{P}(2 ;\{1\} \shuffle\{3\} \shuffle\{P_{45}\} ; n=6)$
  and $V_{C}\left(\left\{4,5, P_{45}\right\}\right)$}
\end{figure}

\begin{figure}[h]
  \centering
  \includegraphics[width=.25\textwidth]{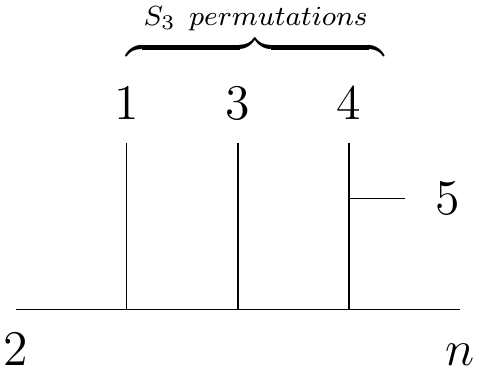}
  \caption{\label{fig:10}   The corresponding effective Feynman diagram}
\end{figure}

The second case is the  contraction of the edge  $\{4, 5\}$ to a node
denoted by $P_{45}$. With this contraction, the Cayley tree is
separated to two parts as shown in the Figure \ref{fig:8}.
The left graph is the standard star graph while the right graph is
the standard Hamiltonian graph. Thus as shown in the Figure
\ref{fig:9} by using our effective vertexes,  the left graph gives
$V_{P}(2 ;\{1\} \shuffle\{3\} \shuffle\{P_{45}\} ; n=6)$, which  can
be expanded to six cubic Feynman diagrams, while the right graph gives
$V_{C}\left(\left\{4,5, P_{45}\right\}\right)$, which is just one cubic
Feynman diagram.
Using the propagator $P_{45}$ to glue these two effective vertexes
together, we get an effective Feynman diagram  as shown in the
Figure \ref{fig:10}
with the following expression
\bea V_{P}\left(2 ;\{1\} \shuffle\{3\} \shuffle\left\{P_{45}\right\}
; n=6\right) \frac{1}{P_{45}^{2}} V_{C}\left(\left\{4,5,
P_{45}\right\}\right)~~~\label{fig-10-1}\eea
Since the left part gives $6$ diagrams while the right part, just
one, the effective Feynman diagram \eqref{fig-10-1} gives six cubic
Feynman diagrams. When we add $12$ diagrams coming  from
\eqref{fig-7-1}, we do get total $18$ diagrams as done using the
iterative construction.

\begin{figure}[h]
  \centering
  \includegraphics[width=0.40\textwidth]{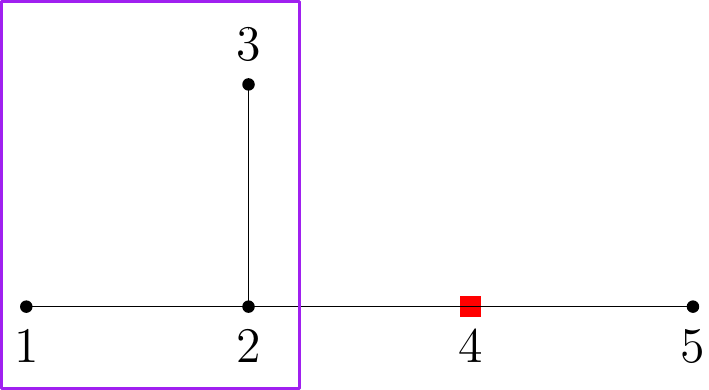}
  \caption{\label{fig:11} The node 4 as the marked point.}
\end{figure}

\begin{figure}[h]
  \centering
  \includegraphics[width=.30\textwidth]{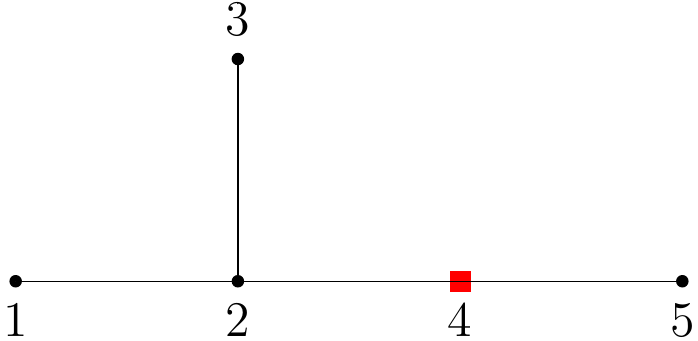}
  \qquad
  \qquad
  \qquad
  \centering
  \includegraphics[width=.30\textwidth]{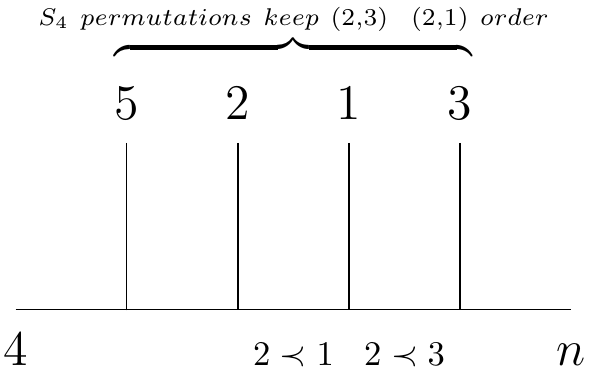}
  \caption{\label{fig:12}  $V_{P}(4 ;\{5\} \shuffle \{2,\{3 \shuffle 1\}\}; n)$ and the corresponding effective Feynman diagram}
\end{figure}

\begin{figure}[h]
  \centering
  \begin{tabular}[t]{m{0.30\linewidth}m{0.01\linewidth}m{0.30\linewidth}m{0.30\linewidth}}
    \specialrule{0pt}{5pt}{15pt}
    \includegraphics[width=.27\textwidth]{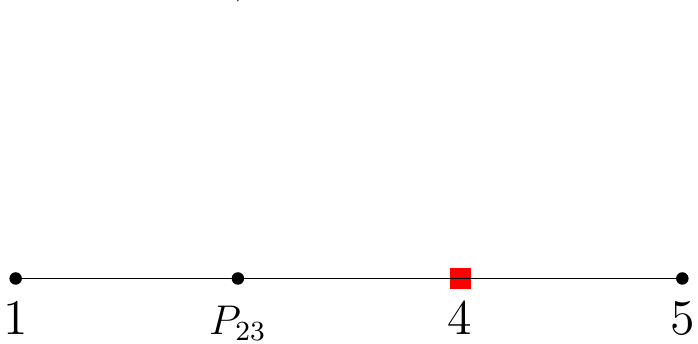} & & \includegraphics[width=.25\textwidth]{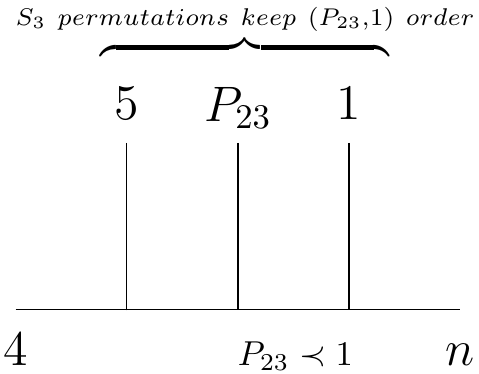} & $V_{P}(4 ;\{5\} \shuffle\{P_{23},1\}  ; n)$\\
    \specialrule{0pt}{5pt}{15pt}
    \includegraphics[width=.20\textwidth]{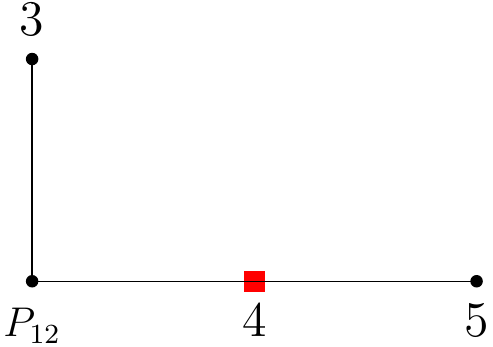} & & \includegraphics[width=.25\textwidth]{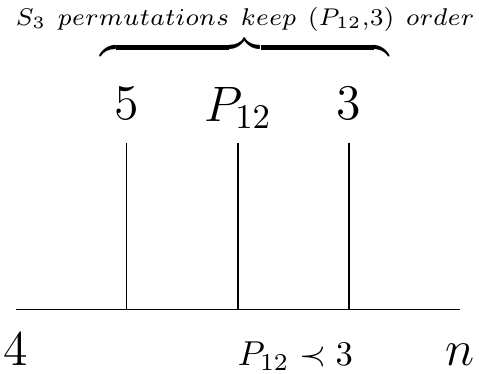} & $V_{P}(4 ;\{5\} \shuffle\{P_{12},3\}  ; n)$\\
    \specialrule{0pt}{5pt}{15pt}
    \includegraphics[width=.13\textwidth]{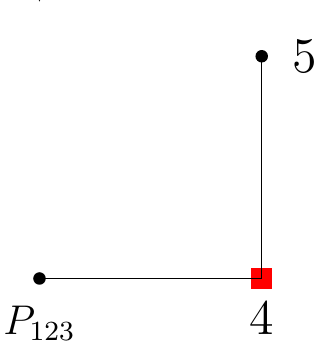} & & \includegraphics[width=.25\textwidth]{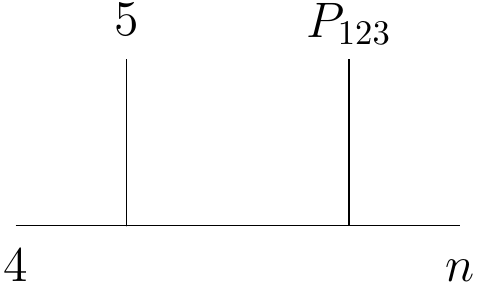} & $V_{P}(4 ;\{5\} \shuffle\{P_{123}\}  ; n)$\\
    \end{tabular}
  \caption{\label{fig:13} The structure after contractions and the effective Feymann vertexes  }
\end{figure}

\begin{figure}[h]
  \centering
  \begin{tabular}[h]{b{0.15\linewidth}m{0.20\linewidth}m{0.30\linewidth}}
    $P_{23}$ & \includegraphics[width=.10\textwidth]{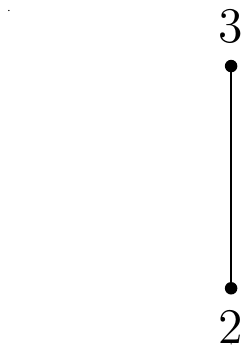} & \includegraphics[width=.20\textwidth]{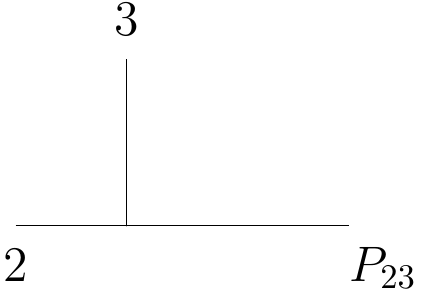}\\
    $P_{12}$ & \includegraphics[width=.10\textwidth]{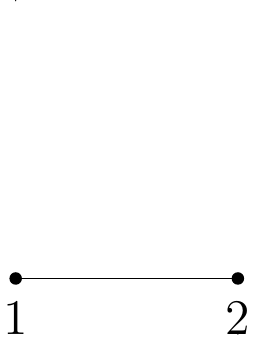} & \includegraphics[width=.20\textwidth]{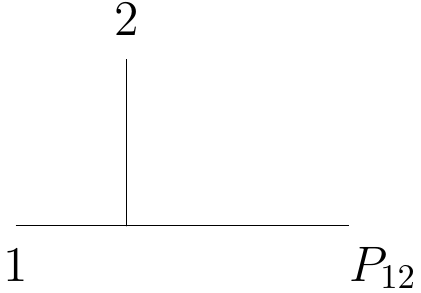}\\
    \specialrule{0pt}{5pt}{15pt}
    $P_{123}$ & \includegraphics[width=.10\textwidth]{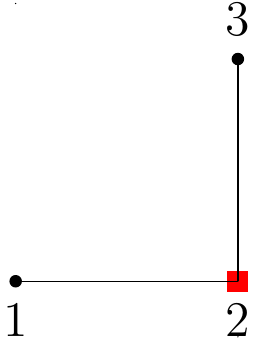} & \includegraphics[width=.25\textwidth]{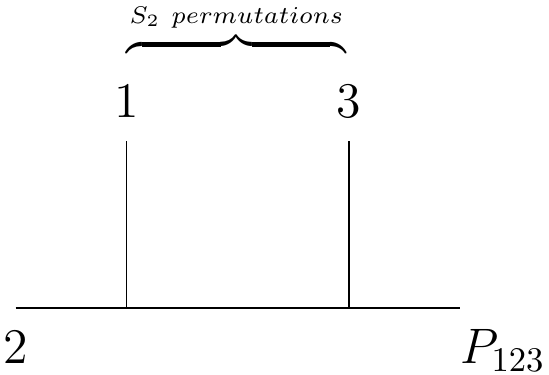} \\
  \end{tabular}
  \caption{\label{fig:14}  The contracted part  the  corresponding sub-Feymann diagrams}
\end{figure}

\begin{figure}[h]
  \centering
  \begin{tabular}[t]{m{0.30\linewidth}m{0.50\linewidth}}
    \specialrule{0pt}{5pt}{15pt}
    \includegraphics[width=.25\textwidth]{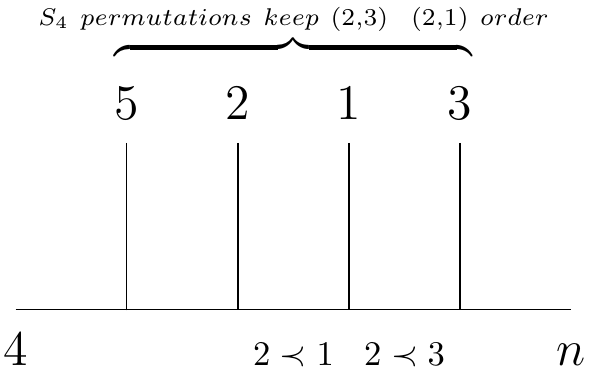} & $V_{P}(4 ;\{5\} \shuffle \{2,\{3 \shuffle 1\}\}; n)$\\
    \specialrule{0pt}{5pt}{15pt}
    \includegraphics[width=.22\textwidth]{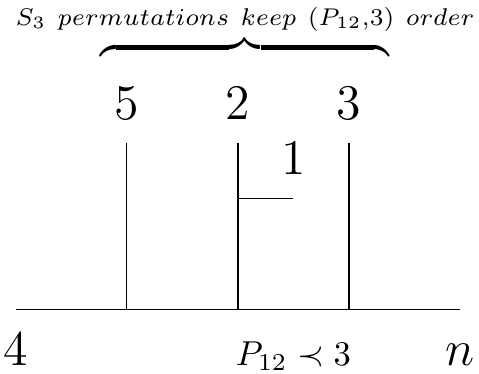} & $V_{P}(4 ;\{5\} \shuffle\{P_{12},3\}  ; n) \frac{1}{P_{12}^{2}} V_{C}(\{1,2, P_{12}\}) $\\
    \specialrule{0pt}{5pt}{15pt}
    \includegraphics[width=.22\textwidth]{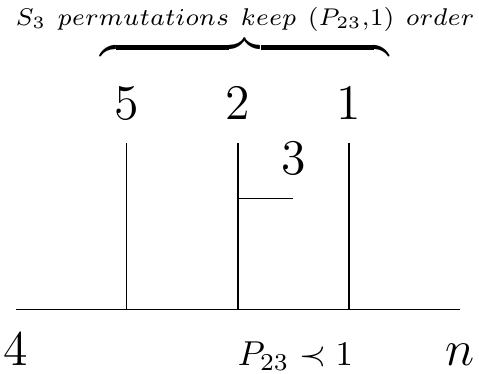} & $V_{P}(4 ;\{5\} \shuffle\{P_{23},1\}  ; n) \frac{1}{P_{23}^{2}} V_{C}(\{2,3, P_{23}\})$\\
    \specialrule{0pt}{5pt}{15pt}
    \includegraphics[width=.22\textwidth]{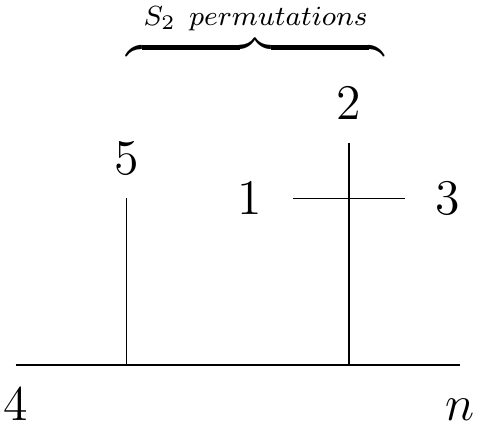} & $V_{P}(4 ;\{5\} \shuffle\{P_{123}\}  ; n) \frac{1}{P_{123}^{2}} V_{C}(\{1,2,3, P_{123}\})$\\
    \end{tabular}
  \caption{\label{fig:15}  The full Effective Feynman diagram representation}
\end{figure}

In above construction of effective Feynman diagrams, we have used the node $2$ as the starting (marked)
point. In general, we can choose any point to start with the whole construction. Let us redo this example
with node $4$ as the staring point as shown in the Figure \ref{fig:11}.
Again we need to consider the subgraphs: one is just a node $5$ and another one, with nodes $1,2,3$.
For the later one, we consider various contractions as above. Now there are two edges, so we have
following four types of contractions: (a) no contraction at all; (b) the edge $\{2,3\}$ has been contracted;
(c) the edge $\{1,2\}$ has been contracted; (d) all edges $\{1,2\},\{2,3\}$ have been contracted.
For the case (a), using the effective vertex of star graph of node $4$, we get the contribution (we have also given the counting when expanding the effective Feynman diagram)
\bea V_{P}(4 ;\{5\} \shuffle \{2,\{3 \shuffle 1\}\};
n=6)~,~~~~~\#=\frac{4!}{3}=8~~\label{fig-12-1}\eea
as shown in the Figure \ref{fig:12}. It is worth to notice the substructure in the shuffle algebra
$\{2,\{3 \shuffle 1\}\}$ to keep the orderings between $\{2,3\}$ and $\{2,1\}$ to the node $4$.
This effective Feynman diagram contain  $\frac{4!}{3}=8$ cubic Feynman diagrams.
For the case (b), (c) and (d) with  contracted edge, we have separated the Cayley tree's to two
parts: the first part is given in the Figure \ref{fig:13} and the second part is given in  the Figure \ref{fig:14}, where the corresponding effective Feynman vertexes have also been written down.
Having known the two parts, we use the propagator to connect them and write down the final effective Feynman diagrams as  shown in the Figure \ref{fig:15}
\bea (b) & = &  V_{P}(4 ;\{5\} \shuffle\{P_{23},1\}  ; n) {1\over P_{23}^2} V_C(\{P_{23},2,3\}),~~~~\#= {3!\over 2}=3  \nn
(c) & = & V_{P}(4 ;\{5\} \shuffle\{P_{12},3\}  ; n){1\over P_{12}^2} V_C(\{P_{12},1,2\})~~~~\#= {3!\over 2}=3 \nn
(d) & = & V_{P}(4 ;\{5\} \shuffle\{P_{123}\}  ; n) {1\over
P_{123}^2} V_P(2;\{1\}\shuffle \{3\}; P_{123}~~~~\#= 2\times
2=4~~~\label{fig-12-2} \eea
It is worth to notice that for the case (d), the part in the Figure \ref{fig:13} can be considered as the star graph of node $2$ with infinity node $P_{123}$ or as the Hamiltonian graph with effective vertex
$V_C(\{1,2,3,P_{123}\})$. When we add all effective Feynman diagrams together, we do get $18$ Feynman diagrams.

Having done above two calculations, now we can summarize the algorithm:
\begin{itemize}

\item (A) At the first step, we choose a marked node, for example, $k$.

\item (B) Then we remove all edges connecting to the marked node $k$. Now the Cayley tree is separated to node $k$ and
several subgraphs, which we can denote as $K_1,K_2,..,K_t$.

\item (C) For each subgraph $K_i$, we need to consider its all substructures by all possible contractions of edges in the subgraph. For a given contraction, we will generate following data: (1) First we shrink all contracted edges in the subgraph $K_i$ (so nodes at the two ends of the edge will be merged to a single node) to generate a new graph $\W K_i$; (2) Secondly, all edges having been contracted will become several disconnected sub-Cayley tree, which we will denote as $K_{i,j}$.

\item (D) Having data from the step (C), now we have an effective vertex $V_P$, which is given by
reconnecting the node $k$ with all $\W K_i$. The roughly expression will be
\bea V_P(k; \W K_1\shuffle \W K_2 \shuffle ... \shuffle \W K_t;
n)~~~\label{VPk-gen}  \eea
where for each $\W K_i$ in the shuffle algebra \eqref{VPk-gen}, possible embedded structure will appear as such given in \eqref{fig-12-1}.

\item (E) For each sub-Carley tree $K_{i,j}$, we can repeat the steps from (A) to (D) to get the corresponding effective vertex $V_{K_{i,j}}$. Then we connect vertex $V_{K_{i,j}}$ to the merged node in $\W K_{i}$ by
    corresponding propagator to construct the effective Feynman diagram.

\item (F) Iterating above steps, we will obtain all effective Feynman diagrams for any Cayley tree.

\end{itemize}
One should use above two calculations to get more clear picture about our algorithm. Some remarks for the algorithm are following:
\begin{itemize}

\item (a) Although in general we will get the $V_P$ vertex in above algorithm, when the sub-Cayley tree is just a line (i.e., the Hamiltonian graph), we do not need to consider its various substructure by contractions. For this simple case, we just need to use the color-ordered effective vertex $V_C$ as explained below the equation \eqref{fig-12-2}.

\item (b) Our algorithm is also recursive, but top-down style.

\item (c) From above two calculations, we see that stating from different marked nodes, we can have different organizations, i.e.,
different sets of effective Feynman diagrams. Thus a good choice of the marked node is very important to get more compact expressions.

\end{itemize}
%
\subsection{Enumerate Feynman diagrams from an effective Feynman diagram}

Since we have used the concept of effective vertex, each effective
Feynman diagram will code several cubic Feynman diagrams when expanding each effective
vertex, thus
the counting of these cubic Feynman diagrams is a very important
check for our algorithm. In this subsection, we will consider this
problem.

For an effective Feynman diagram, its counting $N_F$ is given by
\bea N_F=\prod_{i=1}^t n_{i}~~~\label{EF-count-1}\eea
where $t$ is the number of effective vertexes in the diagram and for
each effective vertex, $n_i$ is the number of its expansion to cubic
Feynman sub-diagrams. Since effective vertex can only be two types,
i.e., either the $P$-type or the $C$-type, we discuss them one by
one.

For the $C$-type effective vertex, the
$|V_{C}\{(l_1,l_2,...,l_n)\}|$ enumerates all the cubic Feynman
trees  respecting  the colour order of the list
$\{(l_1,l_2,...,l_n)\}$. According to \cite{gao2017labelled}, it is the $n$-th
Catalan number ${\rm Cat}_{n}$, which is  given directly in terms of
binomial coefficients by
\bea {\rm Cat}_{n}=\frac{1}{n+1}\left(\begin{array}{c}{2 n} \\
{n}\end{array}\right)=\frac{(2 n) !}{(n+1) ! n !}=\prod_{k=2}^{n}
\frac{n+k}{k} \quad \text { for } n \geq 0~~~\label{Catalan} \eea
The first few Catalan numbers are given by
\bea {\rm Cat}_{2}& = & 1,~~~{\rm Cat}_{3}=2,~~~{\rm
Cat}_{4}=5,~~~{\rm Cat}_{5}=14,~~~{\rm Cat}_{6}=42\nn
{\rm Cat}_{7} & = & 132,~~~{\rm Cat}_{8}=429,~~~{\rm
Cat}_{9}=1430\eea

For the $P$-type vertex, the counting is not so trivial, since in
general, the shuffle algebra contains various substructure. However,
as we will see, the counting can be done level by level. Let us
start with a typical situation, i.e., two arbitrary ordering list
$\alpha=\{\alpha_1,\alpha_2,\ ...\ ,\alpha_m \}$ and
$\beta=\{\beta_1,\beta_2,\ ...\ ,\beta_n \}$.  From
\eqref{shuffle-def}, we see that the shuffle $\alpha \shuffle \beta$
contains all possible permutations of the list $\alpha \cup \beta$,
which preserve the relative ordering in $\alpha$ and $\beta$
respectively. Thus the counting is given by
\bea \left| \alpha \shuffle \beta
\right|=\frac{(m+n)!}{m!n!},~~~\label{shuffle-count-1}\eea
This counting can be easily generalized to multiple lists, thus we
have
\bea \left| V_P(o;\alpha_1\shuffle\a_2\shuffle...\shuffle \a_k;
n)\right|={(\sum_{i=1}^k |\a_i|)!\over \prod_{j=1}^{k}
|\a_j|!}~~~\label{shuffle-count-2}\eea
when all list $\a_i$'s do not have substructure. When the $V_P$ has
some substructures, we could  recursively count the number.  For
example, for
 $$V_P(1;\{2,3\shuffle 4\} \shuffle \{5,
6\shuffle7\};8)$$
there are two levels of  shuffle algebra. For the first layer $\{2,3
\shuffle 4 \}\shuffle\{5, 6\shuffle7\}$, using the formula
\eqref{shuffle-count-2}, we get the counting
$\frac{(3+3)!}{3!3!}=20$. For the second  shuffle layer, i.e.,
$\{3\shuffle 4\}$ and $\{ 6\shuffle7\}$, each of them gives the
number  $\frac{(1+1)!}{1!1!}=2$. Putting them together, we get the
counting  $20\times(2\times2)=80$.

\begin{figure}[h]
  \centering
  \includegraphics[width=.45\textwidth]{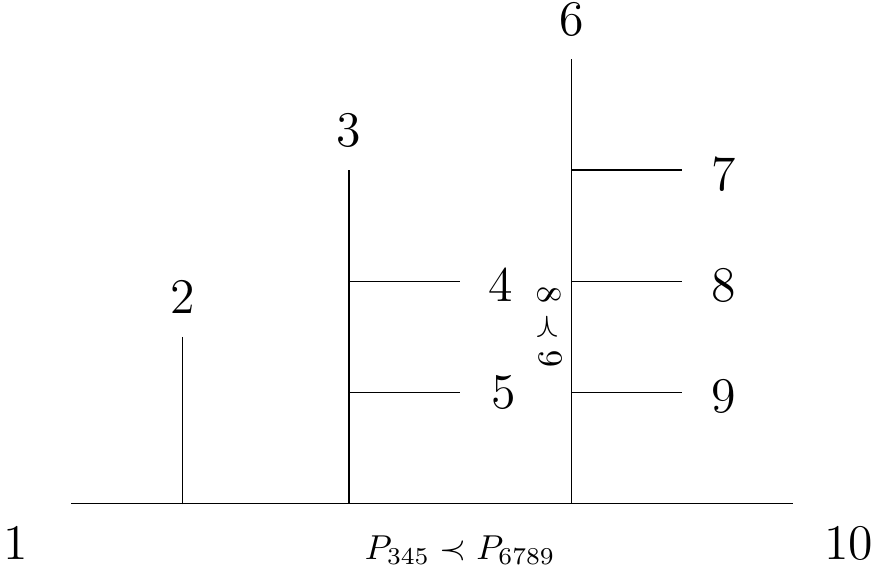}
  \caption{\label{fig:count-1}  $V_P(1;\{2\}\shuffle\{P_{345},P_{6789}\};10)\frac{1}{P_{345}}V_P(3;\{4\}\shuffle \{5\};P_{345})\frac{1}{P_{6789}}V_P(6;\{7\}\shuffle \{8,9\};P_{6789})$}
\end{figure}
\begin{figure}[h]
  \centering
  \includegraphics[width=.45\textwidth]{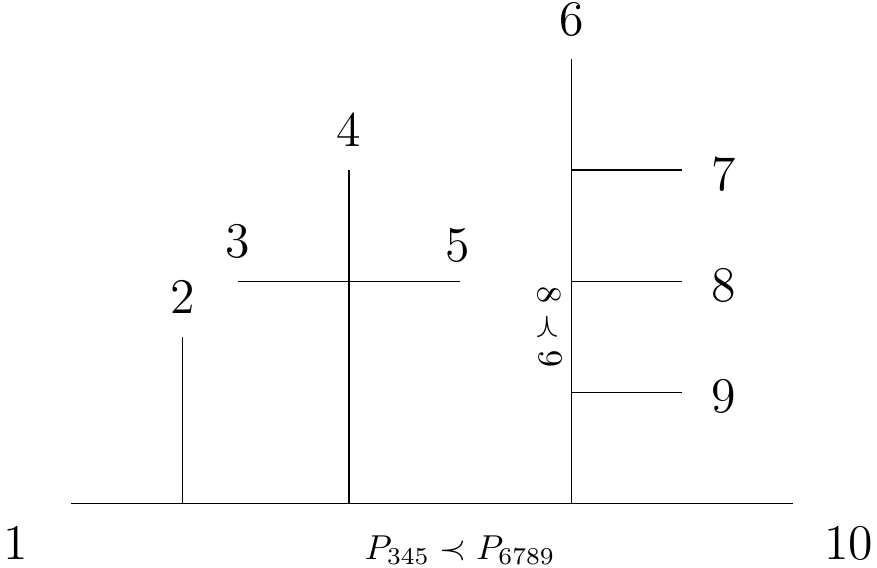}
  \caption{\label{fig:count-2}  $V_P(1;\{2\}\shuffle\{P_{345},P_{6789}\};10)\frac{1}{P_{345}}V_C(\{3,4,5,P_{345}\})
\frac{1}{P_{6789}}V_P(6;\{7\}\shuffle
\{8,9\};P_{6789})$}
\end{figure}

Having known the counting of each effective vertex, we can get the
counting for any effective Feynman diagram.  For the effective
Feynman diagram given in the Figure \ref{fig:count-1}, its
expression is given by
\bea
V_P(1;\{2\}\shuffle\{P_{345},P_{6789}\};10)\frac{1}{P_{345}}V_P(3;\{4\}\shuffle \{5\};P_{345})\frac{1}{P_{6789}}V_P(6;\{7\}\shuffle \{8,9\};P_{6789})~~~~~\label{count-1} \eea
thus the counting is given by
\bea & & \left| V_P(1;\{2\}\shuffle\{P_{345},P_{6789}\};10)\right|
\times \left| V_P(3;\{4\}\shuffle \{5\};P_{345})\right| \times
\left|V_P(6;\{7\}\shuffle \{8,9\};P_{6789})\right|
\nn
& = & \frac{(1+2)!}{2}\times \frac{(1+1)!}{1}\times
\frac{(1+2)!}{2}=18 ~~~\label{count-1-1}\eea
For the example with mixing $V_C$ and $V_P$ types in the Figure
\ref{fig:count-2}, its expression is given by
\bea
V_P(1;\{2\}\shuffle\{P_{345},P_{6789}\};10)\frac{1}{P_{345}}V_C(\{3,4,5,P_{345}\})
\frac{1}{P_{6789}}V_P(6;\{7\}\shuffle
\{8,9\};P_{6789})~~~\eea
and the counting is given by
\bea
\frac{(1+2)!}{2}\times\frac{1}{2+1}\left(\begin{array}{c}{4} \\
{2}\end{array}\right)\times\frac{(1+2)!}{2} = 18 \eea

\subsection{Another example}

Having presented the general algorithm for the construction of
effective Feynman diagrams for arbitrary Cayley tree's, in this
subsection, we give another example for the Cayley tree in the
Figure \ref{fig:ano-exa}:
\begin{figure}[h]
  \centering
  \includegraphics[width=.35\textwidth]{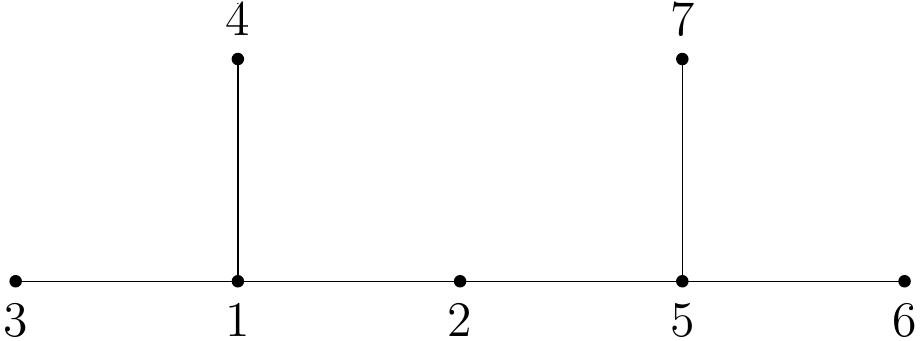}
  \caption{  $C_8\{\{1, 2\},\{1, 3\},\{1, 4\}, \{2, 5\}, \{5, 6\}, \{5, 7\}\}$~~~\label{fig:ano-exa}}
\end{figure}

\begin{figure}[h]
  \centering
  \begin{tabular}[t]{m{0.20\linewidth}m{0.21\linewidth}m{0.25\linewidth}m{0.25\linewidth}}
    \specialrule{0pt}{5pt}{15pt}
    $0-contraction$:& & \includegraphics[width=.25\textwidth]{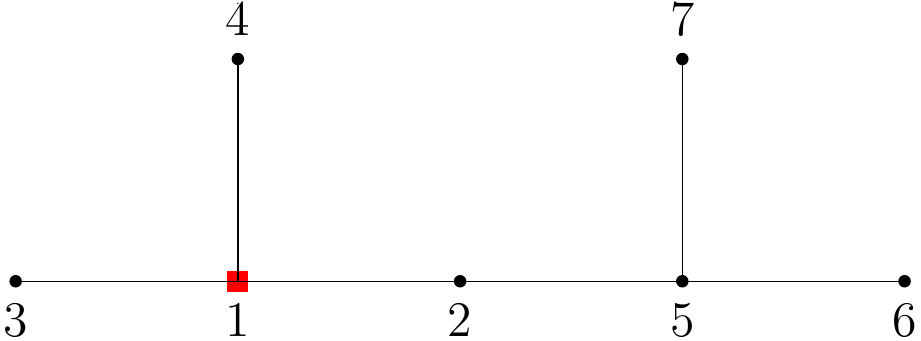} &  \\
    \specialrule{0pt}{5pt}{15pt}
    $1-contraction$:& \includegraphics[width=.20\textwidth]{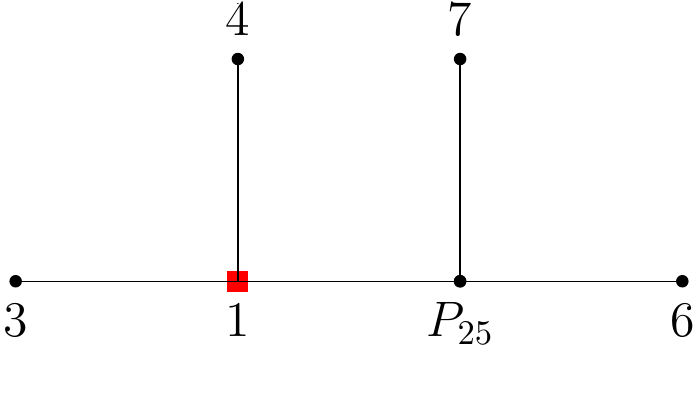} &\includegraphics[width=.25\textwidth]{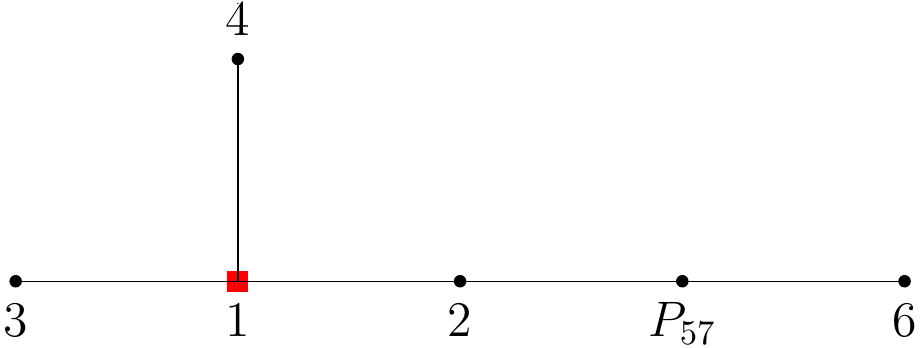} &\includegraphics[width=.20\textwidth]{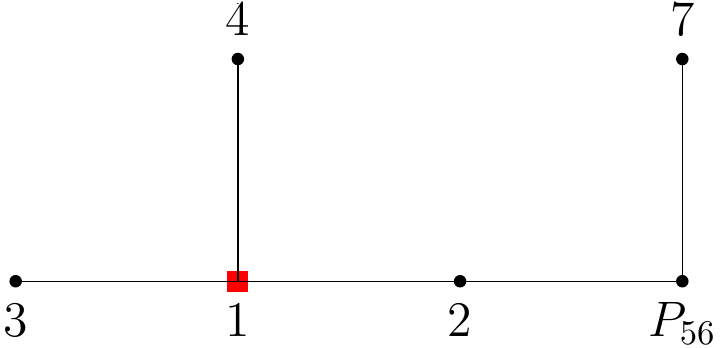}  \\
    \specialrule{0pt}{5pt}{15pt}
    $2-contraction$:&\includegraphics[width=.20\textwidth]{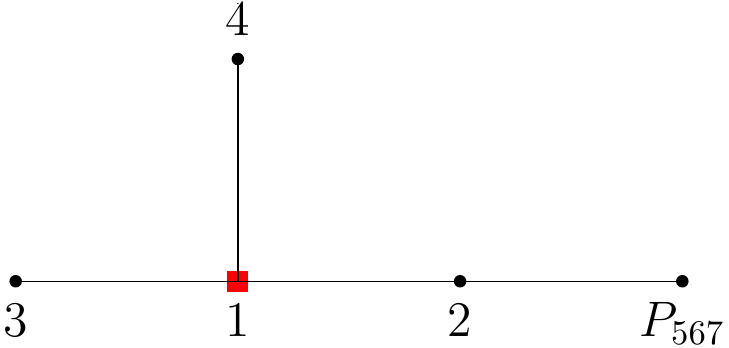} & \includegraphics[width=.20\textwidth]{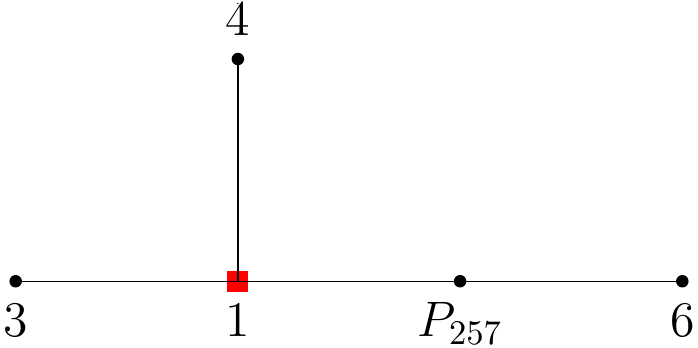}&  \includegraphics[width=.15\textwidth]{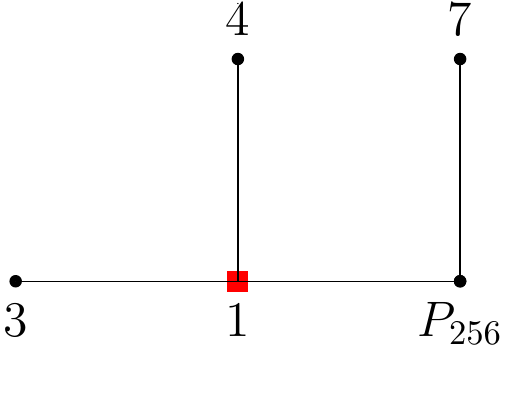}\\
    \specialrule{0pt}{5pt}{15pt}
    $3-contraction$:& & \includegraphics[width=.15\textwidth]{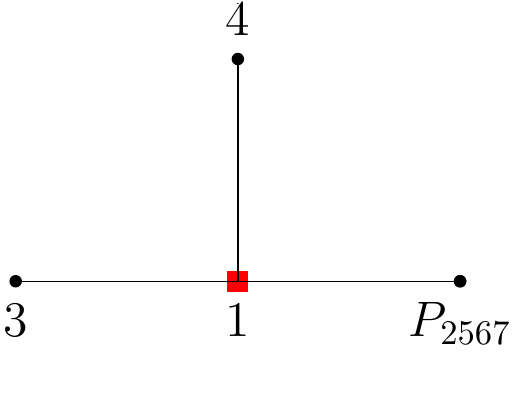} &  \\
    \end{tabular}
  \caption{ The sub-Cayley tree's with various contractions ~~~\label{fig:ano-VP1} }
\end{figure}

If we pick the node $1$ to be the marked point, the subgraphs
related to it will be following three: $\{3\}, \{4\}, \{2,5,6,7\}$.
Only the last one need to consider various contractions. There are
three edges  $\{\{2, 5\}, \{5, 6\}, \{5, 7\}\}$ in this subtree,
thus there are $2^3=8$ different contractions\footnote{The
contraction of each edge is independent of each other. Thus for the
marked point $o$, there are $2^m$ different contractions with
$m=n-2-v_o$ where $v_0$ is the number of edges connecting to the
marked point $o$.}. After contraction and gluing back to node $1$,
we get these eight graphes in the Figure \ref{fig:ano-VP1}. The
corresponding effective $V_P$ vertex \eqref{VPk-gen} can be found in
the Figure \ref{T-ano}.

\begin{figure}[h]
  \centering
  \includegraphics[width=.20\textwidth]{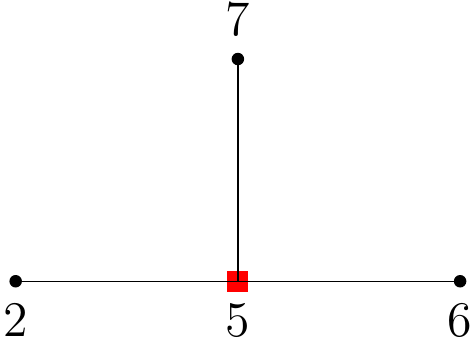}
  \caption{ The last contraction $P_{2567}$.~~~~\label{fig:ano-star}}
\end{figure}

\begin{figure}[h]
  \centering
  \begin{tabular}[t]{m{0.17\linewidth}m{0.24\linewidth}m{0.24\linewidth}m{0.24\linewidth}}
    \specialrule{0pt}{5pt}{15pt}
    $0-contraction$:& & \includegraphics[width=.25\textwidth]{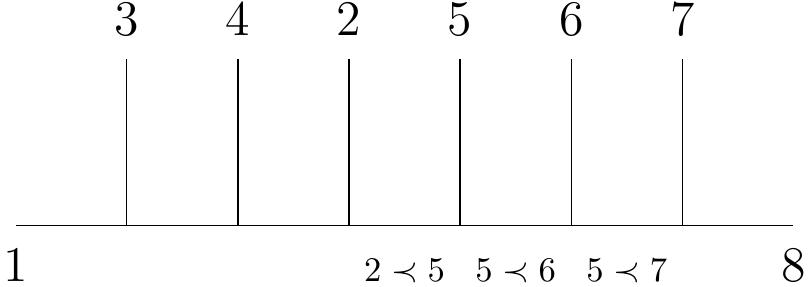} &  \\
    \specialrule{0pt}{5pt}{15pt}
    $1-contraction$:& \includegraphics[width=.25\textwidth]{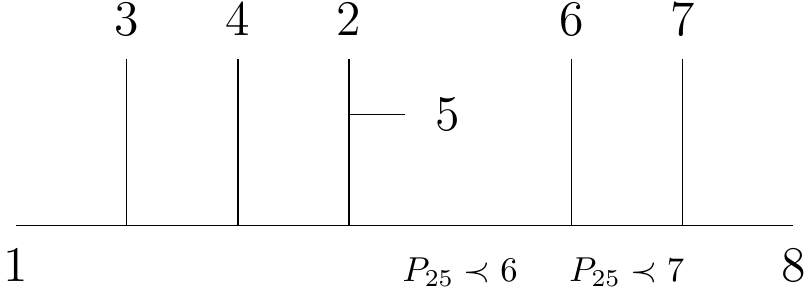} &\includegraphics[width=.25\textwidth]{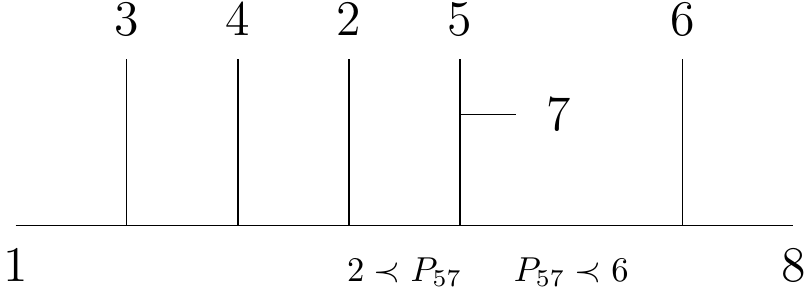} &\includegraphics[width=.25\textwidth]{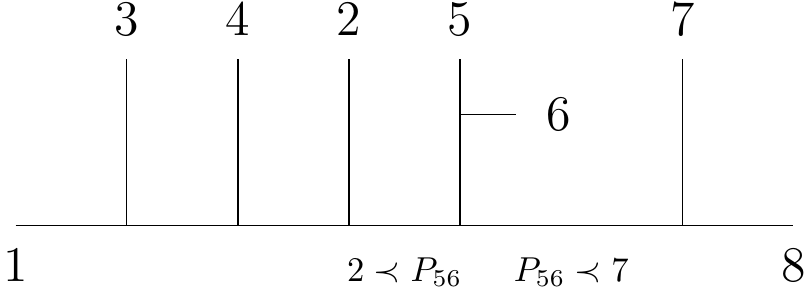}  \\
    \specialrule{0pt}{5pt}{15pt}
    $2-contraction$:&\includegraphics[width=.25\textwidth]{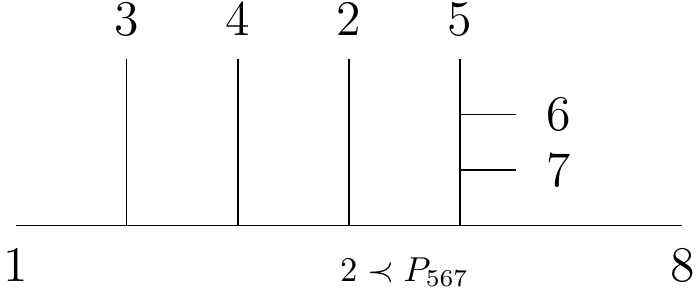} & \includegraphics[width=.25\textwidth]{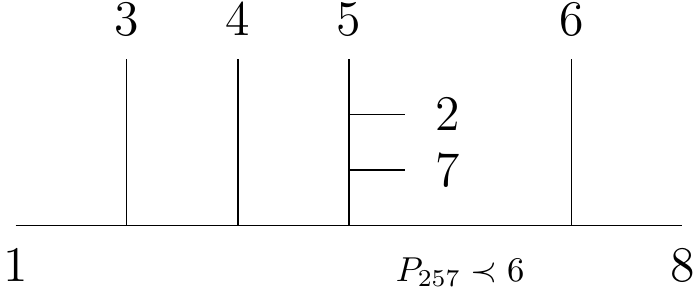}&  \includegraphics[width=.25\textwidth]{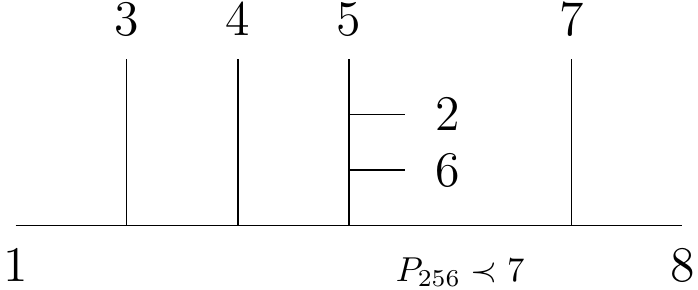}\\
    \specialrule{0pt}{5pt}{15pt}
    $3-contraction$:& & \includegraphics[width=.20\textwidth]{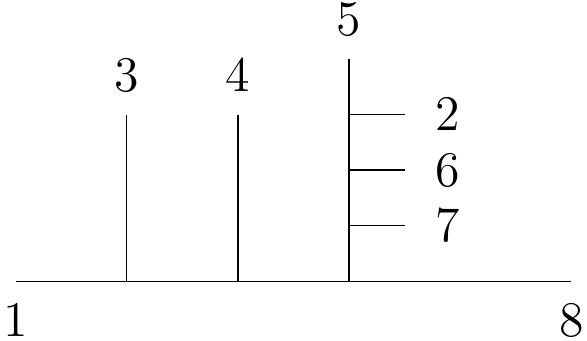} &  \\
    \end{tabular}
  \caption{The full Effective Feynman diagram representation~~~~\label{fig:ano-F} }
\end{figure}

For seven cases with nontrivial contractions in the Figure
\ref{fig:ano-VP1}, six of them are just the Hamiltonian lines, so
the corresponding effective vertexes are just the $V_C$ types
\bea & &
V_C(\{P_{25},2,5\}),~~~V_C(\{P_{57},5,7\}),~~~~V_C(\{P_{56},5,6\}),\nn
& &
V_C(\{P_{567},7,5,6\}),~~~V_C(\{P_{257},2,5,7\}),~~~~V_C(\{P_{256},2,5,6\})\eea
The $V_C$-type vertexes in the second line of above expression can
also written as the $V_P$-type vertexes as given in the Figure
\ref{T-ano}. The last contraction gives a star graph in the Figure
\ref{fig:ano-star}.
Thus the corresponding effective vertex is $V_P(5; 2\shuffle
6\shuffle 7; P_{2567})$.  After putting these two effective vertexes
together, we get the effective Feynman diagrams shown in the Figure
\ref{fig:ano-F}.
The corresponding algebraic expression and the counting for  each
effective Feynman is given in the Figure \ref{T-ano}.
\begin{figure}[h]
  \centering

  \begin{tabular}[t]{m{0.55\linewidth}m{0.27\linewidth}}

  \hline
  \specialrule{0pt}{5pt}{5pt}
  {\rm effective~Feynman~diagram} & {\rm counting} \\
  \specialrule{0.1pt}{5pt}{5pt}

  \specialrule{0pt}{5pt}{15pt}
  $V_P(1; 3 \shuffle 4 \shuffle \{ 2,\{ 5, 6\shuffle7 \} \};8)$ & $\frac{(1+1+4)!}{1!1!4!}\times\frac{(1+1)!}{1!1!}=60$ \\

  \specialrule{0pt}{5pt}{15pt}
  $V_P(1; 3 \shuffle 4 \shuffle \{ P_{25}, 6\shuffle7 \} ;8) \frac{1}{P_{25}} V_{C}(2,5) $ & $\frac{(1+1+3)!}{1!1!3!}\times\frac{(1+1)!}{1!1!}=40$ \\
  \specialrule{0pt}{5pt}{15pt}
  $V_P(1; 3 \shuffle 4 \shuffle \{2, \{P_{57},6\} \} ;8) \frac{1}{P_{57}} V_{C}(5,7) $ & $\frac{(1+1+3)!}{1!1!3!}=20$ \\
  \specialrule{0pt}{5pt}{15pt}
  $V_P(1; 3 \shuffle 4 \shuffle \{2, \{P_{56},7\} \} ;8) \frac{1}{P_{56}} V_{C}(5,6) $ & $\frac{(1+1+3)!}{1!1!3!}=20$ \\

  \specialrule{0pt}{5pt}{15pt}
  $V_P(1; 3 \shuffle 4 \shuffle \{ 2, P_{567} \} ;8) \frac{1}{P_{567}} V_{P}(5;6\shuffle7;P_{567}) $ & $\frac{(1+1+2)!}{1!1!2!}\times\frac{(1+1)!}{1!1!}=24$ \\
  \specialrule{0pt}{5pt}{15pt}
  $V_P(1; 3 \shuffle 4 \shuffle \{P_{257},6 \} ;8) \frac{1}{P_{257}} V_{P}(5;2\shuffle7;P_{257}) $ & $\frac{(1+1+2)!}{1!1!2!}\times\frac{(1+1)!}{1!1!}=24$ \\
  \specialrule{0pt}{5pt}{15pt}
  $V_P(1; 3 \shuffle 4 \shuffle \{P_{256},7 \} ;8) \frac{1}{P_{256}} V_{P}(5;2\shuffle6;P_{256}) $ & $\frac{(1+1+2)!}{1!1!2!}\times\frac{(1+1)!}{1!1!}=24$ \\

  \specialrule{0pt}{5pt}{15pt}
  $V_P(1; 3 \shuffle 4 \shuffle P_{2567} ;8) \frac{1}{P_{2567}} V_{P}(5;2\shuffle6\shuffle7;P_{2567}) $ & $\frac{(1+1+1)!}{1!1!1!}\times\frac{(1+1+1)!}{1!1!1!}=36$ \\
  \specialrule{0.1pt}{15pt}{5pt}
\end{tabular}

\caption{ Effective Feymann
diagrams and their enumaration~~~~~~\label{T-ano} }
\end{figure}
The overall counting for the Cayley tree is  $248$, which is
consistent with the results obtained by recursion or integration rule
 in \cite{baadsgaard2015integration,baadsgaard2015scattering,gao2017labelled}.

Having discussed the effective Feynman diagrams starting from the
node $1$, we can do similar calculations for other nodes. For our
current example, because the symmetry of Cayley tree, all nodes can
be divided into three  categories:
\bea  \{2\} \quad \{1,5\} \quad \{3,4,6,7\} \eea
The second category has been done and now we move to another two
categories. For the first category (i.e., node $2$ as the starting point), there are $2^{6-2}=16$
contractions. Each contraction gives only one effective Feynman
diagram. Thus we have following four group of diagrams. The first
group is
\bea & & V_P(2;\{1,3\shuffle 4\}\shuffle\{5,6\shuffle
7\};8),~~~\#={(3+3)!\over 3! 3!}\times 2\times 2=80 \nn
& & V_P(2;\{1,3\shuffle 4\}\shuffle\{P_{56}, 7\};8){1\over P_{56}^2}
V_C(\{P_{56},5,6\}),~~~\#={(3+2)!\over 3! 2!}\times 2=20 \nn
& & V_P(2;\{1,3\shuffle 4\}\shuffle\{P_{57}, 6\};8){1\over P_{57}^2}
V_C(\{P_{57},5,7\}),~~~\#={(3+2)!\over 3! 2!}\times 2=20 \nn
& & V_P(2;\{1,3\shuffle 4\}\shuffle  P_{567};8){1\over P_{567}^2}
V_C(\{P_{567},7,5,6\}),~~~\#={(3+1)!\over 3! }\times 2\times
2=16~~~~ \eea
The second group is
\bea & & V_P(2;\{P_{13}\shuffle 4\}\shuffle\{5,6\shuffle
7\};8){1\over P_{13}^2} V_C(\{P_{13},1,3\}),~~~\#={(2+3)!\over 2!
3!}\times 2=20 \nn
& & V_P(2;\{P_{13}\shuffle 4\}\shuffle\{P_{56}, 7\};8){1\over
P_{13}^2} V_C(\{P_{13},1,3\}){1\over P_{56}^2}
V_C(\{P_{56},5,6\}),~~~\#={(2+2)!\over 2! 2!}=6 \nn
& & V_P(2;\{P_{13}\shuffle 4\}\shuffle\{P_{57}, 6\};8){1\over
P_{13}^2} V_C(\{P_{13},1,3\}){1\over P_{57}^2}
V_C(\{P_{57},5,7\}),~~~\#={(2+2)!\over 2! 2!}=6 \nn
& & V_P(2;\{P_{13}\shuffle 4\}\shuffle  P_{567};8){1\over P_{13}^2}
V_C(\{P_{13},1,3\}){1\over P_{567}^2}
V_C(\{P_{567},7,5,6\}),~~~\#={(2+1)!\over 2! }\times 2=6 \nn
& &
\eea
The third group is
\bea
& & V_P(2;\{P_{14}\shuffle 3\}\shuffle\{5,6\shuffle
7\};8){1\over P_{14}^2} V_C(\{P_{14},1,4\}),~~~\#={(2+3)!\over 2!
3!}\times 2=20 \nn
& & V_P(2;\{P_{14}\shuffle 3\}\shuffle\{P_{56}, 7\};8){1\over
P_{14}^2} V_C(\{P_{14},1,4\}){1\over P_{56}^2}
V_C(\{P_{56},5,6\}),~~~\#={(2+2)!\over 2! 2!}=6 \nn
& & V_P(2;\{P_{14}\shuffle 3\}\shuffle\{P_{57}, 6\};8){1\over
P_{14}^2} V_C(\{P_{14},1,4\}){1\over P_{57}^2}
V_C(\{P_{57},5,7\}),~~~\#={(2+2)!\over 2! 2!}=6 \nn
& & V_P(2;\{P_{14}\shuffle 3\}\shuffle  P_{567};8){1\over P_{14}^2}
V_C(\{P_{14},1,4\}){1\over P_{567}^2}
V_C(\{P_{567},7,5,6\}),~~~\#={(2+1)!\over 2! }\times 2=6 \nn
& & 
\eea
The fourth group is
\bea
& & V_P(2;P_{134}\shuffle\{5,6\shuffle 7\};8){1\over P_{134}^2}
V_C(\{P_{134},3,1,4\}),~~~\#={(1+3)!\over  3!}\times 2\times 2=16
\nn
& & V_P(2;P_{134}\shuffle\{P_{56}, 7\};8){1\over P_{134}^2}
V_C(\{P_{134},3,1,4\}){1\over P_{56}^2}
V_C(\{P_{56},5,6\}),~~~\#={(1+2)!\over  2!}\times 2=6 \nn
& & V_P(2;P_{134}\shuffle\{P_{57}, 6\};8){1\over P_{134}^2}
V_C(\{P_{134},3,1,4\}){1\over P_{57}^2}
V_C(\{P_{57},5,7\}),~~~\#={(1+2)!\over  2!}\times 2=6 \nn
& & V_P(2;P_{134}\shuffle  P_{567};8){1\over P_{134}^2}
V_C(\{P_{134},3,1,4\}){1\over P_{567}^2}
V_C(\{P_{567},7,5,6\}),~~~\#=(1+1)!\times 2\times 2=8 \nn
& & 
\eea
Adding them up, we get total $248$ cubic Feynman diagrams.

For the third category, for example, the staring node $3$, the
situation  is a bit more complicated. There are $2^5=32$
contractions. But each contraction may give more than one effective
Feynman diagrams, for example, the contractions of  $P_{124567}$ and
$P_{12567}$. If we choose the maximal valency node as the marked
point at each recursive step of the subgraph, we would eventually
conclude that at least $41$ effective Feynman Diagrams are needed to
represent all  Feynman diagram in the third category.

The above discussion tells   us that to get more compact effective
Feynman diagram representation,  we should follow two principles
throughout the whole recursive process:
\begin{itemize}
    \item (1) At each step, select the maximal valency vertex as the starting point.
    \item (2) If there are multiple maximal valency vertexes, choose the one
    that the remaining subgraph have more disconnected branches after removing it and its adjacent edges.
\end{itemize}
The first one guarantees that there are minimal contracted sub-Cayley
trees at each step. The second principle could reduce the number
of Feynman subtrees  for each effective vertex.

\subsection{More examples}

\begin{figure}[h]
  \centering
  \includegraphics[width=.35\textwidth]{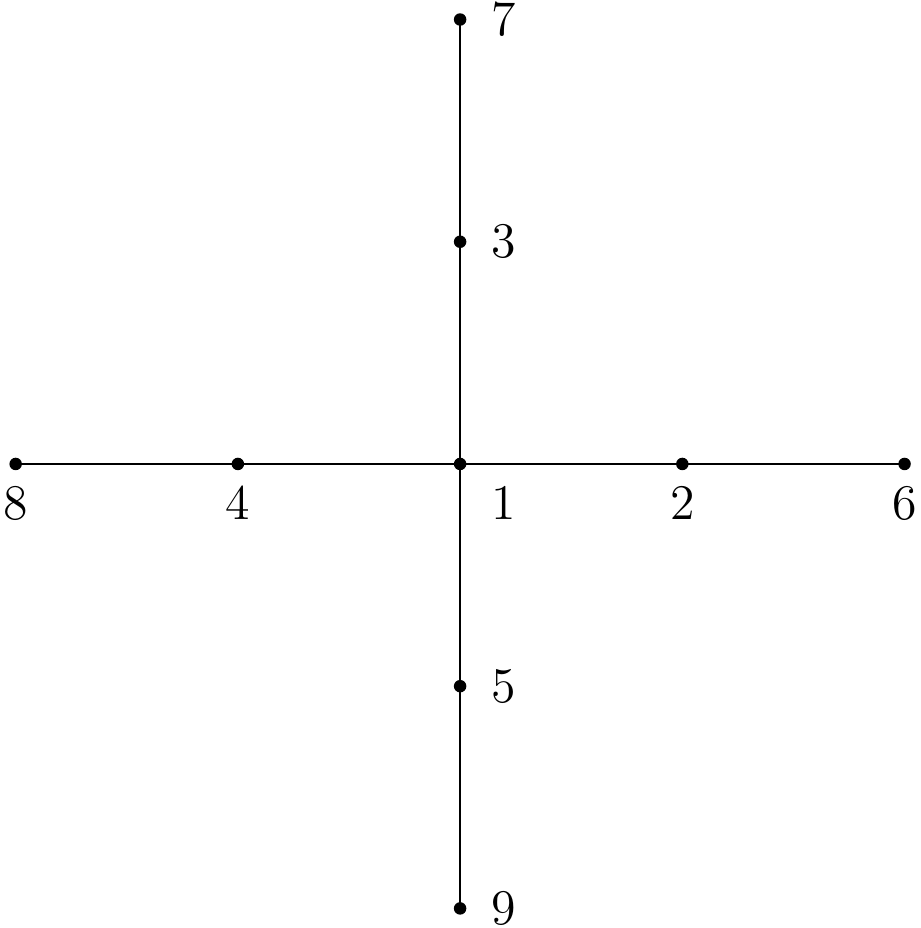}
  \caption{  $C_{10}\{\{1, 2\},\{1, 3\},\{1, 4\}, \{1, 5\}, \{3, 7\}, \{2, 6\},
  \{4, 8\},\{5, 9\}\}$~~~\label{fig:n=10}}
\end{figure}

\begin{figure}[h]
  \centering

  \begin{tabular}[t]{m{0.40\linewidth}m{0.27\linewidth}c}

  \hline
  \specialrule{0pt}{5pt}{5pt}
  {\rm EFD} & {\rm counting} & {\rm $\#$~of~EFD}  \\
  \specialrule{0.1pt}{5pt}{5pt}

  \specialrule{0pt}{5pt}{10pt}
  \includegraphics[width=.35\textwidth]{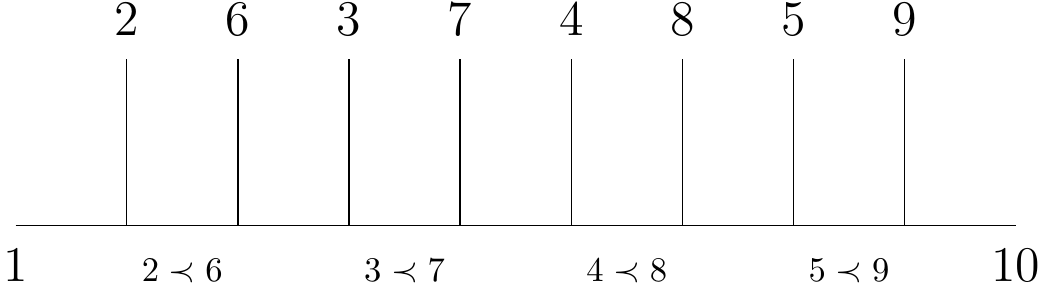} & $\frac{(2+2+2+2)!}{2!2!2!2!}=2520$ & 1\\
  \specialrule{0pt}{5pt}{10pt}
  \includegraphics[width=.35\textwidth]{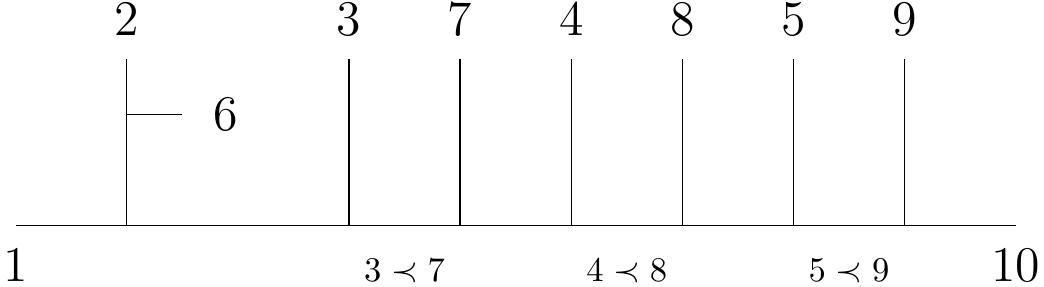} & $\frac{(1+2+2+2)!}{1!2!2!2!}=630$ & 4\\
  \specialrule{0pt}{5pt}{10pt}
  \includegraphics[width=.35\textwidth]{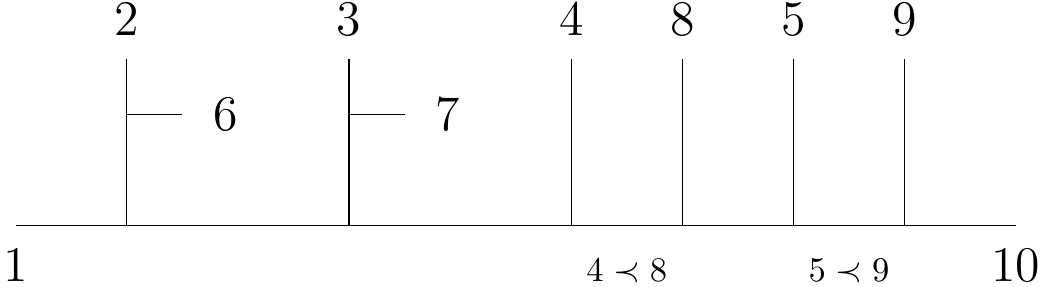} & $\frac{(1+1+2+2)!}{1!1!2!2!}=180$ & 6\\
  \specialrule{0pt}{5pt}{10pt}
  \includegraphics[width=.35\textwidth]{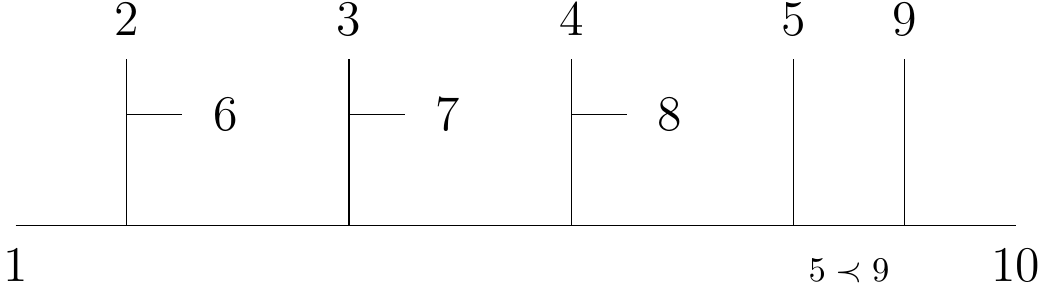} & $\frac{(1+1+1+2)!}{1!1!1!2!}=60$ & 4\\
  \specialrule{0pt}{5pt}{10pt}
  \includegraphics[width=.35\textwidth]{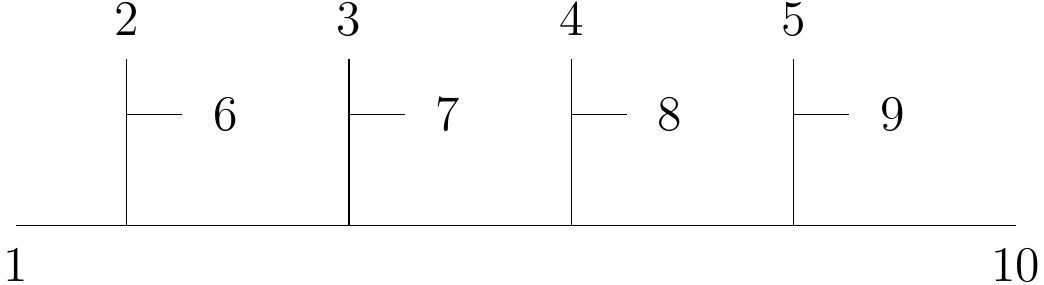} & $\frac{(1+1+1+1)!}{1!1!1!1!}=24$ & 1\\

  \specialrule{0.1pt}{15pt}{5pt}
\end{tabular}

\caption{ Enumerate the Effective Feymann diagrams
~~~\label{fig:n=10-F}}
\end{figure}

In this subsection, we give more examples to show how the effective
Feynman diagrams can very compactly encode all cubic Feynman
diagrams coming from a given Cayley tree, especially this Cayley
tree has symmetric structure. Let us start from a  $n=10$ Cayley
tree given in the Figure \ref{fig:n=10}.
It gives $6384$ cubic Feynman diagrams, which is hard to write all of them down
and get the picture of pole structures. However, with the
symmetric structure, if we choose the node $1$ as the starting
point, we can see that all $2^4=16$ possible contractions give only
$16$ effective Feynman diagrams. Furthermore, these effective
Feynman diagrams can be divided into five types: (1) no contraciton;
(1) one contraction; (3) two contractions; (4) three contractions
(5) four contractions. The typical effective Feynman diagrams and
their  counting have been given in the Figure \ref{fig:n=10-F}.

\begin{figure}[h]
  \centering
  \includegraphics[width=.35\textwidth]{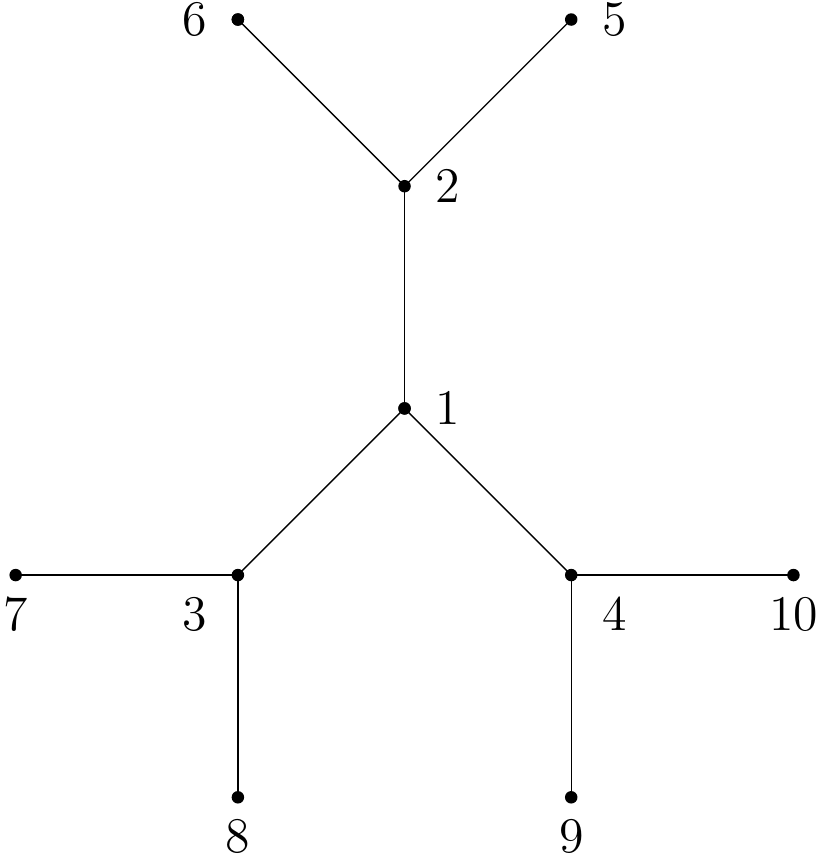}
  \caption{ $C_{11}\{\{1, 2\},\{1, 3\},\{1, 4\}, \{3, 7\}, \{3, 8\},\{2, 5\},
  \{2, 6\},\{4, 9\},\{4, 10\}\}$~~~~\label{fig:n=11} }
\end{figure}

\begin{figure}[h]
  \centering

  \begin{tabular}[t]{m{0.40\linewidth}m{0.40\linewidth}c}

  \hline
  \specialrule{0pt}{5pt}{10pt}
  {\rm EFD} & {\rm counting} & {\rm $\#$~of~EFD} \\\\
  \specialrule{0.1pt}{0pt}{5pt}
  \specialrule{0pt}{5pt}{10pt}
  \includegraphics[width=.35\textwidth]{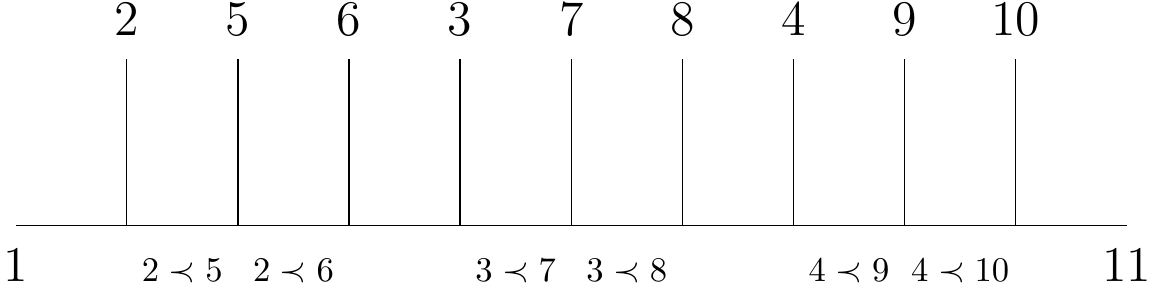} & $\frac{(3+3+3)!}{3!3!3!} \times2!\times2!\times2!=13440$ & 1 \\
  \specialrule{0pt}{5pt}{10pt}
  \includegraphics[width=.35\textwidth]{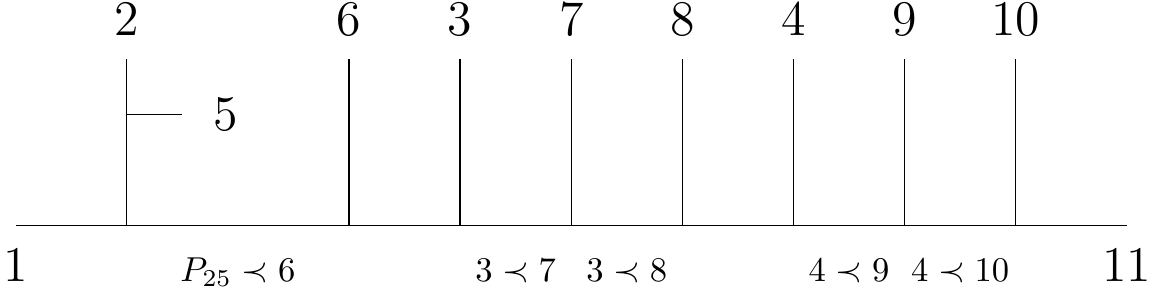} & $\frac{(2+3+3)!}{2!3!3!} \times2!\times2!=2240$ & 6 \\
  \specialrule{0pt}{5pt}{10pt}
  \includegraphics[width=.35\textwidth]{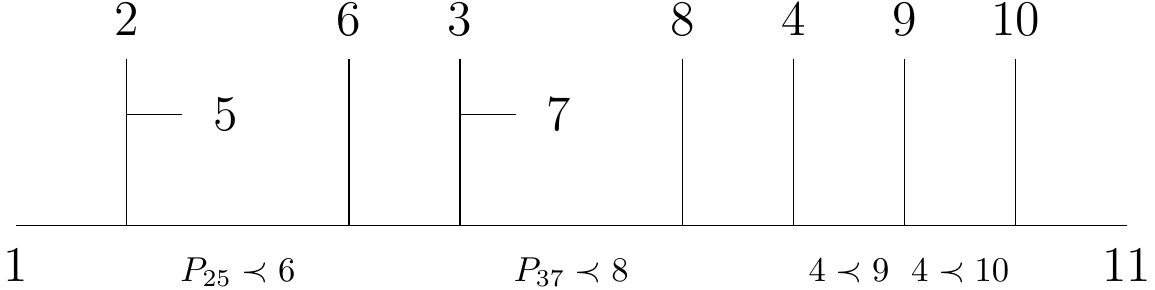} & $\frac{(2+2+3)!}{2!2!3!} \times2!=420$ & 12 \\
  \specialrule{0pt}{5pt}{10pt}
  \includegraphics[width=.35\textwidth]{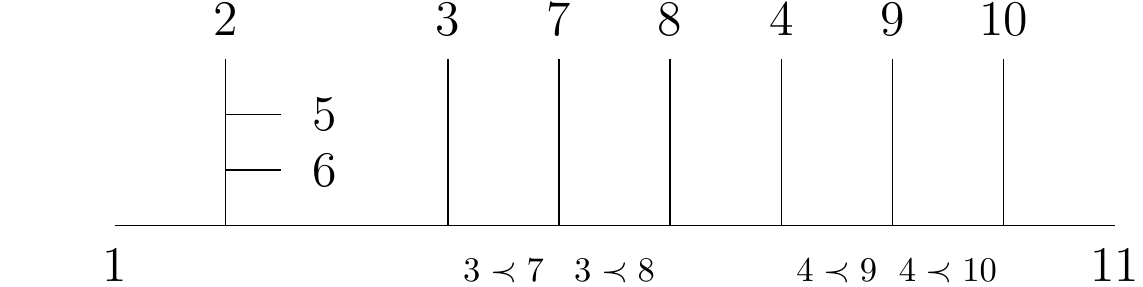} & $\frac{(1+3+3)!}{1!3!3!} \times2!\times2!\times2!=1120$ & 3 \\
  \specialrule{0pt}{5pt}{10pt}
  \includegraphics[width=.35\textwidth]{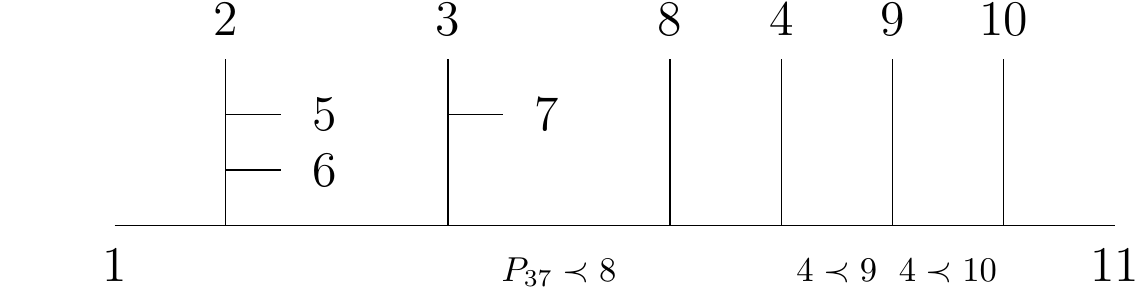} & $\frac{(1+2+3)!}{1!2!3!} \times2!\times2!=240$ & 12 \\
  \specialrule{0pt}{5pt}{10pt}
  \includegraphics[width=.35\textwidth]{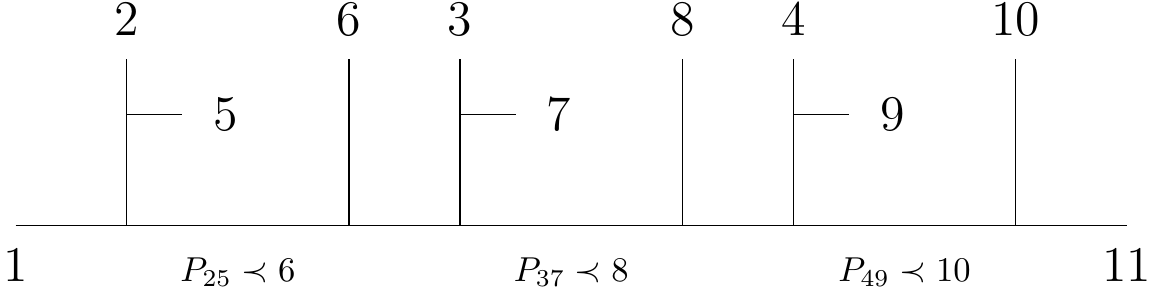} & $\frac{(2+2+2)!}{2!2!2!} =90$ & 8 \\
  \specialrule{0pt}{5pt}{10pt}
  \includegraphics[width=.35\textwidth]{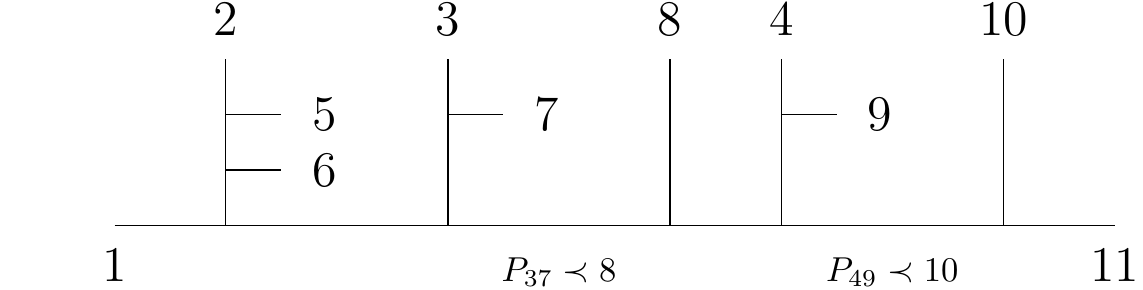} & $\frac{(1+2+2)!}{1!2!2!} \times2!=60$ & 12 \\
  \specialrule{0pt}{5pt}{10pt}
  \includegraphics[width=.35\textwidth]{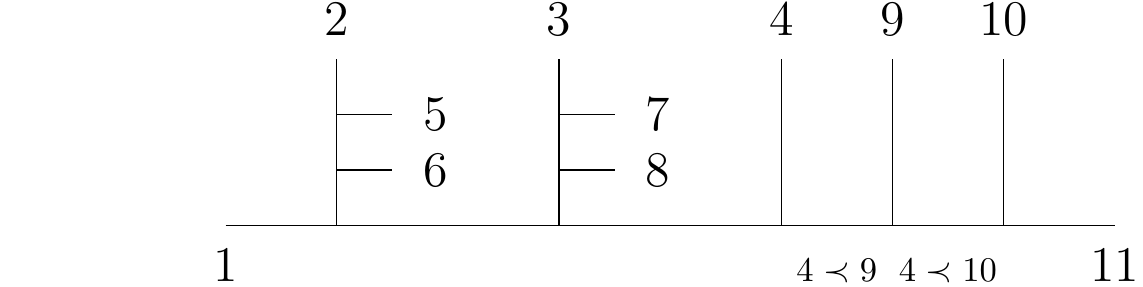} & $\frac{(1+1+3)!}{1!1!3!} \times2!\times2!\times2!=160$ & 3\\
  \specialrule{0pt}{5pt}{10pt}
  \includegraphics[width=.35\textwidth]{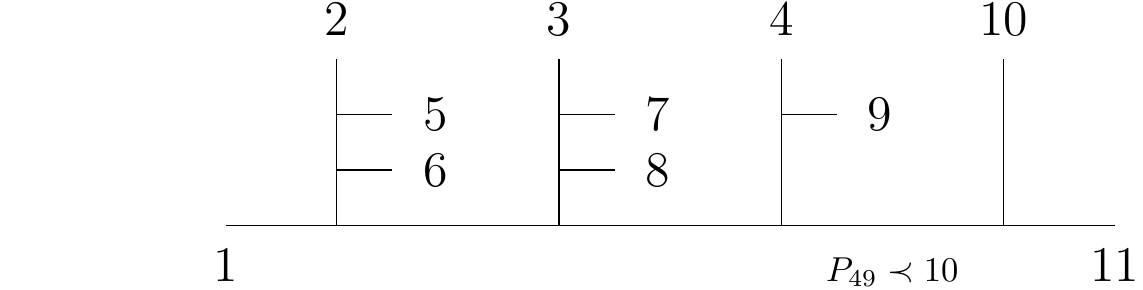} & $\frac{(1+1+2)!}{1!1!2!} \times2!\times2!=48$ & 6 \\
  \specialrule{0pt}{5pt}{10pt}
  \includegraphics[width=.35\textwidth]{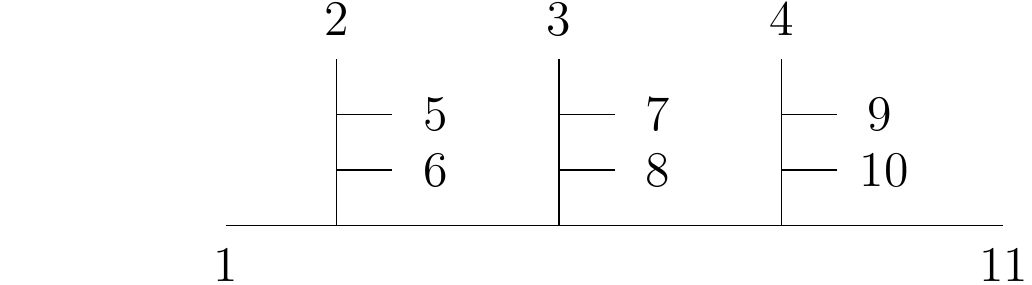} & $\frac{(1+1+1)!}{1!1!1!}\times2!\times2!\times2!=48$ & 1 \\

  \specialrule{0.1pt}{15pt}{5pt}
\end{tabular}

\caption{ Enumerate the Effective Feymann diagrams
~~~~\label{fig:n=11-F}}
\end{figure}

Now let us consider another Cayley tree, which is given in the
Figure \ref{fig:n=11}.
Choosing the node $1$ as the staring marked point, we have $2^6=64$
contractions and $32$ effective Feynman diagrams\footnote{When using the $V_C$
instead of $V_P$ for line structure, we can reduce the number of effective Feynman diagrams.}. These $32$
diagrams can be divided into ten types: (1) one without any
contraction; (2) six of one contraction; (3) twelve of two
contractions at the different branches; (4) three of two
contractions at the same branch; (5) twelve of three contractions at
the two different branches; (6) eight  of three contractions at the
three different branches; (7) twelve of four contractions at the
three different branches; (8) three of four contractions at the two
different branches; (9) six of five contractions at the three
different branches; (10) one of six contractions. The typical
effective Feynman diagrams and their counting are given in the
Figure \ref{fig:n=11-F}. These effective diagrams have coded $40416$
cubic Feynman diagrams.

\clearpage

\subsection{Application of effective Feynman diagrams}

In previous  subsections, we have shown how to use the picture of
effective Feynman diagrams to compactly code pole structures of all
cubic Feynman diagrams coming from a given Cayley tree. In this
subsection, we will show some applications of the new picture.

The first application is to a geometric object, the so called
"Polytope of Feynman diagrams", which is defined for  a collection
of cubic Feynman diagrams of $n$ points  as following
\begin{itemize}

\item (1) Each vertex of this polytope corresponds to a cubic
Feynman diagrams (so there are $(n-3)$ poles).

\item (2) Two vertexes will be connected by an edge when and only
when they share same $(n-4)$ poles.

\item (3) All vertexes on a surface share same $(n-5)$ poles.

\item (4) In general, vertexes of a dimension $r$ surface share same
$(n-3-r)$ poles.

\end{itemize}
Above construction of polytope has used the  {\bf bottom-up}
approach. Our definition of effective vertexes has used an opposite
approach, i.e., the {\bf top-down} method. For example, for the
CHY-integrand $({\rm PT}(\{1,2,...,n\}))^2$, all cubic Feynman
diagrams are represented by a single effective vertex
$V_C(\{1,2,...,n\})$. This single vertex corresponds the
$(n-3)$-dimension polytope, the so called "associahedron".

\begin{figure}[h]
     \quad \quad \includegraphics[width=0.8\textwidth]{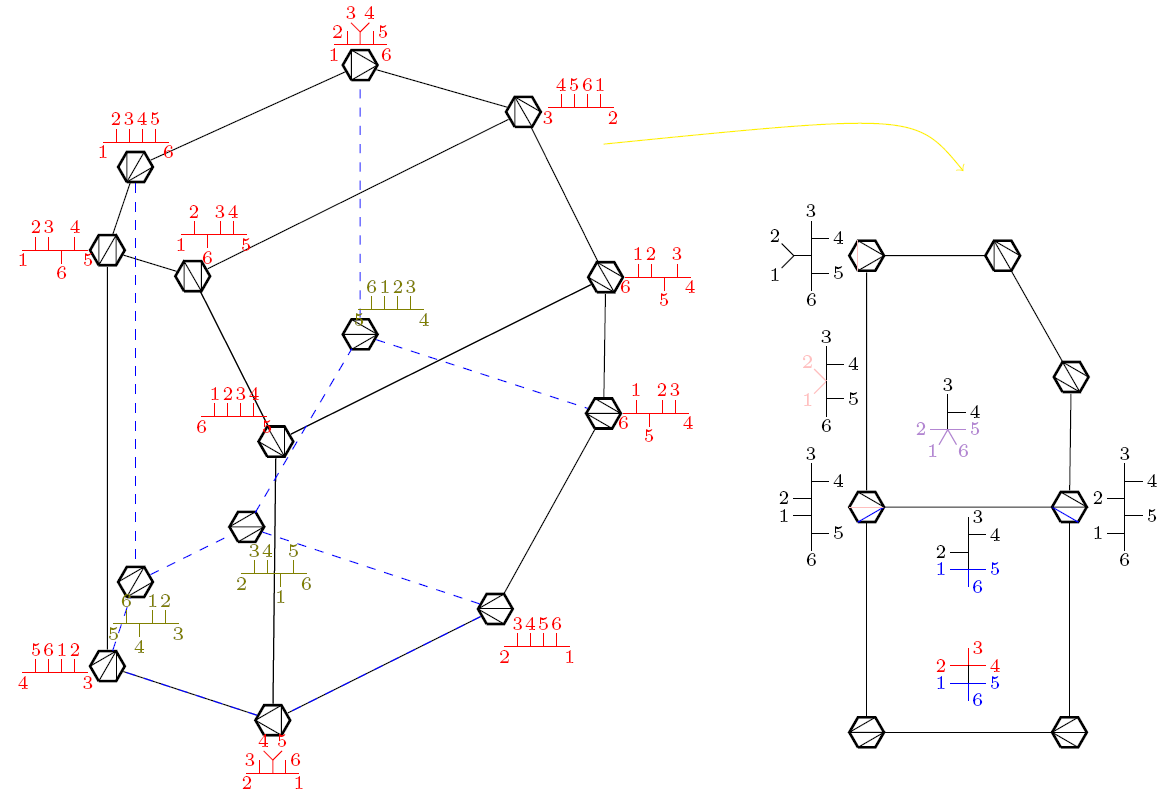}
     \caption{\label{fig:plytop1} Associahedron of $({\rm PT}(\{1,2,3,4,5,6\}))^2$}
 \end{figure}

Its codimension one boundary correspond to fix a given
pole, which corresponds to split the single $V_C$-type effective
vertex to two $V_C$-type effective vertexes connected by this given
pole, i.e,
\bea & & V_C(\{1,2,3,4,5,6\}) \nn &\to & \left\{ \begin{array}{ll}
V_C(\{1,2,P_{12}\}){1\over s_{12}} V_C(\{P_{12},3,4,5,6\}) &~~~
{\rm 6~cases:} s_{12}, s_{23}, s_{34},s_{45},S_{56}, s_{61}\\
V_C(\{1,2,3,P_{123}\}){1\over s_{123}} V_C(\{P_{123},4,5,6\}) &~~~
{\rm 3~cases:} s_{123}, s_{234}, s_{345}
\end{array}\right.~~~~~~\label{PT6-split}\eea
Thus there are $9$ faces. By counting each effective Feynman diagram
in \eqref{PT6-split}, we see that $6$ faces have five edges and five
vertexes while $3$ faces have four edges and four vertexes. We can
split effective vertex further to get the representation of edges
from results in \eqref{PT6-split}\footnote{Such a picture has been discussed in
\cite{huang2018permutation}.}.

The same splitting picture holds for the $V_P$-type effective
vertex. For the Cayley tree given in the Figure \ref{fig:7}, the
whole polytope is given by two effective Feynman diagrams:
\bea F_{A} &= &V_P(2; \{1\}\shuffle\{3\}\shuffle\{P_{45}\};n){1\over
P_{45}^2} V_C(\{4,5,P_{45}\}) \nn
F_B &= & V_P(2; \{1\}\shuffle \{3\}\shuffle \{4,5\};
n)~~~~\label{fig:7-AB}\eea
Since the effective diagram $F_A$ has a fixed pole $s_{45}$, it
defines a two-dimension surface instead of three-dimension volumn.
For the effective diagram $F_B$,
when considering the relative orderings we have following types of
splitting
\bea & & F_B=V_P(2; \{1\}\shuffle \{3\}\shuffle \{4,5\}; n) \nn &\to
&  \left\{ \begin{array}{ll}
V_P(2; \{1\};P_{12}){1\over s_{12}} V_P(P_{12};  \{3\}\shuffle \{4,5\}; n\} &~~~
{\rm 3~cases:} s_{21}, s_{23}, s_{24}\\
V_P(2; \{1\}\shuffle\{3\};P_{123}){1\over s_{123}} V_P(P_{123};  \{4,5\}; n\} &~~~ {\rm 4~cases:}
s_{213}, s_{214}, s_{234}, s_{245} \\
V_P(2; \{1\}\shuffle \{3\}\shuffle \{4\};P_{1234}){1\over
s_{1234}}V_P(P_{1234};\{5\}; n)&~~~ {\rm 3~cases:}s_{2134},
s_{2145}, s_{2345} \end{array} \right.~~~~~~\label{fig:12-split}\eea
Adding together, we find the polytope has  $11$ two-dimension
surfaces. Each surface is defined by an effective Feynman diagram.
Counting the effective Feynman diagrams,  we can find that there are
$4$ surfaces with four edges, $5$ surfaces with five edges and $2$
surfaces with six edges. Which two surfaces share an edge can also
be easily identified.

\begin{figure}
    \quad \quad \includegraphics[width=0.8\textwidth]{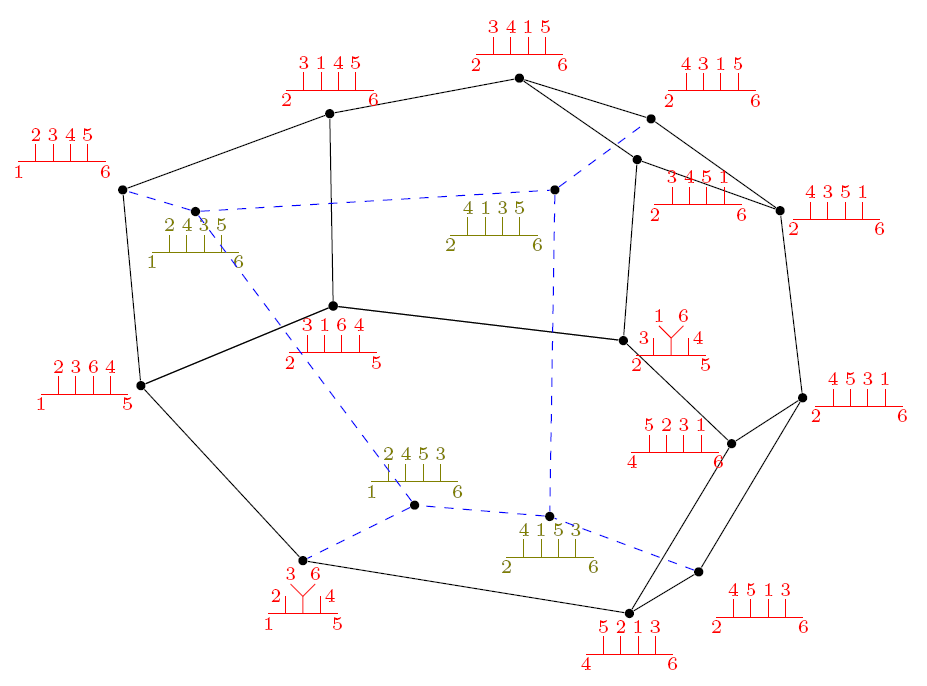}
    \caption{\label{fig:plytop2} The polytope of the Next-to-Star graph with n=6}
\end{figure}

The second application is   for CHY-integrands given by the
multiplication of two different Cayley functions. Again we can draw
the effective Feynman diagrams of these two Cayley functions to find
the common sub-diagrams. Let us consider one example, i.e., two star
graphes of $n=6$, one with the $2$ at the center and another one,
$4$ in the center. To find the common sub-diagrams, it is better to
take the same node as the starting point in the whole construction.
For the star graph with the center of $2$, the effective Feynman
diagram is given by
\bea V_P(2; 1\shuffle 3\shuffle 4\shuffle 5;6)\eea
For the star graph with the center of $4$, when taking the node $2$
as the starting point, we have following effective Feynman diagrams:
\bea & & (0):V_P(2; \{4,1\shuffle 3\shuffle 5\};6); ~~(1):V_P(2;
\{P_{41}, 3\shuffle 5\};6){1\over s_{41}} V_C(\{P_{41},1,4\})\nn
& & V_P(2; \{P_{43}, 1\shuffle 5\};6){1\over s_{43}}
V_C(\{P_{43},3,4\}),~~~V_P(2; \{P_{45}, 3\shuffle 1\};6){1\over
s_{45}} V_C(\{P_{45},5,4\});\nn
& & (2):V_P(2; \{P_{413}, 5\};6){1\over s_{413}}
V_C(\{P_{413},1,4,3\}),~~~V_P(2; \{P_{415}, 3\};6){1\over s_{415}}
V_C(\{P_{415},1,4,5\}),\nn
& & V_P(2; \{P_{435}, 1\};6){1\over s_{435}}
V_C(\{P_{435},3,4,5\});~~~(3):V_P(2; P_{4135};6){1\over
s_{4135}}V_P(4;1\shuffle 3\shuffle 5\};P_{4135})~~~~~~~~\eea
Comparing above two sets of effective Feynman diagrams, we see
immediately the common sub effective Feynman diagram is given
by\footnote{There is a sign to be determined when CHY-integrands
given by the multiplication of two different Cayley functions. It
has been discussed in the paper \cite{cardona2016cross}.}
\bea V_P(2; \{4,1\shuffle 3\shuffle 5\};6)\eea
which contains six cubic Feynman diagrams.

\clearpage

\section{Pick up Poles}
\label{Pick up Poles}

\begin{figure}[h]
    \centering
    \raisebox{-3cm}{\includegraphics[width=.35\textwidth]{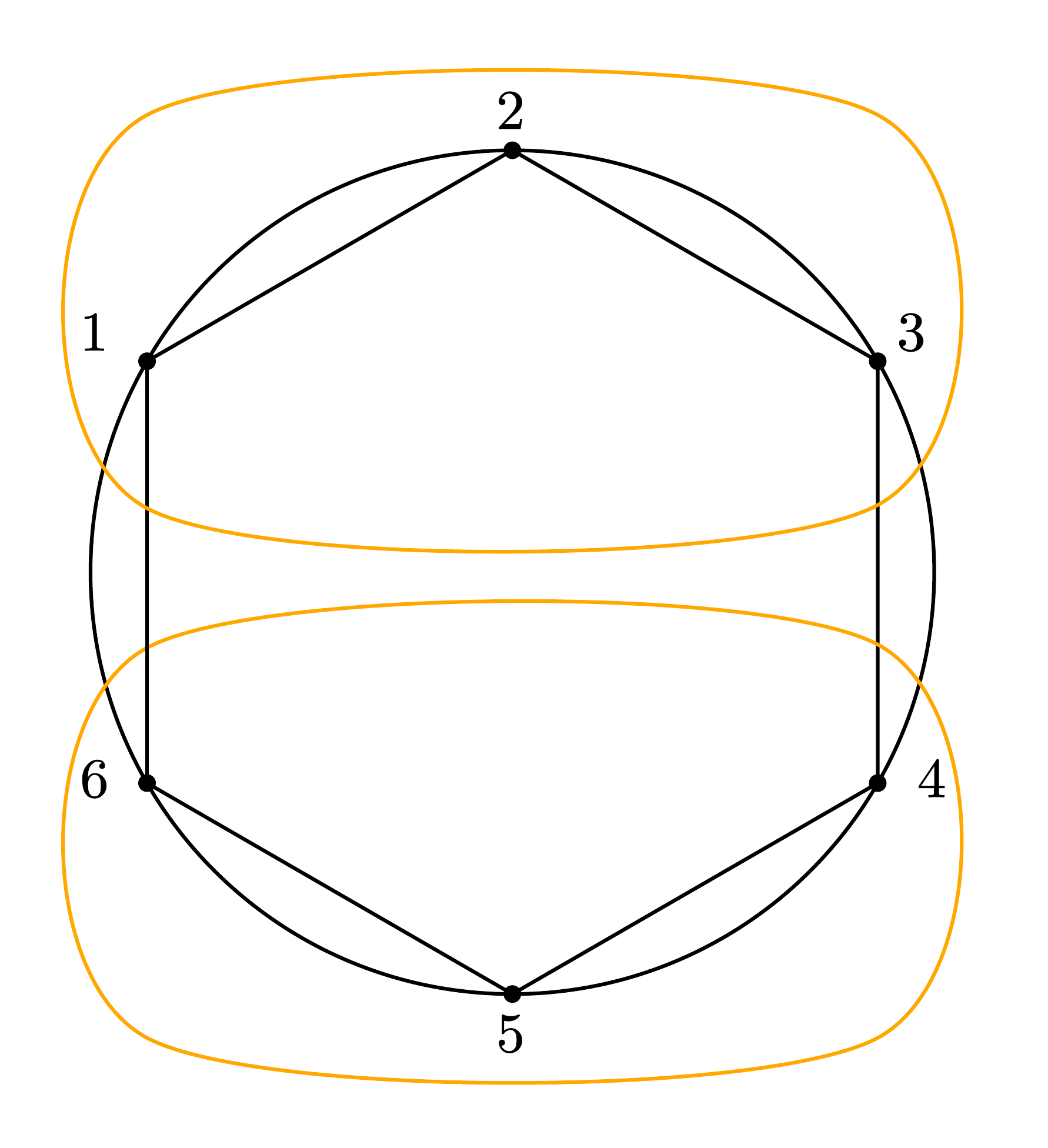}}
    \qquad
    $\Longrightarrow$
    \qquad
    \centering
    \raisebox{-1.2cm}{\includegraphics[width=.25\textwidth]{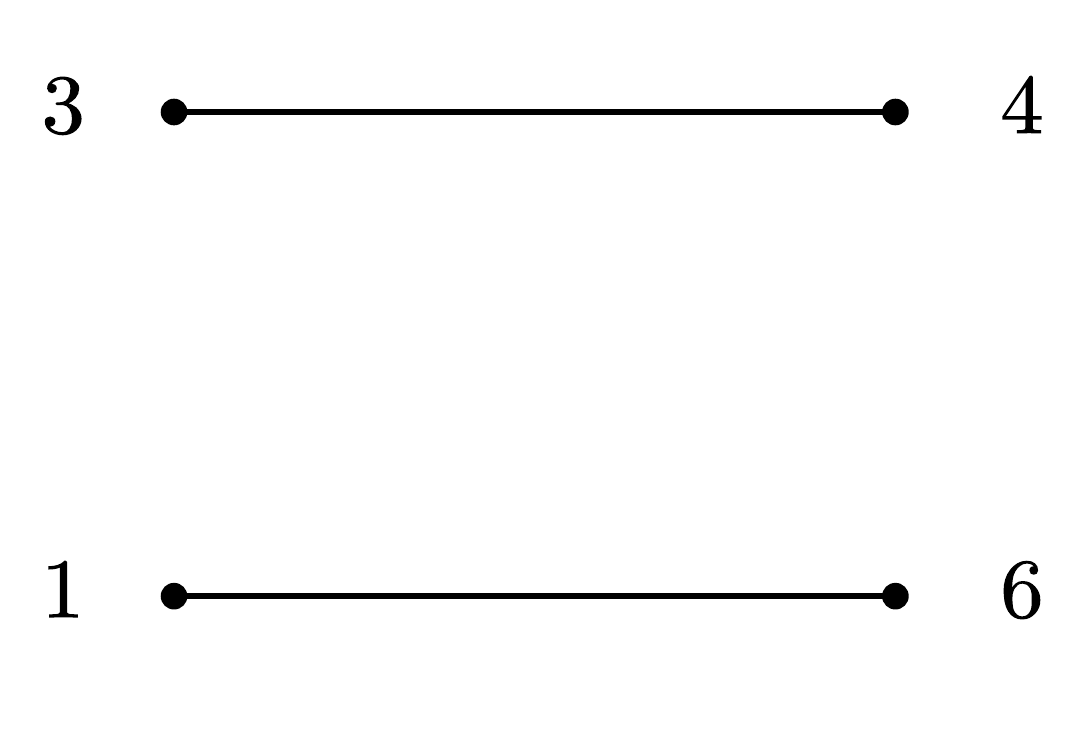}}
    \caption{\label{fig:i}  $Links[\{1,2,3\},\{4,5,6\}]$ of integrand $\frac{1}{z_{12}^{2} z_{23}^{2} z_{34}^{2} z_{45}^{2} z_{56}^{2} z_{61}^{2}}$ }\label{pick3-L.pdf}
\end{figure}

For a bi-adjoint scalar theory defined by two PT-factors, there is a way to extract a subset of all Feynman diagrams containing a particular pole structure. It is given in  \cite{feng2016chy} by using following
 cross-ratio factor
\begin{equation}
    \mathcal{P}^{ac}_{bd}:=\frac{[a c][b d]}{[a d][b c]} \quad, \quad[a b]:=z_{a b}~~~\label{Cross-def}
\end{equation}
To be more explicitly, to pick up a specific pole $\frac{1}{s_A}$ from a given CHY-integrand composed by two
ordered PT-factors, the subset $A$ must be contiguous subset in each PT-factor. Let us focus on just one of
the PT-factor, there $a$ and $b$ to be the first and last elements of the subset $A$ respectively, while $c$ and $d$ are nearest elements of $a$ and $b$ respectively in the complement subset $\overline{A}$. Multiplying the cross-ratio factor \eref{Cross-def} will reduce the index of possible poles containing nodes $a,c$ or nodes $b,d$ through numerators $[a c], [b d]$, while keeping the index of pole $s_A$ invariant. Although one can see that
the denominator $[ad]$ and $[bc]$ from \eref{Cross-def} will increase the index of poles containing nodes $a,d$ or nodes $b,c$, but
since in the original PT-factor, there are no factors $[ad]$ and $[bc]$ in the denominator, their damage is under control. By these observations, one can show that multiplying \eref{Cross-def} will remove all
Feynman diagrams having non-compatible poles, so we are left with diagrams all having  the pole $s_A$.
For example, with the CHY-integrand
   $ \frac{1}{z_{12}^{2} z_{23}^{2} z_{34}^{2} z_{45}^{2} z_{56}^{2} z_{61}^{2}}
$
 we produce following fourteen Feynmann diagrams up to a sign
\begin{equation}
  \begin{aligned}
  & \frac{1}{s_{12} s_{34} s_{56}}  +\frac{1}{s_{12} s_{56} s_{123}}+\frac{1}{s_{23}s_{56} s_{123}}+\frac{1}{s_{12} s_{34} s_{126}}+\frac{1}{s_{16} s_{34}s_{126}}+\\
  & \frac{1}{s_{16} s_{23} s_{156}} +\frac{1}{s_{16} s_{34}s_{156}}+\frac{1}{s_{23} s_{56} s_{156}}+\frac{1}{s_{34} s_{56}s_{156}}+\frac{1}{s_{16} s_{23} s_{45}}+\\
  & \frac{1}{s_{12} s_{123}s_{45}}  +\frac{1}{s_{23} s_{123} s_{45}}+\frac{1}{s_{12} s_{126}s_{45}}+\frac{1}{s_{16} s_{126} s_{45}}~.
  \end{aligned}
\end{equation}
To pick out all  items containing $s_{123}$, first we  split the whole set into the pole set $A=\{1,2,3\}$ and its complement $\O A=\{4,5,6\}$ as in the Figure \ref{pick3-L.pdf}.  We see that  there are
lines   connecting subsets $A$ and $\O A$. For later convenience, we will define the
set ${\bf{\rm Links}}[A, \O A]$ as the collections of lines connecting the set $A$ and $\O A$. Each line will be represented by two nodes: one is in $A$ and another one, in $\O A$. Furthermore, we should distinguish the solid line(corresponding the factor $[ab]$ in the denominator) and dashed line(corresponding the factor $[ab]$ in the numerator) by $\underline{\{a,b\}}$ and $\overline{\{a,b\}}$ respectively. Using this notation, we have
${\rm Links}[\{1,2,3\},\{4,5,6\}]=\{ \underline{\{1,6\}},\underline{\{3,4\}}\}$. Now using the pair in the Links set, we can construct  a single cross-ratio
$\mathcal{P}^{16}_{34}=\frac{z_{16} z_{34}}{z_{14} z_{36}}$.

One can easily check that by multiplying such cross-ratio factor to the original CHY-integrand
\begin{equation}
    \frac{1}{z_{12}^{2} z_{23}^{2} z_{34}^{2} z_{45}^{2} z_{56}^{2} z_{61}^{2}} \mathcal{P}^{16}_{34}=
    \frac{1}{z_{12}^{2} z_{14} z_{16} z_{23}^{2} z_{34} z_{36} z_{45}^{2} z_{56}^{2}}
\end{equation}
we produce only four terms all containing the pole $s_{123}$
\begin{equation}
\frac{1}{s_{12} s_{45} s_{123}}+\frac{1}{s_{23} s_{45} s_{123}}+\frac{1}{s_{12}
   s_{56} s_{123}}+\frac{1}{s_{23} s_{56} s_{123}}
\end{equation}

We are interesting in this problem is because in many situations, such as the soft limit and collinear limit, we want to know the singular behavior of a given amplitude. These singular behaviors are connected with particular poles, thus how to isolating contributions from these poles becomes important to many studies. In \cite{feng2016chy}, by removing these singular contributions, the two-loop CHY-integrand of the planar bi-adjoing scalar theory has been constructed. In \cite{huang2018permutation}, the technique of picking poles has been  used  to study the symmetry properties of different PT-integrands.
In \cite{feng2020one}, the same technique has been used to the contraction of the one-loop CHY-integrand for general
bi-adjoing scalar theory.

In these mentioned applications, the pole picking is constraint to the CHY-integrands of two PT-factors. However,
as presented in \cite{gao2017labelled}, a large class defined by the "labelled tree", which is the natural generalization of
PT-factor, has been introduced. Thus it is curious that if the picking pole technique can be generalized to
these more general situations. In this section, we would explore the possibility. We will give an algorithm to pick up a particular pole for the most general CHY-integrands, which do not contain any higher order pole.
We must emphasize, unlike the case of PT-factors, where the picking pole algorithm can be rigorously proved, the
general algorithm present in this section is based on many tested examples and we could not give the proof

Although we have talked only picking a given pole in this section, in fact, the procedure can continue to pick up more and more poles. When multiplying the cross-ratio factor, we get a new CHY-integrand and then from it, we continue to pick up another new pole. In other words, by iterating our algorithm, one can
 pick up a series of poles as long as they are compatible .

Another interesting point of picking pole technique is following. In the previous part, we have talked about the
effective Feynman diagrams and the corresponding geometric picture, i.e., polytope.
For the square Cayley integrand, all  Feynman diagrams constitute a high-dimensional polytope.
The process of picking out a particular pole corresponds precisely to the operation of projecting from a high-dimensional volume onto a specific face.

\subsection{The General algorithm}

Having discussed our purpose in this section, let us elaborate how to attack the problem.
Unlike the bi-adjoint scalar theory, the general CHY-integrand of weight four contain both denominators and
numerators and in general, amplitudes will depend on both, so the constructed cross-ratio factor of picking pole
should depend on both too. Another important fact is that the index of a given pole is determined by
links inside these nodes, thus the cross-ratio factor should not affect the index, which means lines used in the
construction should come from  these lines connecting the set $A$ and its complement $\O A$. With this reasoning,
one can see why we define the linking set ${\rm Links}[A, \O A]$ in previous paragraphes.

Knowing the linking set ${\rm Links}[A, \O A]$, in principle we can construct a cross-ratio factor like those in
\eref{Cross-def} for any two linking lines, which contain four nodes\footnote{If two nodes are the same point,
the cross-ratio factor is automatically one. }. However, in general case, there are two types of linking lines
(i.e., the solid line representing the factor in denominator and the dashed line representing the factor in the numerator), it is
understandable their roles should be different. Thus when we construct the cross-ratio factor, we should put their role into count. We will meet three situations. In the first situation where both linking lines are solid lines,
for example, $\underline{\{a,c\}}$ and $\underline{\{b,d\}}$,  we can use  \eref{Cross-def} to define
the cross-ratio factor by our experience from the bi-adjoint scalar theory. For later convenience, we will call them {\bf pure primary cross-ratio factor}. In the second situation where both linking lines are dashed lines, for example, $\overline{\{a,c\}}$ and $\overline{\{b,d\}}$, the right definition of  cross-ratio factor is not so straightforward. Naively one can use the definition \eref{Cross-def}, but now the role of solid lines and dashed lines are same. In fact, there is another definition, i.e., the inverse of \eref{Cross-def}, which we will write as
\begin{equation}
     \bar{\mathcal{P}}^{ac}_{bd} \equiv \frac{z_{ad} z_{bc}}{z_{ac} z_{bd}}=\left({\mathcal{P}}^{ac}_{bd}\right)^{-1}
    ~~~\label{Def-dcs}
\end{equation}
In the new definition \eref{Def-dcs}, the linking lines $\overline{\{a,c\}}$ and $\overline{\{b,d\}}$ are in denominator, so the difference between two kinds of lines is manifest. Although we feel the definition
\eref{Def-dcs} is better, for all examples studied by us according to the algorithm presented below, the cross-ratio factor \eref{Def-dcs} will not appear.

The third situation is most tricky one, where one linking line is solid and another one, dashed, for example,
$\overline{\{a,b\}}$ and $\underline{\{e,f\}}$. Now, which definition, i.e., \eref{Cross-def} and \eref{Def-dcs},
we should use. In fact, in both definitions, the role of lines are same, which is not we want. We want an expression where the factor $[ab]$ is in the numerator while the factor $[ef]$ is in the denominator. However, using this pair, it is impossible to reach the goal. A way to solve the difficulty is to involve another linking line, for example, $\overline{\{c,d\}}$ and define following combination
\bea \mathcal{P}^{\overline{ab}}_{\underline{ef};\overline{cd}}\equiv \mathcal{P}^{{ab}}_{cd}\bar{\mathcal{P}}^{cd}_{{ef}}=\frac{z_{{ab}} z_{cd}}{z_{ad} z_{cb}}  \frac{z_{ed} z_{cf}}{z_{{ef}} z_{cd}}= \frac{z_{{ab}} z_{ed} z_{cf}}{z_{{ef}} z_{ad} z_{cb}}
    ~~~\label{Def-mcs1}\eea
We will call $\mathcal{P}^{\overline{ab}}_{\underline{ef};\overline{cd}}$ the {\bf mixed primary cross-ratio factor}. There is another possibility, i.e., involving $\underline{\{c,d\}}$. Using this choice, we will get the inverse of
\eref{Def-mcs1} essentially. The reason we define \eref{Def-mcs1} is that it does appear in the construction
as shown in later examples. We must point out, although the line $\overline{\{c,d\}}$ acts as an intermediate variable, we do need to impose the condition $a≠c,~ e≠c$ and $b≠d,~ f≠d$.

Up to now, we have construct the (pure/mixed) primary cross-ratio factors, i.e., the \eref{Cross-def}, \eref{Def-mcs1} and their inverses, by picking two or three lines from the linking set ${\rm Links}[A, \O A]$.
For general CHY-integrands, there will be many linking lines and we will have many primary cross-ratio factors.
To pick up  a given pole, we need to determine how to combine them. Based on many examples studied by us, now we present our algorithm in the following:
\begin{itemize}

\item (1) Given a CHY-integrand, draw the corresponding 4-regular graph, where the factor in the denominator is represented by solid line while the factor in the numerator is represented by dashed line.

\item (2) To pick up the pole $s_A$, we divide all nodes of the graph into two subsets, i.e., the $A$ and its complement $\O A$. Now there are lines connecting the subset $A$ and $\O A$. Collecting them (by removing duplications) we form the linking set ${\rm Links}[A, \O A]$.

\item (3) From the  linking set ${\rm Links}[A, \O A]$, we generate four collections of primary cross-ratio factors: (I) using two solid lines according to the formula \eref{Cross-def}; (II) using two dashed lines according to the formula \eref{Def-dcs}; (III) Using two solid lines and one dashed line according to the formula \eref{Def-mcs1}. We want to emphasize that in the definition \eref{Def-mcs1}, the role of two solid lines are different, so we should construct two primary cross-ratio factors; (IV) Using two dashed lines and one solid line according to the inverse of the formula \eref{Def-mcs1}.

\item (4) Now we use above four collections to construct the wanted cross-ratio factor, which
picks up a particular pole. We start from the type (I) only. Assuming there are $N_I$ primary
cross-ratio factors, we multiply them together to  form the initial test pick-factor $\bm{\mathcal{P}}_0$. Now we need to have some criterions to see if it is the right answer. The first criterion is that
 \begin{itemize}
    \item  \textbf{Criterion I:} The new CHY-integrand, i.e, the multiplication of original CHY-integrand and the pick-factor, should not contain any new poles or higher order poles comparing with  the original CHY-integrand.
    \end{itemize}
It is worth to emphasize that when we say it does not create new poles or higher poles, we are just calculate the {\bf  pole index} of a subset $A_i$ as defined in \eref{index-1}
 If the $\bm{\mathcal{P}}_0$ satisfies the Criterion I, it is the right answer (see examples given in
\ref{ss1}).

\item (5) If the $\bm{\mathcal{P}}_0$ in the above item does not satisfy the Criterion I, we should consider factors obtained by removing one primary cross-ratio factor from $\bm{\mathcal{P}}_0$. In other words, among $N_I$ primary
cross-ratio factors in the collection (I), we choose arbitrary $(N_I-1)$ of them and multiply them together. There will be $N_I$ different choices. For each choice, we check if it satisfies the Criterion I. If it violates the Criterion I, we just discard it. Otherwise, we keep it.

Assuming that there are $1\leq m_1\leq N_I$ left combinations in the list, now we need to impose the second criterion:
  \begin{itemize}
    \item  \textbf{Criterion II:}  The pick-factor should remove all incompatible poles of the original integrand.
    \end{itemize}
More explicitly, for each pick-factor in the list, we calculate the remaining poles (by just
calculating the pole index) after
multiplying it to the original CHY-integrand. From it, we can find the kept poles for each pick-factor. We claim that the right pick-factors are those that all incompatible poles have been removed. There may be several combinations satisfying the Criterion II. Each one will be the right choice of pick-factors (see examples in \ref{ss2}).

\item (6) If there is no combination satisfying the Criterion I by removing just one primary cross-ratio factor from $\bm{\mathcal{P}}_0$, i.e., $m_1=0$ in previous item. We continue to consider the situation by removing two primary cross-ratio factors from $\bm{\mathcal{P}}_0$. Again we consider various allowed combinations and check them with the Criterion I. If there are combinations left after this checking, we impose the Criterion II on these left combination.  If there is no combination after this checking, we continue to remove
    three primary cross-ratio factors from $\bm{\mathcal{P}}_0$ and checking with Criterion I and II.

    Now one sees the iterating procedure. At the end with primary cross-ratio factors in the collection (I), we will face two situations: either finding the right pick-factor according to our algorithm (see examples in \ref{ss3}), or either there is no combination satisfying both Criterions.

\item (7) When we can not find the right pick-factor just from the collection (I), we need to enlarge our pool by including the collection (III). For the enlarged collection, we repeat
    the procedure from the item (4) to item (6). However, since in general the number of element in the collection (III) is large and the final answer contains very small number of factors from the collection (III), based on experience with many examples we find it is more effective doing following way:
    \begin{itemize}

    \item (7a) First we collect all cross-ratio factors obtained from the multiplication of the primary cross-ratio factors in the collection (I), which   satisfy the Criterion I\footnote{It is possible that the combination in collection (I), which does not satisfy the Criterion I, will give the right pick-factor when multiplying some factors from the collection (III). The more effective algorithm presented here has overlooked this possibility. Thus if it does not work, we should go back to the more complete algorithm, i.e., we repeat
    the procedure from the item (4) to item (6) for the collections (I) plus (III).  }. We denote them as ${\cal T}_{i;k}$ where $i$ is the number of primary cross-ratio factors in the multiplication and $k$ distinguishes different combinations with same number $i$. The allowed choices of $k$ will be denoted as $N_{i}$.

    \item (7b) Secondly, we denote elements in the collection (III) by $a_i, i=1,...,M_{III}$.

    \item (7c) Now we consider following combination
    \bea {\cal T}_{i;k} \prod_{1\leq i_1<...<i_{m}\leq M_{III}} a_{i_1}...a_{i_m}~~~\label{rule-7b}\eea
    We check cross-ratio factor by the Criterion I and II in \eref{rule-7b} according to following ordering. When searching through the combination \eref{rule-7b}, there are four variables. The first one is  the changing of the index $i$. The second one is  the changing of the index $k$. The third one is  the number $m$ and the fourth one is the different choices of a given number $m$. In other words, there are four nested loops in the searching and one can arrange the searching orderings of four variables according to his preference.

    Our choice is to start from the smallest number $m=1$ and the largest number $i$, then we change $k=1,...,N_i$ and different $i_1$. If there are combinations satisfy both criterions, we find the right pick-factor. If there is no combinations satisfy both, we go to the next largest number $i$ and search for all combinations of $k$ and $i_1$. If after going through all $i$ with fixed $m=1$ we do not find the right pick-factor, we move to the case $m=2$ and start from largest $i$ to smallest $i$ again. In other words, for the four nested loops. The outmost loop is $m=1$ to $m= M_{III}$. The next loop is from largest $i$ to smallest $i$. The third loop is the different choice of $k$ and the innermost loop is different combinations of $a_{i_1}...a_{i_m}$ with given $m$.

    \end{itemize}

\item (8) Although by our explicit examples, there is no need to include the collections (II) and (IV), but when the step (4) to step (7) fail, this possibility should be considered.

\end{itemize}

Having presented the algorithm above, let us give the motivations
of the construction without the  rigorous proof. First, since all primary cross-ratio factors are constructed from links between $A$ and $\O A$,  multiplying them to the original CHY-integrand will not change the linking number of the subset $A$ and any other subsets $B\subset A$. In other words, any compatible pole of $s_A$ in the original theory will be kept. A good point is that  at the same time, these cross-ratio factors will remove some incompatible poles. Naively, since each factor will remove part of
incompatible poles, multiplying them together will remove all incompatible poles. However, while
getting rid old incompatible poles, it could produce new incompatible poles. Thus we should impose
the Criterion I. Furthermore, if we do not multiply all factor together, it will have the chance that some
incompatible poles are left, thus we need to impose the Criterion II to remove this possibility (see example in the Figure \ref{table1}).

Before ending the subsection, let us give some technical remarks. As in general discussion in the previous paragraph, to pick a given pole, we need to remove all incompatible poles by multiplying some cross-ratio factors while not add new poles (i.e., the Criterion I and II). To reach the goal, some understanding of primary cross-ratio factors is very useful. For the primary factor in the collection (I) and (II), the general form is
\bea {[ab][cd]\over [ad][cb]}\eea
From this form, we see that this factor will remove poles containing nodes in any one of following two groups
\bea \{a,b\};~~~~\{c,d\}~~~\label{Cha-1}\eea
and has the potential to add new poles containing containing nodes in any one of following two groups
\bea \{a,d\};~~~~\{c,b\}~~~\label{Cha-2}\eea
For the primary factor in the collection (III) and (IV), the general form is
\bea {[ab][cf][ed]\over [ad][cb][ef]}\eea
From this form, we see that this factor will remove poles containing  nodes in any one of following six groups
\bea \{a,b\};~~ \{c,f\}; ~~\{e,d\};~~ \{a,b,c,f\};~~\{a,b,e,d\};~~
\{c,f,e,d\}~~~\label{Cha-3} \eea
while has the potential to add new poles containing in any one of following six groups
\bea \{a,d\};~~ \{c,b\}; ~~\{e,f\};~~ \{a,d,c,b\};~~\{a,d,e,f\};~~
\{c,b,e,f\} ~~~\label{Cha-4}\eea
Using information \eref{Cha-1}, \eref{Cha-2}, \eref{Cha-3} and  \eref{Cha-4}, we can speed up the check with the Criterion I and II through our algorithm laid above.

\subsection{Examples}

In this part, we will present various examples to demonstrate our algorithm.

\subsubsection{The simplest case  \label{ss1}}

The simplest case is that we have used only the collection (I) and multiplications of all its factors satisfies the Criterion I. As we have claimed, the obtained cross-ratio factor is the right answer.  We will present two examples. The first example is the Hamilton star graph
\begin{equation}
    \mathcal{I}^{star}_{6}= \frac{z_{16}^{4}}{z_{12}^{2} z_{13}^{2} z_{14}^{2} z_{15}^{2} z_{26}^{2} z_{36}^{2} z_{46}^{2} z_{56}^{2}}
\end{equation}
It gives a basic effective vertex as discussed in the first part of the paper. This effective vertex can be expanded to $4!=24$ cubic Feynman diagrsms
\begin{equation}
    \begin{aligned}
       & \frac{1}{s_{12}s_{36}s_{124}}+\frac{1}{s_{14}s_{36}s_{124}}+\frac{1}{s_{12}s_{56}s_{124}}+\frac{1}{s_{14}s_{56}s_{124}}+\frac{1}{s_{12}s_{36}s_{125}}+\\
       & \frac{1}{s_{15}s_{36}s_{125}}+\frac{1}{s_{12}s_{46}s_{125}}+\frac{1}{s_{15}s_{46}s_{125}}+\frac{1}{s_{13}s_{26}s_{134}}+\frac{1}{s_{14}s_{26}s_{134}}+\\
       & \frac{1}{s_{13}s_{56}s_{134}}+\frac{1}{s_{14}s_{56}s_{134}}+\frac{1}{s_{13}s_{26}s_{135}}+\frac{1}{s_{15}s_{26}s_{135}}+\frac{1}{s_{13}s_{46}s_{135}}+\\
       & \frac{1}{s_{15}s_{46}s_{135}}+\frac{1}{s_{14}s_{26}s_{145}}+\frac{1}{s_{15}s_{26}s_{145}}+\frac{1}{s_{14}s_{36}s_{145}}+\frac{1}{s_{15}s_{36}s_{145}}+\\
       & \frac{1}{s_{12}s_{46}s_{123}}+\frac{1}{s_{13}s_{46}s_{123}}+\frac{1}{s_{12}s_{56}s_{123}}+\frac{1}{s_{13}s_{56}s_{123}}
    \end{aligned}
\end{equation}

\begin{figure}[h]
    \centering
    \raisebox{-2.8cm}{\includegraphics[width=.35\textwidth]{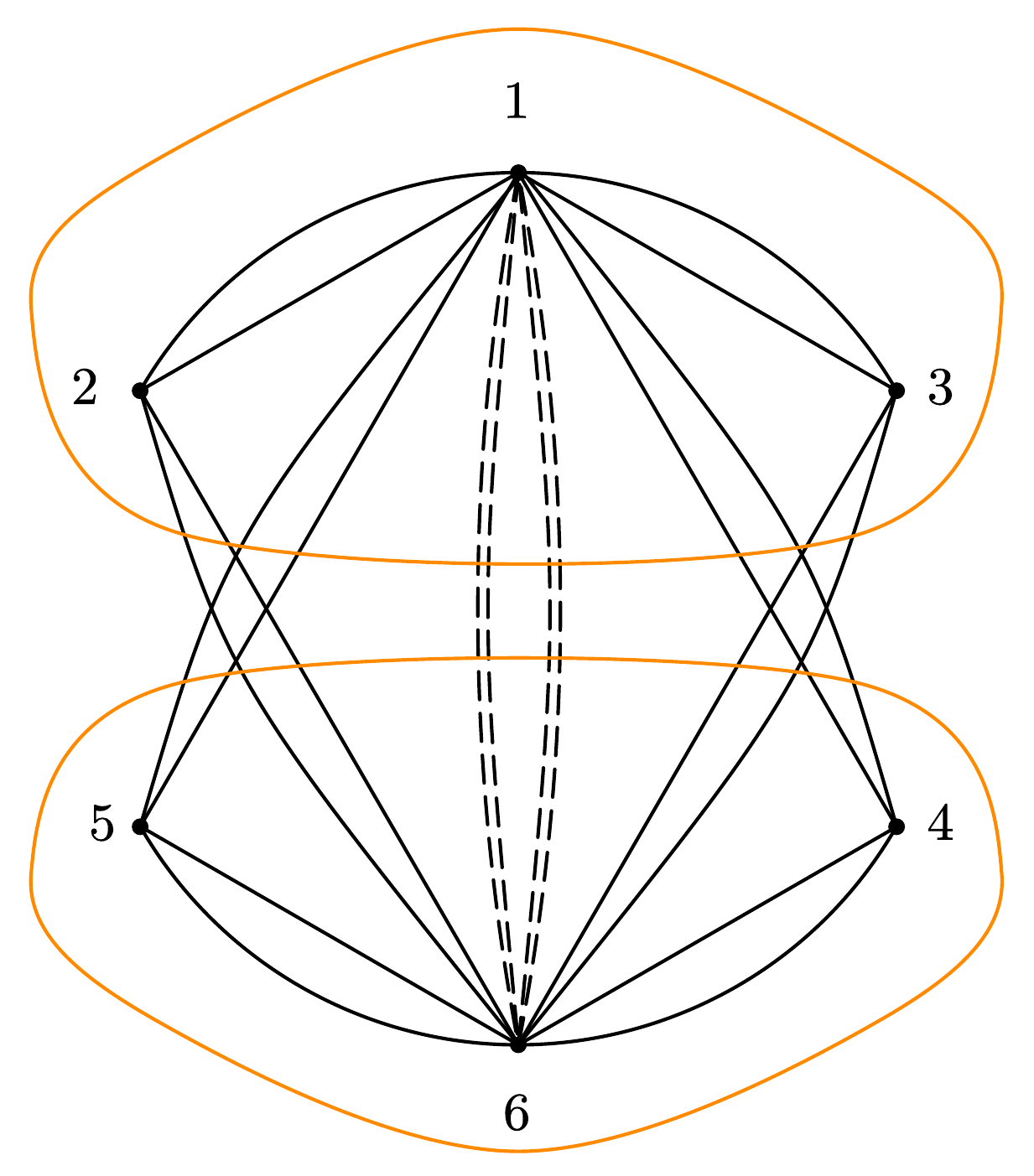}}
    \qquad
    $\Longrightarrow$
    \qquad
    \centering
    \raisebox{-2cm}{\includegraphics[width=.25\textwidth]{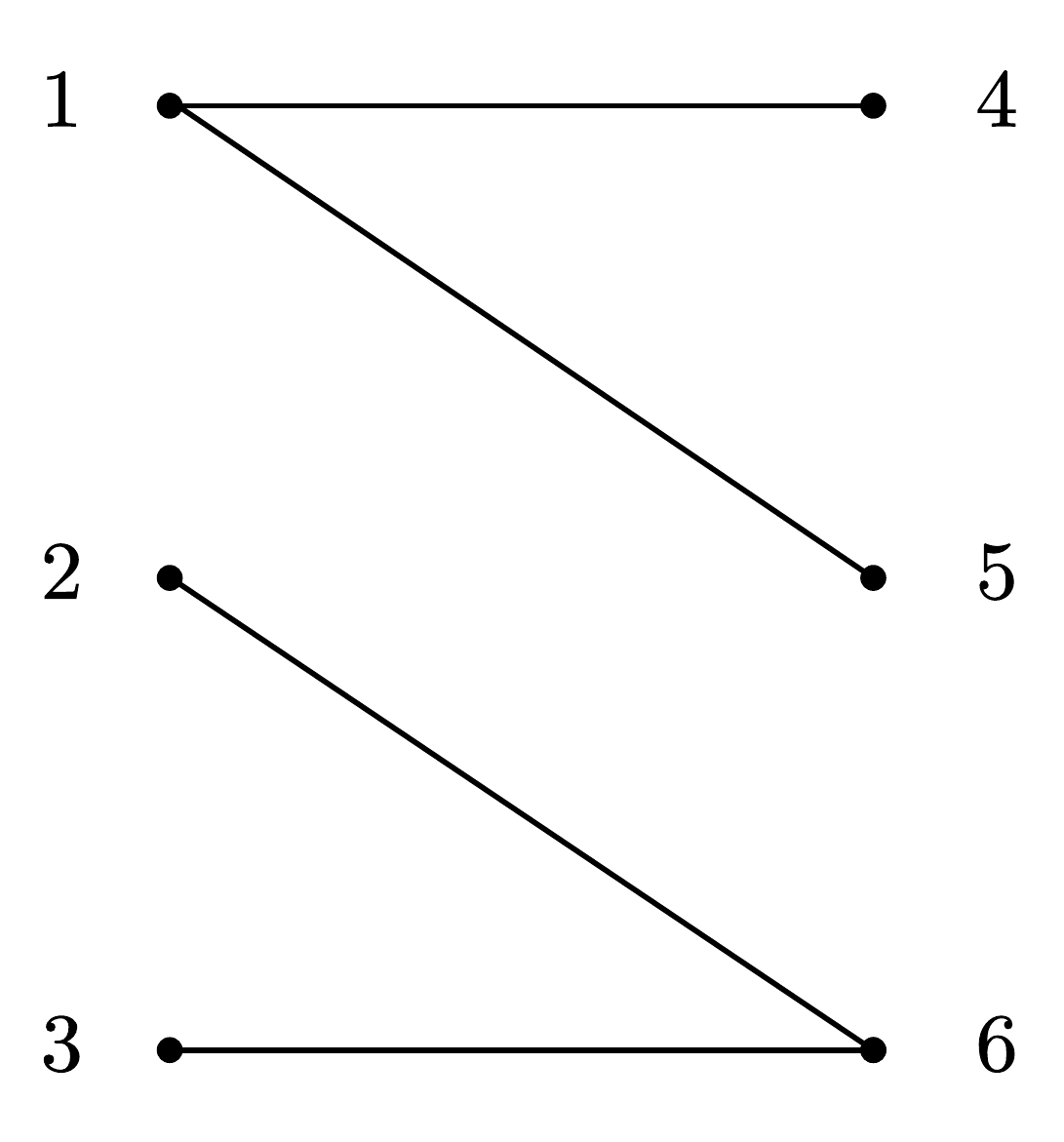}}
    \caption{\label{fig:i} The partition of the integrand $\frac{z_{16}^{4}}{z_{12}^{2} z_{13}^{2} z_{14}^{2} z_{15}^{2} z_{26}^{2} z_{36}^{2} z_{46}^{2} z_{56}^{2}}$ and the corresponding links between two parts }~~~\label{figsection4//pick4-L}
\end{figure}

Let us try to pick the pole  $s_{123}$. The corresponding node-set would be $A=\{1,2,3\},\O A=\{4,5,6\}$ as shown in the Figure \ref{figsection4//pick4-L}.
There are four solid lines connecting $A, \O A$, i.e.,  $
\{1,4\}$, $\{1,5\}$, $\{2,6\}$ and $\{3,6\}$. From them, we can construct six primary cross-ratio factors, but two of them are just trivial one:
\begin{equation}
    \begin{aligned}
    \mathcal{P}^{14}_{26}=\frac{z_{14} z_{26}}{z_{16} z_{24}} \quad \mathcal{P}^{14}_{36}=\frac{z_{14} z_{36}}{z_{16} z_{34}} \quad \mathcal{P}^{14}_{15}=1 \\
    \mathcal{P}^{15}_{26}=\frac{z_{15} z_{26}}{z_{16} z_{25}} \quad \mathcal{P}^{15}_{36}=\frac{z_{15} z_{36}}{z_{16} z_{35}} \quad \mathcal{P}^{26}_{36}=1
    \end{aligned}
    \label{ex_tr_1}
\end{equation}
When multiplying them together, we get $\bm{\mathcal{P}}_0=\frac{z_{14}^{2} z_{15}^{2} z_{26}^{2} z_{36}^{2}}{z_{16}^{4} z_{24} z_{25} z_{34} z_{35}}$. It is easy to check that $\bm{\mathcal{P}}_0$ satisfies the Criterion I, so it is the right factor. One can easily check that
the new CHY-integrand
\begin{equation}
    \mathcal{I}^{star}_{6} \frac{z_{14}^{2} z_{15}^{2} z_{26}^{2} z_{36}^{2}}{z_{16}^{4} z_{24} z_{25} z_{34} z_{35}} = \frac{1}{z_{12}^{2} z_{13}^{2} z_{24} z_{25} z_{34} z_{35} z_{46}^{2} z_{56}^{2}}
\end{equation}
will produce the amplitudes
\begin{equation}
    \frac{1}{s_{12} s_{46} s_{123}}+\frac{1}{s_{13} s_{46}s_{123}}+\frac{1}{s_{12} s_{56} s_{123}}+\frac{1}{s_{13}s_{56} s_{123}}
\end{equation}
which does pick up all terms containing the pole $s_{123}$. We verify that for all the star graphs with $n\leq 8$ the algorithm is right.

Another example is following. The CHY-integrand\footnote{In fact, this CHY-integrand is obtained from the star graph $\mathcal{I}_{6}=\frac{z_{26}^2}{z_{12}^2 z_{13}^2 z_{24}^2 z_{25}^2 z_{36}^2 z_{46}^2 z_{56}^2}$ by multiplying the cross-ratio factor $\frac{z_{13} z_{16} z_{24} z_{25}}{z_{14} z_{15} z_{23} z_{26}}$ of picking up pole $s_{12}$ according to our algorithm. }
\begin{equation}
    \mathcal{I}^{'}_{6}=\frac{z_{16} z_{26}}{z_{12}^2 z_{13} z_{14} z_{15} z_{23} z_{24} z_{25} z_{36}^2 z_{46}^2 z_{56}^2}
\end{equation}
gives the amplitude
\begin{equation}
-\frac{1}{s_{12} s_{56} s_{123}}-\frac{1}{s_{12} s_{36} s_{124}}-\frac{1}{s_{12} s_{56} s_{124}}-\frac{1}{s_{12} s_{36} s_{125}}-\frac{1}{s_{12} s_{46} s_{125}}-\frac{1}{s_{12} s_{46} s_{123}}
\end{equation}
where pole $s_{12}$ appears in every term.
\begin{figure}[h]
    \centering
    \raisebox{-2.5cm}{\includegraphics[width=.35\textwidth]{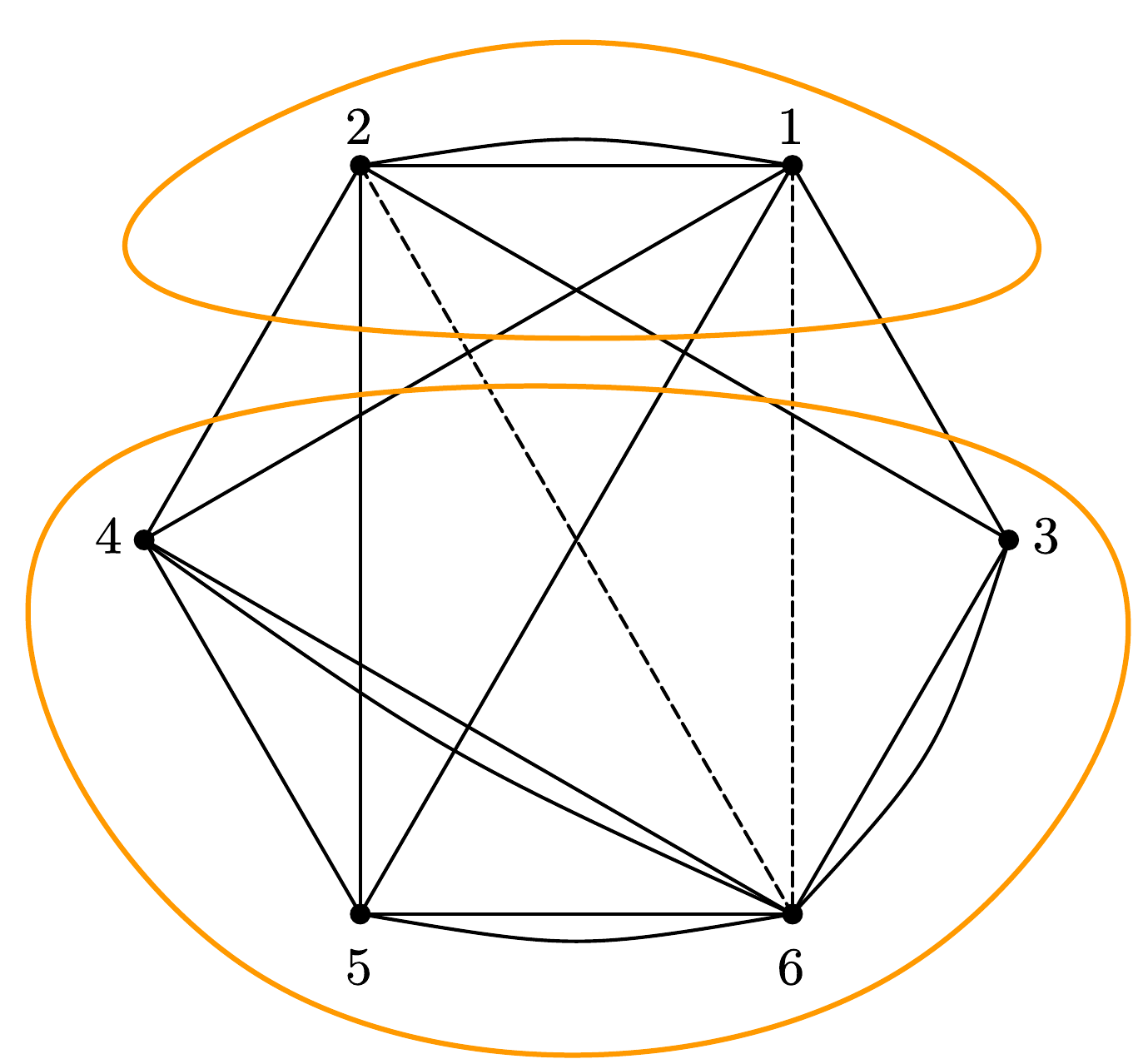}}
    \qquad
    $\Longrightarrow$
    \qquad
    \centering
    \raisebox{-1.8cm}{\includegraphics[width=.25\textwidth]{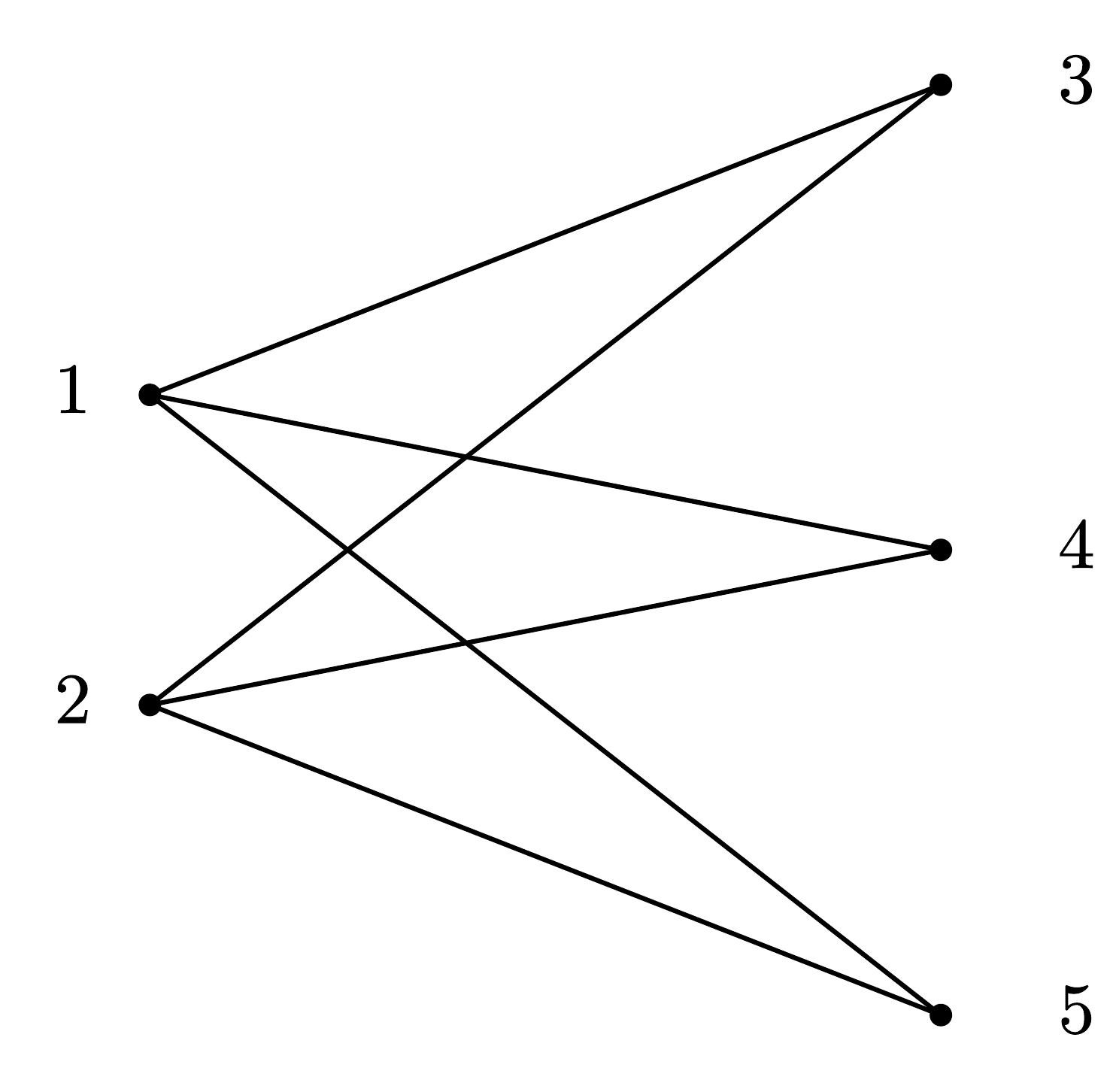}}
    \caption{\label{fig:i}  $Links[\{1,2\},\{3,4,5,6\}]$ of integrand $\frac{z_{16} z_{26}}{z_{12}^2 z_{13} z_{14} z_{15} z_{23} z_{24} z_{25} z_{36}^2 z_{46}^2 z_{56}^2}$ }
    ~~~\label{figsection4//pick9L}
\end{figure}
Now we want to  pick pole $s_{12}$. From the Figure \ref{figsection4//pick9L}, we see that
there are six solid lines connecting $A=\{1,2\}, \O A=\{3,4,5,6\}$.  Among $15$ primary cross-ratio
factors of collection (I), only following six are nontrivial.
\bea
    \mathcal{P}^{13}_{24}& = & \frac{z_{13} z_{24}}{z_{14} z_{23}},~~~ \mathcal{P}^{13}_{25}=\frac{z_{13} z_{25}}{z_{15} z_{23}},~~~
    \mathcal{P}^{14}_{23}=\frac{z_{14} z_{23}}{z_{13} z_{24}} \nn
    \mathcal{P}^{14}_{25}&= & \frac{z_{14} z_{25}}{z_{15} z_{24}},~~~
    \mathcal{P}^{15}_{23}=\frac{z_{15} z_{23}}{z_{13} z_{25}},~~~ \mathcal{P}^{15}_{24}=\frac{z_{15} z_{24}}{z_{14} z_{25}}\eea
Now according to our algorithm, we multiply them together. Amazingly, we get
\begin{equation}
\begin{aligned}
\mathcal{P}^{13}_{24} \mathcal{P}^{13}_{25} \mathcal{P}^{14}_{23} \mathcal{P}^{14}_{25} \mathcal{P}^{15}_{23} \mathcal{P}^{15}_{24} = 1 \\
\end{aligned}
\end{equation}
Although $1$ is a trivial factor at the right hand side, the left hand side is nontrivial combination. The result tells us that to pick up terms containing the pole $s_{12}$, the cross-ratio factor we need to multiply is just one as we have expected.

\subsubsection{The next simplest examples~~\label{ss2}}

In this part, we present some examples where pick-pole factors are multiplication of primary cross-ratio factor in the collection (I), but not with the maximum number. In this case, we need
to impose the Criterion I and II.

The first example is the CHY-integrand
\begin{equation}
    \mathcal{I}_{6}^{[1]}= \frac{1}{z_{12}^{2} z_{23}^{2} z_{34}^{2} z_{56}^{2} z_{45} z_{61} z_{46} z_{15}}
\end{equation}
which produce following five cubic Feynman diagrams
(all terms with common pole $s_{56}$)
\begin{equation}
    \frac{1}{s_{12} s_{34} s_{56}}+\frac{1}{s_{12} s_{123} s_{56}}+\frac{1}{s_{23} s_{123} s_{56}}+\frac{1}{s_{23} s_{156} s_{56}}+\frac{1}{s_{34} s_{156} s_{56}}
\end{equation}
To pick pole $s_{12}$, first we get three solid lines connecting $A=\{1,2\},\O A=\{3,4,5,6\}$ as
$\{2,3\}$,$\{1,6\}$ and $\{1,5\}$. From it, we get the primary cross-ratio factors:
\bea
\mathcal{P}^{23}_{16}=\frac{z_{16} z_{23}}{z_{13} z_{26}}, \quad \mathcal{P}^{23}_{15}=\frac{z_{15} z_{23}}{z_{13} z_{25}}, \quad \mathcal{P}^{16}_{15}=1
\eea
Multiplying them together we get
    $\mathcal{P}^{23}_{16}\mathcal{P}^{23}_{15} = \frac{z_{15} z_{16} z_{23}^{2}}{z_{13}^{2} z_{25} z_{26}}$. It is easy to see that it violates the Criterion I because of the appearance new pole $s_{13}$. Now we remove one primary cross-ratio factor and get two choices
\begin{equation}
    \mathcal{I}[1]_{s_{12}}=\mathcal{I}_{6}^{[1]} \mathcal{P}^{23}_{16}, \quad \mathcal{I}[2]_{s_{12}}=\mathcal{I}_{6}^{[1]} \mathcal{P}^{23}_{15}
\end{equation}
It can be checked that both satisfy the Criterion II and are right answer. Indeed they
produce the same amplitude
\begin{equation}
    \frac{1}{s_{12} s_{34} s_{56}}+\frac{1}{s_{12} s_{123} s_{56}}
\end{equation}

The next example  is more tricky. The CHY-integrand
\begin{equation}
    \mathcal{I}_{6}^{[2]}= \frac{z_{46}^2}{z_{13} z_{14} z_{16}^2 z_{24} z_{25} z_{26}^2 z_{34}^2 z_{36} z_{45}^2 z_{56}}
    \label{EX_I3}
\end{equation}
produces following six cubic Feynman diagrams:
\begin{equation}
    \frac{1}{s_{26} s_{34} s_{126}}+\frac{1}{s_{16} s_{45} s_{126}}+\frac{1}{s_{26} s_{45} s_{126}}+\frac{1}{s_{26} s_{34} s_{134}}+\frac{1}{s_{16} s_{45} s_{136}}+\frac{1}{s_{16} s_{34} s_{126}}
\end{equation}
containing $\{s_{16},s_{26},s_{34},s_{45},s_{126},s_{134},s_{136}\}$ seven poles.
\begin{figure}[h]
    \centering
    \raisebox{-2cm}{\includegraphics[width=.35\textwidth]{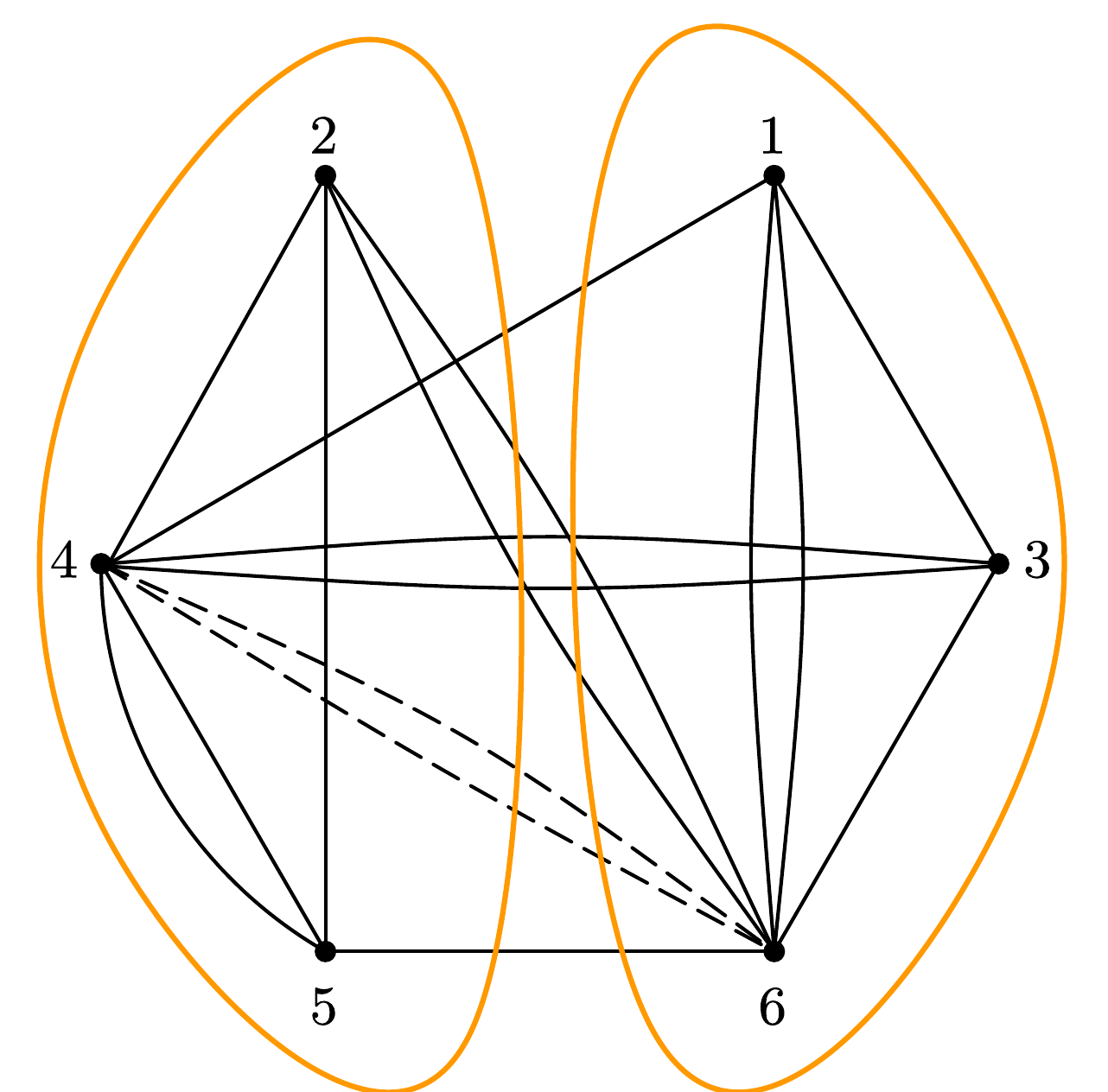}}
    \qquad
    $\Longrightarrow$
    \qquad
    \centering
    \raisebox{-1.8cm}{\includegraphics[width=.25\textwidth]{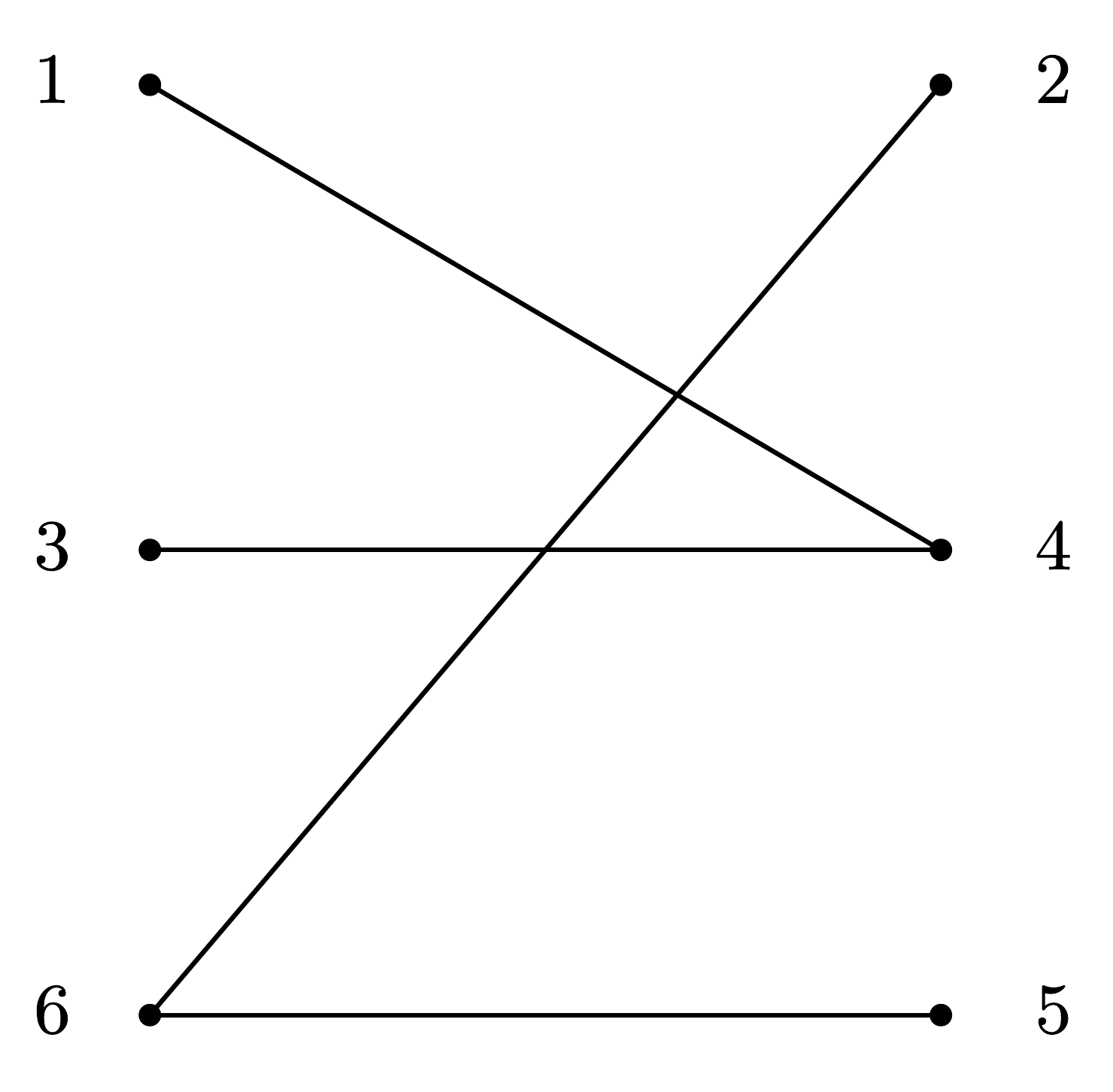}}
    \caption{\label{fig:i}  The partition of the CHY-integrand $\frac{z_{46}^2}{z_{13} z_{14} z_{16}^2 z_{24} z_{25} z_{26}^2 z_{34}^2 z_{36} z_{45}^2 z_{56}}$ and the corresponding lines connecting the two groups }~~~\label{figsection4//pick6L}
\end{figure}
Now we want to pick pole $s_{136}$. From the Figure \ref{figsection4//pick6L} we see there are for solid lines $\{1,4\}$, $\{1,6\}$, $\{3,4\}$ and $\{5,6\}$ connecting to $A=\{1,3,6\},\O A=\{2,4,5\}$, which produce four primary cross-ratio factors
\bea
   \mathcal{P}^{14}_{26}=\frac{z_{14} z_{26}}{z_{12}z_{46}},~~~
   \mathcal{P}^{14}_{56}=\frac{z_{14} z_{56}}{z_{15}z_{46}},~~~
   \mathcal{P}^{26}_{34}=\frac{z_{26} z_{34}}{z_{23}z_{46}},~~~
   \mathcal{P}^{34}_{56}=\frac{z_{34} z_{56}}{z_{35}z_{46}}
   \eea
It is easy to see that multiplying four together, we will get $\frac{z_{14}^2 z_{26}^2 z_{34}^2 z_{56}^2}{z_{12} z_{15} z_{23} z_{35} z_{46}^4}$, which will produce new pole $s_{46}$ when multiplying the original CHY-integrand. Thus by our algorithm, we need to consider combinations with only three primary cross-ratio factors. There are four choices. When
multiplying with $\mathcal{I}_{6}^{[3]}$, the kept poles are listed in following  Figure \ref{table1}.
\begin{figure}[h]
    \centering

    \begin{tabular}[t]{m{0.55\linewidth}m{0.35\linewidth}}

    \hline
    \specialrule{0pt}{5pt}{5pt}
    $\mathcal{I}$ & {\rm Kept~Poles} \\
    \specialrule{0.1pt}{5pt}{5pt}
    \specialrule{0pt}{5pt}{15pt}

    $\mathcal{I}_{6}^{[2]}\mathcal{P}^{14}_{26} \mathcal{P}^{14}_{56} \mathcal{P}^{26}_{34}=\frac{z_{14}}{z_{12} z_{13} z_{15} z_{16}^2 z_{23} z_{24} z_{25} z_{34} z_{36} z_{45}^2 z_{46}}$ & $\{s_{16},s_{45},s_{136}\} $ \\
    \specialrule{0pt}{5pt}{15pt}
    $\mathcal{I}_{6}^{[2]}\mathcal{P}^{14}_{26} \mathcal{P}^{14}_{56} \mathcal{P}^{34}_{56}=\frac{z_{14} z_{56}}{z_{12} z_{13} z_{15} z_{16}^2 z_{24} z_{25} z_{26} z_{34} z_{35} z_{36} z_{45}^2 z_{46}}$ & $\{s_{16},s_{45},s_{126},s_{136}\} $ \\
    \specialrule{0pt}{5pt}{15pt}
    $\mathcal{I}_{6}^{[2]}\mathcal{P}^{14}_{26} \mathcal{P}^{26}_{34} \mathcal{P}^{34}_{56}=\frac{1}{z_{12} z_{13} z_{16}^2 z_{23} z_{24} z_{25} z_{35} z_{36} z_{45}^2 z_{46}}$ & $\{s_{16},s_{45},s_{136}\} $ \\
    \specialrule{0pt}{5pt}{15pt}
    $\mathcal{I}_{6}^{[2]}\mathcal{P}^{14}_{56} \mathcal{P}^{26}_{34} \mathcal{P}^{34}_{56}=\frac{z_{56}}{z_{13} z_{15} z_{16}^2 z_{23} z_{24} z_{25} z_{26} z_{35} z_{36} z_{45}^2 z_{46}}$ & $\{s_{16},s_{45},s_{136}\} $ \\
    \specialrule{0pt}{5pt}{15pt}
    \specialrule{0.1pt}{5pt}{5pt}
    \end{tabular}
\caption{\label{fig:i} All the combinations without add new poles of (\ref{EX_I2})}
\label{table1}
\end{figure}
We see that all four choices satisfy the Criterion I, but when checking the kept poles,
one of them contains incompatible pole $s_{126}$. Thus by our Criterion II, the right answer is other three combinations. One can easily check that all three combination do produce the same amplitude $\frac{1}{s_{16} s_{45} s_{136}}$ as we expected.

\subsubsection{More complicated examples~~~\label{ss3}}

In this subsubsection, we present examples with removing more than just one primary cross-ratio factors in the collection (I). The first example is the familiar bi-adjoint scalar theory with
the CHY-integrand coming from the product of two PT-factors
\begin{equation}
    \mathcal{I}_{6}^{[3]}= \frac{1}{z_{13}^{2} z_{14} z_{16} z_{24} z_{25}^{2} z_{26} z_{34} z_{36} z_{45}  z_{56}}
   \label{EX_I2}
\end{equation}
It provide following two Feynman diagrams, where the only difference between the two is poles $s_{134}$ and $s_{136}$:
\begin{equation}
    \frac{1}{s_{13} s_{25} s_{134}}+\frac{1}{s_{13}  s_{25}  s_{136}}
\end{equation}
For this example, in previous literature \cite{feng2016chy}, we have given the algorithm to pick up a particular pole. Now we will see how the old algorithm is included in our new algorithm.

To pick the pole $s_{134}$, first we find four solid lines connecting $A=\{1,3,4\},\O A=\{2,5,6\}$, i.e.,
$\{1,6\}$,$\{4,2\}$,$\{3,6\}$ and $\{4,5\}$. From it, we get four nontrivial primary cross-ratio factors:
\bea
\mathcal{P}^{16}_{24}=\frac{z_{16}z_{24}}{z_{12}z_{46}},~~~
\mathcal{P}^{24}_{36}=\frac{z_{24}z_{36}}{z_{23}z_{46}},~~~
\mathcal{P}^{16}_{45}=\frac{z_{16}z_{45}}{z_{15}z_{46}},~~~
\mathcal{P}^{36}_{45}=\frac{z_{36}z_{45}}{z_{35}z_{46}}~~\label{2pi-122}
\eea
Multiplying them together we get
    $\mathcal{P}^{16}_{24}  \mathcal{P}^{16}_{45}  \mathcal{P}^{24}_{36} \mathcal{P}^{36}_{45} =\frac{z_{16}z_{24}z_{36}z_{45}}{z_{12}z_{13}^2 z_{14}z_{15}z_{23}z_{25}^2 z_{26}z_{34}z_{35}z_{46}^4 z_{56}}$. It has added the new poles $s_{123}$,$s_{125}$,$s_{135}$ and $s_{146}$, which violate the Criterion I. Next, we need to remove one of four primary cross-ratio factors in turn to check with our criterion. We have shown such process in the figure \ref{table2}. From it we could see that it is not enough to remove just one primary cross-ratio factor. To get the pick-factor without adding new poles, we need to remove three factors. In the end to satisfy the Criterion I we can only keep one primary cross-ratio factor
    in \eref{2pi-122}.  One can check that each of them produces the same right amplitude
\begin{equation}
    \frac{1}{s_{13} s_{25} s_{134}}
\end{equation}

\begin{figure}[h]
    \centering

    \begin{tabular}[t]{m{0.55\linewidth}m{0.35\linewidth}}

    \hline
    \specialrule{0pt}{5pt}{5pt}
    $\mathcal{I}$ & {\rm Added~Poles} \\
    \specialrule{0.1pt}{5pt}{5pt}

    \specialrule{0pt}{5pt}{5pt}

 $\mathcal{I}_{6}^{[3]}\mathcal{P}^{16}_{24}  \mathcal{P}^{16}_{45}  \mathcal{P}^{24}_{36} \mathcal{P}^{36}_{45} =\frac{z_{16}z_{24}z_{36}z_{45}}{z_{12}z_{13}^2 z_{14}z_{15}z_{23}z_{25}^2 z_{26}z_{34}z_{35}z_{46}^4 z_{56}}$& $\{s_{123},s_{125},s_{135},s_{146}\}$ \\
 \specialrule{0pt}{0pt}{5pt}
 $\mathcal{I}_{6}^{[3]}\mathcal{P}^{16}_{45}  \mathcal{P}^{24}_{36}  \mathcal{P}^{36}_{45} =\frac{z_{36}z_{45}}{z_{13}^2 z_{14}z_{15}z_{23}z_{25}^2 z_{26}z_{34}z_{35}z_{46}^3 z_{56}}$ & $\{s_{135},s_{146}\}$ \\
 \specialrule{0pt}{0pt}{5pt}
 $\mathcal{I}_{6}^{[3]}\mathcal{P}^{16}_{24}  \mathcal{P}^{16}_{45} \mathcal{P}^{36}_{45} =\frac{z_{16}z_{45}}{z_{12}z_{13}^2 z_{14}z_{15}z_{25}^2 z_{26}z_{34}z_{35}z_{46}^3 z_{56}}$ & $\{s_{125},s_{135}\}$ \\
 \specialrule{0pt}{0pt}{5pt}
 $\mathcal{I}_{6}^{[3]}\mathcal{P}^{16}_{24}  \mathcal{P}^{24}_{36} \mathcal{P}^{36}_{45} =\frac{z_{24}z_{36}}{z_{12}z_{13}^2 z_{14}z_{23}z_{25}^2 z_{26}z_{34}z_{35}z_{46}^3 z_{56}}$ & $\{s_{123},s_{146}\}$ \\
 \specialrule{0pt}{0pt}{5pt}
 $\mathcal{I}_{6}^{[3]}\mathcal{P}^{16}_{24} \mathcal{P}^{16}_{45}  \mathcal{P}^{24}_{36} =\frac{z_{16}z_{24}}{z_{12}z_{13}^2 z_{14}z_{15}z_{23}z_{25}^2 z_{26}z_{34}z_{46}^3 z_{56}}$ & $\{s_{123},s_{125}\} $\\
 \specialrule{0pt}{0pt}{5pt}
 $\mathcal{I}_{6}^{[3]}\mathcal{P}^{16}_{45}  \mathcal{P}^{36}_{45} =\frac{z_{45}}{z_{13}^2 z_{14}z_{15}z_{24}z_{25}^2 z_{26}z_{34}z_{35}z_{46}^2 z_{56}}$ & $\{s_{46},s_{135}\}$ \\
 \specialrule{0pt}{0pt}{5pt}
 $\mathcal{I}_{6}^{[3]}\mathcal{P}^{24}_{36}  \mathcal{P}^{36}_{45} =\frac{z_{36}}{z_{13}^2 z_{14}z_{16}z_{23}z_{25}^2 z_{26}z_{34} z_{35}z_{46}^2 z_{56}}$ & $\{s_{46},s_{146}\}$ \\
 \specialrule{0pt}{0pt}{5pt}
 $\mathcal{I}_{6}^{[3]}\mathcal{P}^{16}_{45}  \mathcal{P}^{24}_{36} =\frac{1}{z_{13}^2 z_{14}z_{15}z_{23}z_{25}^2 z_{26}z_{34}z_{46}^2 z_{56}}$ & $ \{s_{46}\}$ \\
 \specialrule{0pt}{0pt}{5pt}
 $\mathcal{I}_{6}^{[3]}\mathcal{P}^{16}_{24}  \mathcal{P}^{36}_{45} =\frac{1}{z_{12}z_{13}^2 z_{14}z_{25}^2 z_{26}z_{34}z_{35}z_{46}^2 z_{56}}$ & $\{s_{46}\}$ \\
 \specialrule{0pt}{0pt}{5pt}
 $\mathcal{I}_{6}^{[3]}\mathcal{P}^{16}_{24}  \mathcal{P}^{16}_{45} =\frac{z_{16}}{z_{12}z_{13}^2 z_{14}z_{15}z_{25}^2 z_{26}z_{34}z_{36}z_{46}^2 z_{56}}$ & $ \{s_{46},s_{125}\} $\\
 \specialrule{0pt}{0pt}{5pt}
 $\mathcal{I}_{6}^{[3]}\mathcal{P}^{16}_{24}  \mathcal{P}^{24}_{36} =\frac{z_{24}}{z_{12}z_{13}^2 z_{14}z_{23}z_{25}^2 z_{26}z_{34}z_{45}z_{46}^2 z_{56}}$ & $\{s_{46},s_{123}\}$ \\
 \specialrule{0pt}{0pt}{5pt}
 $\mathcal{I}_{6}^{[3]}\mathcal{P}^{36}_{45} =\frac{1}{z_{13}^2 z_{14}z_{16}z_{24}z_{25}^2 z_{26}z_{34}z_{35}z_{46}z_{56}}$ & $\{\}$ \\
 \specialrule{0pt}{0pt}{5pt}
 $\mathcal{I}_{6}^{[3]}\mathcal{P}^{16}_{45} =\frac{1}{z_{13}^2 z_{14}z_{15}z_{24}z_{25}^2 z_{26}z_{34}z_{36}z_{46}z_{56}}$ & $\{\}$ \\
 \specialrule{0pt}{0pt}{5pt}
 $\mathcal{I}_{6}^{[3]}\mathcal{P}^{24}_{36} =\frac{1}{z_{13}^2 z_{14}z_{16}z_{23}z_{25}^2 z_{26}z_{34}z_{45}z_{46}z_{56}}$ & $\{\}$ \\
 \specialrule{0pt}{0pt}{5pt}
 $\mathcal{I}_{6}^{[3]}\mathcal{P}^{16}_{24} =\frac{1}{z_{12}z_{13}^2 z_{14}z_{25}^2 z_{26}z_{34}z_{36}z_{45}z_{46}z_{56}}$ & $\{\}$ \\

    \specialrule{0pt}{5pt}{15pt}
    \specialrule{0.1pt}{5pt}{5pt}
    \end{tabular}
\caption{\label{fig:i} All the combinations without add new poles of (\ref{EX_I2})}
\label{table2}
\end{figure}
Next, we consider another CHY-integrand which do not belong to the category of Cayley tree
\begin{equation}
    \mathcal{I}_{8}^{[4]}= \frac{z_{18}^2 z_{28} z_{36}}{z_{12} z_{15}^2 z_{16}^2 z_{17} z_{23}^2 z_{24} z_{26} z_{34} z_{37} z_{38} z_{45} z_{48} z_{58} z_{68}^2 z_{78}^2}
   \label{EX_I4}
\end{equation}
It gives seven cubic Feynman diagrams
\begin{equation}
    \begin{aligned}
    &\frac{1}{s_{16} s_{23} s_{78} s_{156} s_{234}}
    +\frac{1}{s_{15} s_{23} s_{68} s_{234} s_{678}}
    +\frac{1}{s_{15} s_{23} s_{78} s_{234} s_{678}}
    -\frac{1}{s_{16} s_{78} s_{126} s_{378} s_{1256}}\\
    &-\frac{1}{s_{15} s_{78} s_{156} s_{378} s_{1256}}
    -\frac{1}{s_{16} s_{78} s_{156} s_{378} s_{1256}}
    +\frac{1}{s_{15} s_{23} s_{78} s_{156} s_{234}}
    \end{aligned}
\end{equation}
with an important property, i.e., different terms have relative sign.

\begin{figure}[h]
    \centering

    \begin{tabular}[t]{m{0.7\linewidth}c}

    \hline
    \specialrule{0pt}{5pt}{5pt}
    $\mathcal{I}$ & {\rm \# of Added~Poles} \\
    \specialrule{0.1pt}{5pt}{5pt}

    \specialrule{0pt}{5pt}{5pt}
    $\mathcal{I}^{[4]}\mathcal{P}^{12}_{58} \mathcal{P}^{16}_{58} \mathcal{P}^{17}_{58}=\frac{z_{28} z_{36} z_{58}^2}{z_{15}^2 z_{16} z_{18} z_{23}^2 z_{24} z_{25} z_{26} z_{34} z_{37} z_{38} z_{45} z_{48} z_{56} z_{57} z_{68}^2 z_{78}^2}$ & $2$ \\
    \specialrule{0pt}{0pt}{5pt}
    $\mathcal{I}^{[4]}\mathcal{P}^{16}_{58} \mathcal{P}^{17}_{45} \mathcal{P}^{17}_{58}=\frac{z_{17} z_{28} z_{36} z_{58}}{z_{12} z_{14} z_{15}^2 z_{16} z_{23}^2 z_{24} z_{26} z_{34} z_{37} z_{38} z_{48} z_{56} z_{57}^2 z_{68}^2 z_{78}^2}$ & $2$ \\
    \specialrule{0pt}{0pt}{5pt}
    $\mathcal{I}^{[4]}\mathcal{P}^{12}_{58} \mathcal{P}^{17}_{45} \mathcal{P}^{17}_{58}=\frac{z_{17} z_{28} z_{36} z_{58}}{z_{14} z_{15}^2 z_{16}^2 z_{23}^2 z_{24} z_{25} z_{26} z_{34} z_{37} z_{38} z_{48} z_{57}^2 z_{68}^2 z_{78}^2}$ & $3$ \\
    \specialrule{0pt}{0pt}{5pt}
    $\mathcal{I}^{[4]}\mathcal{P}^{12}_{58} \mathcal{P}^{16}_{58} \mathcal{P}^{17}_{45}=\frac{z_{28} z_{36} z_{58}}{z_{14} z_{15}^2 z_{16} z_{23}^2 z_{24} z_{25} z_{26} z_{34} z_{37} z_{38} z_{48} z_{56} z_{57} z_{68}^2 z_{78}^2}$ & $0$ \\
    \specialrule{0pt}{0pt}{5pt}
    $\mathcal{I}^{[4]}\mathcal{P}^{16}_{45} \mathcal{P}^{16}_{58} \mathcal{P}^{17}_{58}=\frac{z_{28} z_{36} z_{58}}{z_{12} z_{14} z_{15}^2 z_{23}^2 z_{24} z_{26} z_{34} z_{37} z_{38} z_{48} z_{56}^2 z_{57} z_{68}^2 z_{78}^2}$ & $2$ \\
    \specialrule{0pt}{0pt}{5pt}
    $\mathcal{I}^{[4]}\mathcal{P}^{12}_{58} \mathcal{P}^{16}_{45} \mathcal{P}^{17}_{58}=\frac{z_{28} z_{36} z_{58}}{z_{14} z_{15}^2 z_{16} z_{23}^2 z_{24} z_{25} z_{26} z_{34} z_{37} z_{38} z_{48} z_{56} z_{57} z_{68}^2 z_{78}^2}$ & $0$ \\
    \specialrule{0pt}{0pt}{5pt}
    $\mathcal{I}^{[4]}\mathcal{P}^{12}_{58} \mathcal{P}^{16}_{45} \mathcal{P}^{16}_{58}=\frac{z_{28} z_{36} z_{58}}{z_{14} z_{15}^2 z_{17} z_{23}^2 z_{24} z_{25} z_{26} z_{34} z_{37} z_{38} z_{48} z_{56}^2 z_{68}^2 z_{78}^2}$ & $2$ \\
    \specialrule{0pt}{0pt}{5pt}
    $\mathcal{I}^{[4]}\mathcal{P}^{16}_{45} \mathcal{P}^{17}_{45} \mathcal{P}^{17}_{58}=\frac{z_{17} z_{18} z_{28} z_{36} z_{45}}{z_{12} z_{14}^2 z_{15}^2 z_{16} z_{23}^2 z_{24} z_{26} z_{34} z_{37} z_{38} z_{48} z_{56} z_{57}^2 z_{68}^2 z_{78}^2}$ & $5$ \\
    \specialrule{0pt}{0pt}{5pt}
    $\mathcal{I}^{[4]}\mathcal{P}^{16}_{45} \mathcal{P}^{16}_{58} \mathcal{P}^{17}_{45}=\frac{z_{18} z_{28} z_{36} z_{45}}{z_{12} z_{14}^2 z_{15}^2 z_{23}^2 z_{24} z_{26} z_{34} z_{37} z_{38} z_{48} z_{56}^2 z_{57} z_{68}^2 z_{78}^2}$ & $4$ \\
    \specialrule{0pt}{0pt}{5pt}
    $\mathcal{I}^{[4]}\mathcal{P}^{12}_{58} \mathcal{P}^{16}_{45} \mathcal{P}^{17}_{45}=\frac{z_{18} z_{28} z_{36} z_{45}}{z_{14}^2 z_{15}^2 z_{16} z_{23}^2 z_{24} z_{25} z_{26} z_{34} z_{37} z_{38} z_{48} z_{56} z_{57} z_{68}^2 z_{78}^2}$ & $2$ \\
    \specialrule{0pt}{0pt}{5pt}
    $\mathcal{I}^{[4]}\mathcal{P}^{12}_{45} \mathcal{P}^{16}_{58} \mathcal{P}^{17}_{58}=\frac{z_{28} z_{36} z_{58}}{z_{14} z_{15}^2 z_{16} z_{23}^2 z_{24} z_{25} z_{26} z_{34} z_{37} z_{38} z_{48} z_{56} z_{57} z_{68}^2 z_{78}^2}$ & $0$ \\
    \specialrule{0pt}{0pt}{5pt}
    $\mathcal{I}^{[4]}\mathcal{P}^{12}_{45} \mathcal{P}^{12}_{58} \mathcal{P}^{17}_{58}=\frac{z_{12} z_{28} z_{36} z_{58}}{z_{14} z_{15}^2 z_{16}^2 z_{23}^2 z_{24} z_{25}^2 z_{26} z_{34} z_{37} z_{38} z_{48} z_{57} z_{68}^2 z_{78}^2}$ & $5$ \\
    \specialrule{0pt}{0pt}{5pt}
    $\mathcal{I}^{[4]}\mathcal{P}^{12}_{45} \mathcal{P}^{12}_{58} \mathcal{P}^{16}_{58}=\frac{z_{12} z_{28} z_{36} z_{58}}{z_{14} z_{15}^2 z_{16} z_{17} z_{23}^2 z_{24} z_{25}^2 z_{26} z_{34} z_{37} z_{38} z_{48} z_{56} z_{68}^2 z_{78}^2}$ & $4$ \\
    \specialrule{0pt}{0pt}{5pt}
    $\mathcal{I}^{[4]}\mathcal{P}^{12}_{45} \mathcal{P}^{17}_{45} \mathcal{P}^{17}_{58}=\frac{z_{17} z_{18} z_{28} z_{36} z_{45}}{z_{14}^2 z_{15}^2 z_{16}^2 z_{23}^2 z_{24} z_{25} z_{26} z_{34} z_{37} z_{38} z_{48} z_{57}^2 z_{68}^2 z_{78}^2}$ & $7$ \\
    \specialrule{0pt}{0pt}{5pt}
    $\mathcal{I}^{[4]}\mathcal{P}^{12}_{45} \mathcal{P}^{16}_{58} \mathcal{P}^{17}_{45}=\frac{z_{18} z_{28} z_{36} z_{45}}{z_{14}^2 z_{15}^2 z_{16} z_{23}^2 z_{24} z_{25} z_{26} z_{34} z_{37} z_{38} z_{48} z_{56} z_{57} z_{68}^2 z_{78}^2}$ & $2$ \\
    \specialrule{0pt}{0pt}{5pt}
    $\mathcal{I}^{[4]}\mathcal{P}^{12}_{45} \mathcal{P}^{12}_{58} \mathcal{P}^{17}_{45}=\frac{z_{12} z_{18} z_{28} z_{36} z_{45}}{z_{14}^2 z_{15}^2 z_{16}^2 z_{23}^2 z_{24} z_{25}^2 z_{26} z_{34} z_{37} z_{38} z_{48} z_{57} z_{68}^2 z_{78}^2}$ & $5$ \\
    \specialrule{0pt}{0pt}{5pt}
    $\mathcal{I}^{[4]}\mathcal{P}^{12}_{45} \mathcal{P}^{16}_{45} \mathcal{P}^{17}_{58}=\frac{z_{18} z_{28} z_{36} z_{45}}{z_{14}^2 z_{15}^2 z_{16} z_{23}^2 z_{24} z_{25} z_{26} z_{34} z_{37} z_{38} z_{48} z_{56} z_{57} z_{68}^2 z_{78}^2}$ & $2$ \\
    \specialrule{0pt}{0pt}{5pt}
    $\mathcal{I}^{[4]}\mathcal{P}^{12}_{45} \mathcal{P}^{16}_{45} \mathcal{P}^{16}_{58}=\frac{z_{18} z_{28} z_{36} z_{45}}{z_{14}^2 z_{15}^2 z_{17} z_{23}^2 z_{24} z_{25} z_{26} z_{34} z_{37} z_{38} z_{48} z_{56}^2 z_{68}^2 z_{78}^2}$ & $5$ \\
    \specialrule{0pt}{0pt}{5pt}
    $\mathcal{I}^{[4]}\mathcal{P}^{12}_{45} \mathcal{P}^{12}_{58} \mathcal{P}^{16}_{45}=\frac{z_{12} z_{18} z_{28} z_{36} z_{45}}{z_{14}^2 z_{15}^2 z_{16} z_{17} z_{23}^2 z_{24} z_{25}^2 z_{26} z_{34} z_{37} z_{38} z_{48} z_{56} z_{68}^2 z_{78}^2}$ & $4$ \\
    \specialrule{0pt}{0pt}{5pt}
    $\mathcal{I}^{[4]}\mathcal{P}^{12}_{45} \mathcal{P}^{16}_{45} \mathcal{P}^{17}_{45}=\frac{z_{18}^2 z_{28} z_{36} z_{45}^2}{z_{14}^3 z_{15}^2 z_{16} z_{23}^2 z_{24} z_{25} z_{26} z_{34} z_{37} z_{38} z_{48} z_{56} z_{57} z_{58} z_{68}^2 z_{78}^2}$ & $6$ \\
    \specialrule{0pt}{0pt}{5pt}
    \specialrule{0pt}{5pt}{15pt}
    \specialrule{0.1pt}{5pt}{5pt}
    \end{tabular}
\caption{\label{fig:i} All the three combinations of (\ref{I4-dec})}
\label{table32}
\end{figure}

To pick the pole $s_{15}$, first we write down five solid lines connecting $A=\{1,5\},\O A=\{2,3,4,5,6\}$, i.e.,
$\{1,2\}$,$\{1,6\}$,$\{1,7\}$,$\{5,4\}$ and $\{5,8\}$. From it, we get six nontrivial primary cross-ratio factors:
\bea
\begin{aligned}
&\mathcal{P}^{12}_{45}=\frac{z_{12} z_{45}}{z_{14} z_{25}},~~~
\mathcal{P}^{16}_{45}=\frac{z_{16} z_{45}}{z_{14} z_{56}},~~~
\mathcal{P}^{17}_{45}=\frac{z_{17} z_{45}}{z_{14} z_{57}},~~~\\
&\mathcal{P}^{16}_{58}=\frac{z_{16} z_{58}}{z_{18} z_{56}},~~~
\mathcal{P}^{17}_{58}=\frac{z_{17} z_{58}}{z_{18} z_{57}},~~~
\mathcal{P}^{12}_{58}=\frac{z_{12} z_{58}}{z_{18} z_{25}}~~\label{I4-dec}
\end{aligned}
\eea
Multiplying them together we get $\mathcal{P}^{12}_{45} \mathcal{P}^{12}_{58} \mathcal{P}^{16}_{45} \mathcal{P}^{16}_{58} \mathcal{P}^{17}_{45} \mathcal{P}^{17}_{58} =\frac{z_{12}^2 z_{16}^2 z_{17}^2 z_{45}^3 z_{58}^3}{z_{14}^3 z_{18}^3 z_{25}^2 z_{56}^2 z_{57}^2}$. It would added the new poles $s_{25}$,$s_{56}$ and $s_{57}$, thus violates the Criterion I. Next, we need to remove one or more of six primary cross-ratio factors in turn to check with our criterion. To get the pick-factor without adding new poles, we need to remove at least three factors as shown in the Figure \ref{table32}.
For the cross-ratio factors, which satisfy the Criterion I in the Figure \ref{table32}, we  check if they meet Criterion II as shown in the Figure \ref{table33}. Thus they are the wanted pick-factor. Direct calculation shows that they all produces the same right amplitude:
\begin{equation}
    \frac{1}{s_{15} s_{23} s_{68} s_{234} s_{678}}+\frac{1}{s_{15} s_{23} s_{78} s_{234} s_{678}}-\frac{1}{s_{15} s_{78} s_{156} s_{378} s_{1256}}+\frac{1}{s_{15} s_{23} s_{78} s_{156} s_{234}}
\end{equation}
One thing we want to emphasize is that
these pick factors  keep the signs among different terms.

\begin{figure}[h]
    \centering

    \begin{tabular}[t]{m{0.62\linewidth}c}

    \hline
    \specialrule{0pt}{5pt}{5pt}
    $\mathcal{I}$ & {\rm \# of kept~incompatible~poles} \\
    \specialrule{0.1pt}{5pt}{5pt}
    \specialrule{0pt}{5pt}{15pt}

    $\mathcal{I}^{[4]}\mathcal{P}^{12}_{58} \mathcal{P}^{16}_{58} \mathcal{P}^{17}_{45}=\frac{z_{28} z_{36} z_{58}}{z_{14} z_{15}^2 z_{16} z_{23}^2 z_{24} z_{25} z_{26} z_{34} z_{37} z_{38} z_{48} z_{56} z_{57} z_{68}^2 z_{78}^2}$ & $0$ \\
    \specialrule{0pt}{0pt}{5pt}
    $\mathcal{I}^{[4]}\mathcal{P}^{12}_{58} \mathcal{P}^{16}_{45} \mathcal{P}^{17}_{58}=\frac{z_{28} z_{36} z_{58}}{z_{14} z_{15}^2 z_{16} z_{23}^2 z_{24} z_{25} z_{26} z_{34} z_{37} z_{38} z_{48} z_{56} z_{57} z_{68}^2 z_{78}^2}$ & $0$ \\
    \specialrule{0pt}{0pt}{5pt}
    $\mathcal{I}^{[4]}\mathcal{P}^{12}_{45} \mathcal{P}^{16}_{58} \mathcal{P}^{17}_{58}=\frac{z_{28} z_{36} z_{58}}{z_{14} z_{15}^2 z_{16} z_{23}^2 z_{24} z_{25} z_{26} z_{34} z_{37} z_{38} z_{48} z_{56} z_{57} z_{68}^2 z_{78}^2}$ & $0$\\
    \specialrule{0pt}{0pt}{5pt}

    \specialrule{0pt}{5pt}{15pt}
    \specialrule{0.1pt}{5pt}{5pt}
    \end{tabular}
\caption{\label{fig:i} All the combinations without add new poles of (\ref{I4-dec})}
\label{table33}
\end{figure}

\clearpage

\subsection{The pick-factors with mixed type of cross-ratio factors}

In this subsubsection, we will present several examples where pick-factors are constructed using both collection (I) and (III).

\begin{figure}[h]
    \centering
    \raisebox{-2.5cm}{\includegraphics[width=.35\textwidth]{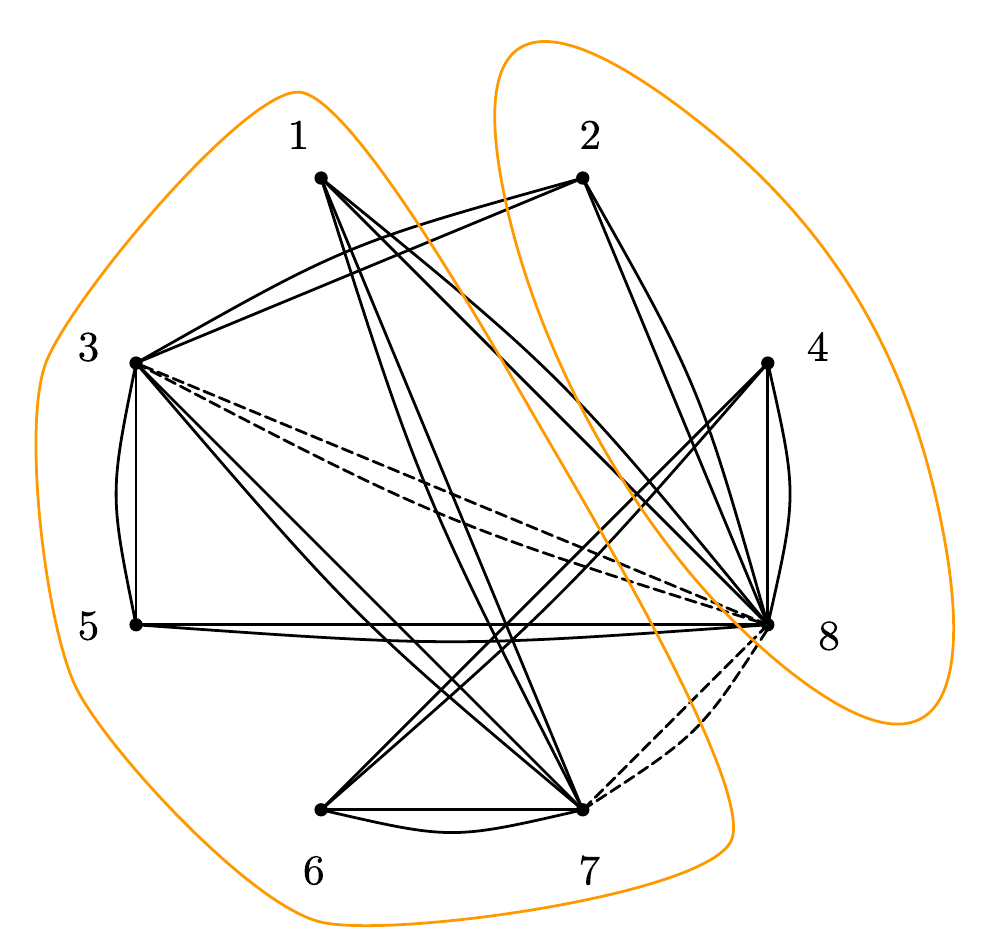}}
    \qquad
    $\Longrightarrow$
    \qquad
    \centering
    \raisebox{-1.8cm}{\includegraphics[width=.25\textwidth]{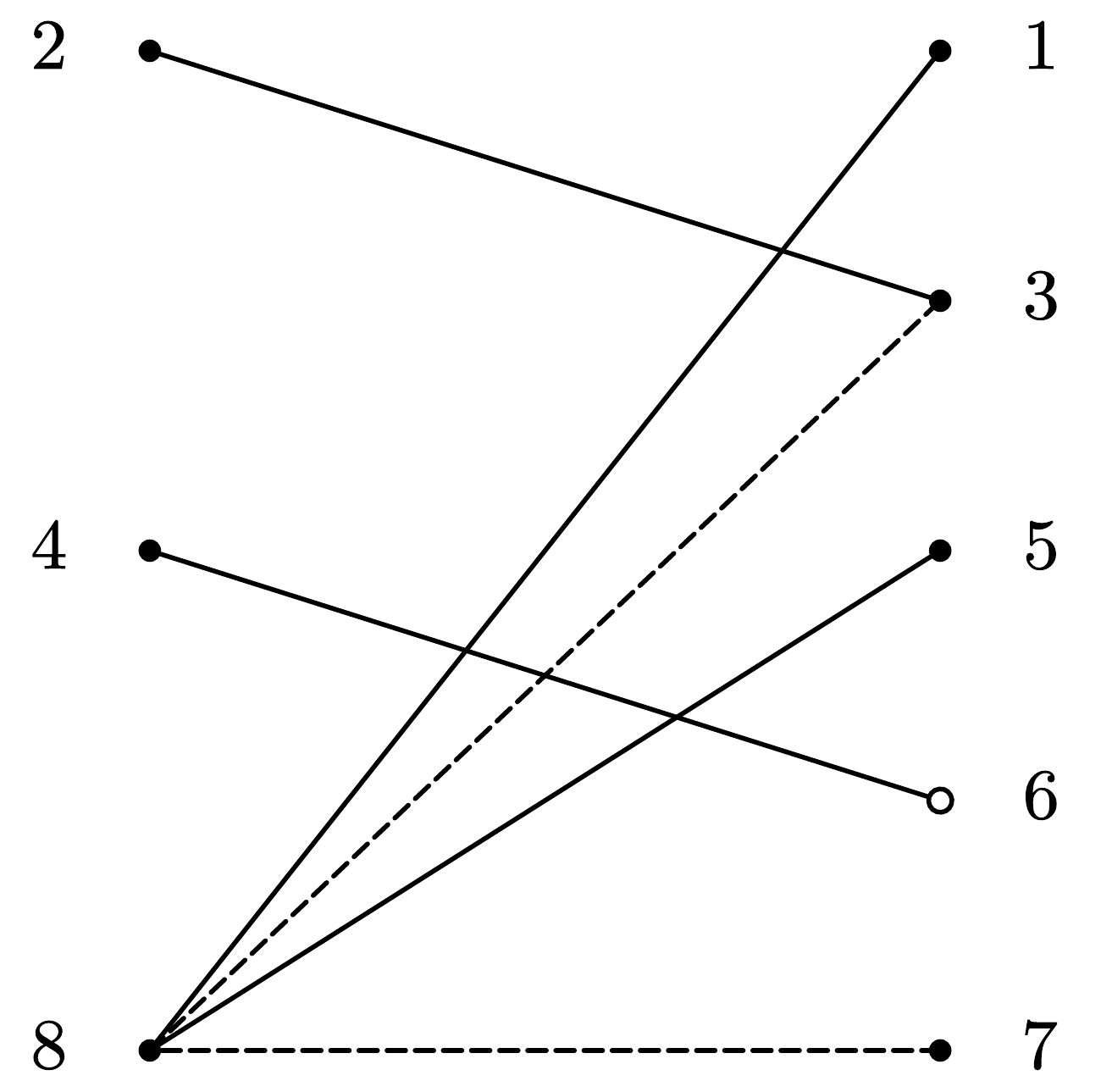}}
    \caption{\label{fig:i}  The partition of the CHY-integrand $\frac{z_{38}^2 z_{78}^2}{z_{17}^2 z_{18}^2 z_{23}^2 z_{28}^2 z_{35}^2 z_{37}^2 z_{46}^2 z_{48}^2 z_{58}^2 z_{67}^2}$ }~~~\label{Com-8-1}
\end{figure}

The first example is the CHY-integrand of $8$-point
\begin{equation}
    \mathcal{I}_{8}= \frac{z_{38}^2 z_{78}^2}{z_{17}^2 z_{18}^2 z_{23}^2 z_{28}^2 z_{35}^2 z_{37}^2 z_{46}^2 z_{48}^2 z_{58}^2 z_{67}^2}
    \label{Ex_I81}
\end{equation}
which will contain  270 cubic Feynman diagrams (it can be studied using the effective Feynman diagram in the first part of the paper). Following poles have appeared in the final result
$$
\begin{aligned}
\{ &s_{17},s_{18},s_{23},s_{28},s_{35},s_{37},s_{46},s_{48},s_{58},s_{67},s_{128},s_{137},s_{148},\\
&s_{158},s_{167},s_{235},s_{237},s_{248},s_{258},s_{357},s_{367},s_{458},s_{467},s_{468},\\
&s_{1237},s_{1248},s_{1258},s_{1357},s_{1367},s_{1458},s_{1467},s_{1468} \}\\
\end{aligned}
$$
Let us pick the pole $s_{248}$ which appears in $36$ terms. From the Figure \ref{Com-8-1}, the linking between the subset
 $\{2,4,8\}$ and its complement  $\{1,3,5,6,7\}$ are solid lines $\{1,8\}$, $\{2,3\}$, $\{4,6\}$, $\{5,8\}$ and the dashed lines $\{3,8\}$, $\{7,8\}$.

Following the algorithm, let us construct the collection (I) with following pure primary cross-ratios
$$
\begin{aligned}
&\mathcal{P}^{18}_{23}=\frac{z_{18} z_{23}}{z_{12} z_{38}} \quad
\mathcal{P}^{18}_{46}=\frac{z_{18} z_{46}}{z_{14} z_{68}} \quad
\mathcal{P}^{23}_{46}=\frac{z_{23} z_{46}}{z_{26} z_{34}} \\
&\mathcal{P}^{23}_{58}=\frac{z_{23} z_{58}}{z_{25} z_{38}} \quad
\mathcal{P}^{46}_{58}=\frac{z_{46} z_{58}}{z_{45} z_{68}}
\end{aligned}
$$
Multiplying all of them together, we will get $\frac{z_{23} z_{46} z_{78}^2}{z_{12} z_{14} z_{17}^2 z_{25} z_{26} z_{28}^2 z_{34} z_{35}^2 z_{37}^2 z_{45} z_{48}^2 z_{67}^2 z_{68}^2}$, which will produce new pole $s_{68}$ when multiplying the original CHY-integrand. Next, we need to remove the pure primary cross-ratio one by one, until there are non new added poles. However, for all these combinations satisfying the Criterion I, they do not satisfy the Criterion II.  We have presented them in Figure \ref{table3} and listed  the number of kept incompatible poles.
\begin{figure}[h]
    \centering

    \begin{tabular}[t]{m{0.62\linewidth}c}

    \hline
    \specialrule{0pt}{5pt}{5pt}
    $\mathcal{I}$ & {\rm \# of kept~incompatible~poles} \\
    \specialrule{0.1pt}{5pt}{5pt}

    \specialrule{0pt}{5pt}{5pt}
    $\mathcal{I}_{8}{\cal T}_{1;1}\equiv\mathcal{I}_{8}\mathcal{P}^{18}_{23}=\frac{z_{38} z_{78}^2}{z_{12} z_{17}^2 z_{18} z_{23} z_{28}^2 z_{35}^2 z_{37}^2 z_{46}^2 z_{48}^2 z_{58}^2 z_{67}^2}$ & $10$ \\
    \specialrule{0pt}{0pt}{5pt}
    $\mathcal{I}_{8}{\cal T}_{1;2}\equiv\mathcal{I}_{8}\mathcal{P}^{23}_{46}=\frac{z_{38}^2 z_{78}^2}{z_{17}^2 z_{18}^2 z_{23} z_{26} z_{28}^2 z_{34} z_{35}^2 z_{37}^2 z_{46} z_{48}^2 z_{58}^2 z_{67}^2}$ & $9$ \\
    \specialrule{0pt}{0pt}{5pt}
    $\mathcal{I}_{8}{\cal T}_{1;3}\equiv\mathcal{I}_{8}\mathcal{P}^{23}_{58}=\frac{z_{38} z_{78}^2}{z_{17}^2 z_{18}^2 z_{23} z_{25} z_{28}^2 z_{35}^2 z_{37}^2 z_{46}^2 z_{48}^2 z_{58} z_{67}^2}$ & $11$ \\
    \specialrule{0pt}{0pt}{5pt}
    $\mathcal{I}_{8}{\cal T}_{1;4}\equiv\mathcal{I}_{8}\mathcal{P}^{18}_{46}=\frac{z_{38}^2 z_{78}^2}{z_{14} z_{17}^2 z_{18} z_{23}^2 z_{28}^2 z_{35}^2 z_{37}^2 z_{46}z_{48}^2 z_{58}^2 z_{67}^2 z_{68}}$ & $12$ \\
    \specialrule{0pt}{0pt}{5pt}
    $\mathcal{I}_{8}{\cal T}_{1;5}\equiv\mathcal{I}_{8}\mathcal{P}^{46}_{58}=\frac{z_{38}^2 z_{78}^2}{z_{17}^2 z_{18}^2 z_{23}^2 z_{28}^2 z_{35}^2 z_{37}^2 z_{45} z_{46}z_{48}^2 z_{58} z_{67}^2 z_{68}}$ & $11$ \\
    \specialrule{0pt}{0pt}{5pt}
    $\mathcal{I}_{8}{\cal T}_{2;1}\equiv\mathcal{I}_{8}\mathcal{P}^{18}_{23} \mathcal{P}^{23}_{58}=\frac{z_{78}^2}{z_{12} z_{17}^2 z_{18} z_{25} z_{28}^2 z_{35}^2z_{37}^2 z_{46}^2 z_{48}^2 z_{58} z_{67}^2}$ & $7$ \\
    \specialrule{0pt}{0pt}{5pt}
    $\mathcal{I}_{8}{\cal T}_{2;2}\equiv\mathcal{I}_{8}\mathcal{P}^{18}_{23} \mathcal{P}^{18}_{46}=\frac{z_{38} z_{78}^2}{z_{12} z_{14} z_{17}^2 z_{23} z_{28}^2 z_{35}^2z_{37}^2 z_{46} z_{48}^2 z_{58}^2 z_{67}^2 z_{68}}$ & $6$ \\
    \specialrule{0pt}{0pt}{5pt}
    $\mathcal{I}_{8}{\cal T}_{2;3}\equiv\mathcal{I}_{8}\mathcal{P}^{18}_{23} \mathcal{P}^{46}_{58}=\frac{z_{38} z_{78}^2}{z_{12} z_{17}^2 z_{18} z_{23} z_{28}^2 z_{35}^2z_{37}^2 z_{45} z_{46} z_{48}^2 z_{58} z_{67}^2 z_{68}}$ & $4$ \\
    \specialrule{0pt}{0pt}{5pt}
    $\mathcal{I}_{8}{\cal T}_{2;4}\equiv\mathcal{I}_{8}\mathcal{P}^{23}_{46} \mathcal{P}^{23}_{58}=\frac{z_{38} z_{78}^2}{z_{17}^2 z_{18}^2 z_{25} z_{26} z_{28}^2 z_{34}z_{35}^2 z_{37}^2 z_{46} z_{48}^2 z_{58} z_{67}^2}$ & $5$ \\
    \specialrule{0pt}{0pt}{5pt}
    $\mathcal{I}_{8}{\cal T}_{2;5}\equiv\mathcal{I}_{8}\mathcal{P}^{18}_{46} \mathcal{P}^{23}_{58}=\frac{z_{38} z_{78}^2}{z_{14} z_{17}^2 z_{18} z_{23} z_{25} z_{28}^2z_{35}^2 z_{37}^2 z_{46} z_{48}^2 z_{58} z_{67}^2 z_{68}}$ & $6$ \\
    \specialrule{0pt}{0pt}{5pt}
    $\mathcal{I}_{8}{\cal T}_{2;6}\equiv\mathcal{I}_{8}\mathcal{P}^{23}_{58} \mathcal{P}^{46}_{58}=\frac{z_{38} z_{78}^2}{z_{17}^2 z_{18}^2 z_{23} z_{25} z_{28}^2z_{35}^2 z_{37}^2 z_{45} z_{46} z_{48}^2 z_{67}^2 z_{68}}$ & $6$ \\
    \specialrule{0pt}{0pt}{5pt}
    $\mathcal{I}_{8}{\cal T}_{3;1}\equiv\mathcal{I}_{8}\mathcal{P}^{18}_{23} \mathcal{P}^{18}_{46} \mathcal{P}^{23}_{58}=\frac{z_{78}^2}{z_{12} z_{14} z_{17}^2z_{25} z_{28}^2 z_{35}^2 z_{37}^2 z_{46} z_{48}^2 z_{58} z_{67}^2 z_{68}}$ & $3$ \\
    \specialrule{0pt}{0pt}{5pt}
    $\mathcal{I}_{8}{\cal T}_{3;2}\equiv\mathcal{I}_{8}\mathcal{P}^{18}_{23} \mathcal{P}^{23}_{58} \mathcal{P}^{46}_{58}=\frac{z_{78}^2}{z_{12} z_{17}^2 z_{18}z_{25} z_{28}^2 z_{35}^2 z_{37}^2 z_{45} z_{46} z_{48}^2 z_{67}^2 z_{68}}$ & $2$ \\

    \specialrule{0pt}{5pt}{15pt}
    \specialrule{0.1pt}{5pt}{5pt}
    \end{tabular}
\caption{\label{fig:i} All the combinations without add new poles of (\ref{Ex_I81})}
\label{table3}
\end{figure}

Since we can not find the right pick-factor from just the collection (I). According to our algorithm we need to included the collection (III) with following $9$ primary cross-ratio factors
\footnote{It is worth to recall that according to our construction in \eref{Def-mcs1}, the  $\mathcal{P}^{23}_{78;18}$ is not allowed.}
\begin{equation}
\begin{aligned}
    &a_1\equiv \mathcal{P}^{18}_{78;23}=\frac{z_{18} z_{27}}{z_{12} z_{78}}
    & &a_2\equiv\mathcal{P}^{23}_{38;46}=\frac{z_{23} z_{68}}{z_{26} z_{38}}
    & &a_3\equiv\mathcal{P}^{46}_{78;23}=\frac{z_{27} z_{38} z_{46}}{z_{26} z_{34} z_{78}} \\
    &a_4\equiv\mathcal{P}^{23}_{78;46}=\frac{z_{23} z_{47} z_{68}}{z_{26} z_{34} z_{78}}
    & &a_5\equiv\mathcal{P}^{58}_{78;23}=\frac{z_{27} z_{58}}{z_{25} z_{78}}
    & &a_6\equiv\mathcal{P}^{18}_{38;46}=\frac{z_{18} z_{34}}{z_{14} z_{38}} \\
    &a_7\equiv\mathcal{P}^{18}_{78;46}=\frac{z_{18} z_{47}}{z_{14} z_{78}}
    & &a_8\equiv\mathcal{P}^{58}_{38;46}=\frac{z_{34} z_{58}}{z_{38} z_{45}}
    & &a_9\equiv\mathcal{P}^{58}_{78;46}=\frac{z_{47} z_{58}}{z_{45} z_{78}}
\end{aligned}
\label{I81-mixed}
\end{equation}
There are $13$'s ${\cal T}_{i;k}$ according to \eref{rule-7b} with $i=3,2,1$ and $N_{3}=2, N_2=6, N_1=5$ respectively. Searching along the nested loops, we find that
\begin{itemize}

\item (1) With $m=1$, we find all $13\times 9$ can not satisfy both Criterion I and II at the same time.
\item (2) With $m=2$, we find all $13\times 36$ can not satisfy both Criterion I and II at the same time.
\item (3) With $m=3$, we tested all $13\times 84$ items.Twelve combinations satisfy both Criterion I and II:

        \begin{tabular}[t]{m{0.05\linewidth}m{0.1\linewidth}cc}
            \specialrule{0pt}{0pt}{10pt}
             & ${\cal T}_{2;1}$ : & $\mathcal{P}^{18}_{23} \mathcal{P}^{23}_{58} \mathcal{P}^{18}_{38;46} \mathcal{P}^{46}_{78;23} \mathcal{P}^{58}_{78;46}$ &
            $\mathcal{P}^{18}_{23} \mathcal{P}^{23}_{58} \mathcal{P}^{18}_{78;46} \mathcal{P}^{46}_{78;23} \mathcal{P}^{58}_{38;46}$ \\
            \specialrule{0pt}{5pt}{5pt}
            & ${\cal T}_{2;2}$ : & $\mathcal{P}^{18}_{23} \mathcal{P}^{18}_{46} \mathcal{P}^{23}_{38;46} \mathcal{P}^{58}_{78;23} \mathcal{P}^{58}_{78;46}$ &
            $\mathcal{P}^{18}_{23} \mathcal{P}^{18}_{46} \mathcal{P}^{23}_{78;46} \mathcal{P}^{58}_{38;46} \mathcal{P}^{58}_{78;23}$ \\
            \specialrule{0pt}{5pt}{5pt}
            & ${\cal T}_{2;3}$ : & $\mathcal{P}^{18}_{23} \mathcal{P}^{46}_{58} \mathcal{P}^{18}_{78;46} \mathcal{P}^{23}_{38;46} \mathcal{P}^{58}_{78;23}$ &
            $\mathcal{P}^{18}_{23} \mathcal{P}^{46}_{58} \mathcal{P}^{18}_{38;46} \mathcal{P}^{23}_{78;46} \mathcal{P}^{58}_{78;23}$ \\
            \specialrule{0pt}{5pt}{5pt}
            & ${\cal T}_{2;4}$ : & $\mathcal{P}^{23}_{46} \mathcal{P}^{23}_{58} \mathcal{P}^{18}_{38;46} \mathcal{P}^{18}_{78;23}\mathcal{P}^{58}_{78;46}$ &
            $\mathcal{P}^{23}_{46} \mathcal{P}^{23}_{58} \mathcal{P}^{18}_{78;23} \mathcal{P}^{18}_{78;46} \mathcal{P}^{58}_{38;46}$ \\
            \specialrule{0pt}{5pt}{5pt}
            & ${\cal T}_{2;5}$ : & $\mathcal{P}^{18}_{46} \mathcal{P}^{23}_{58} \mathcal{P}^{18}_{78;23} \mathcal{P}^{23}_{38;46} \mathcal{P}^{58}_{78;46}$ &
            $\mathcal{P}^{18}_{46} \mathcal{P}^{23}_{58} \mathcal{P}^{18}_{78;23} \mathcal{P}^{23}_{78;46} \mathcal{P}^{58}_{38;46}$ \\
            \specialrule{0pt}{5pt}{5pt}
            & ${\cal T}_{2;6}$ : & $\mathcal{P}^{23}_{58} \mathcal{P}^{46}_{58} \mathcal{P}^{18}_{78;23} \mathcal{P}^{18}_{78;46} \mathcal{P}^{23}_{38;46}$ &
            $\mathcal{P}^{23}_{58} \mathcal{P}^{46}_{58} \mathcal{P}^{18}_{38;46} \mathcal{P}^{18}_{78;23} \mathcal{P}^{23}_{78;46}$ \\
            \specialrule{0pt}{0pt}{10pt}
        \end{tabular}

All of above twelve combinations of primary cross-ratio factors give the same pick-factor:
\begin{equation}
    \frac{z_{18}^2 z_{23}^2 z_{27} z_{46} z_{47} z_{58}^2}{z_{12} z_{14} z_{25} z_{26} z_{38}^2 z_{45} z_{78}^2}
\end{equation}
It gives a new integrand.
\begin{equation}
    \mathcal{I}^{'}_{8}=\frac{z_{27} z_{47}}{z_{12} z_{14} z_{17}^2 z_{25} z_{26} z_{28}^2 z_{35}^2 z_{37}^2 z_{45} z_{46} z_{48}^2 z_{67}^2}
\end{equation}

The Feynman diagrams corresponding to this new integrand contains the only $s_{248}$ terms.
\begin{align*}
    & -\frac{1}{s_{17} s_{35} s_{48} s_{167} s_{248}}   -\frac{1}{s_{28} s_{35} s_{67} s_{167} s_{248}}   -\frac{1}{s_{35} s_{48} s_{67} s_{167} s_{248}}   -\frac{1}{s_{17} s_{28} s_{35} s_{167} s_{248}}  \\
    & -\frac{1}{s_{35} s_{48} s_{67} s_{248} s_{1248}}   -\frac{1}{s_{28} s_{35} s_{248} s_{357} s_{1248}}   -\frac{1}{s_{28} s_{37} s_{248} s_{357} s_{1248}}   -\frac{1}{s_{28} s_{35} s_{67} s_{248} s_{1248}}  \\
    & -\frac{1}{s_{48} s_{67} s_{248} s_{367} s_{1248}}   -\frac{1}{s_{17} s_{28} s_{35} s_{248} s_{1357}}   -\frac{1}{s_{17} s_{35} s_{48} s_{248} s_{1357}}   -\frac{1}{s_{28} s_{67} s_{248} s_{367} s_{1248}}  \\
    & -\frac{1}{s_{37} s_{48} s_{248} s_{357} s_{1248}}   -\frac{1}{s_{28} s_{37} s_{248} s_{367} s_{1248}}   -\frac{1}{s_{37} s_{48} s_{248} s_{367} s_{1248}}   -\frac{1}{s_{35} s_{48} s_{248} s_{357} s_{1248}}  \\
    & -\frac{1}{s_{28} s_{37} s_{137} s_{248} s_{1357}}   -\frac{1}{s_{17} s_{48} s_{137} s_{248} s_{1357}}   -\frac{1}{s_{37} s_{48} s_{137} s_{248} s_{1357}}   -\frac{1}{s_{17} s_{28} s_{137} s_{248} s_{1357}}  \\
    & -\frac{1}{s_{28} s_{37} s_{248} s_{357} s_{1357}}   -\frac{1}{s_{35} s_{48} s_{248} s_{357} s_{1357}}   -\frac{1}{s_{37} s_{48} s_{248} s_{357} s_{1357}}   -\frac{1}{s_{28} s_{35} s_{248} s_{357} s_{1357}}  \\
    & -\frac{1}{s_{28} s_{37} s_{137} s_{248} s_{1367}}   -\frac{1}{s_{17} s_{48} s_{137} s_{248} s_{1367}}   -\frac{1}{s_{37} s_{48} s_{137} s_{248} s_{1367}}   -\frac{1}{s_{17} s_{28} s_{137} s_{248} s_{1367}}  \\
    & -\frac{1}{s_{17} s_{48} s_{167} s_{248} s_{1367}}   -\frac{1}{s_{28} s_{67} s_{167} s_{248} s_{1367}}   -\frac{1}{s_{48} s_{67} s_{167} s_{248} s_{1367}}   -\frac{1}{s_{17} s_{28} s_{167} s_{248} s_{1367}}  \\
    & -\frac{1}{s_{37} s_{48} s_{248} s_{367} s_{1367}}   -\frac{1}{s_{28} s_{67} s_{248} s_{367} s_{1367}}   -\frac{1}{s_{48} s_{67} s_{248} s_{367} s_{1367}}   -\frac{1}{s_{28} s_{37} s_{248} s_{367} s_{1367}}  \\
\end{align*}

\item (4) Although we could stop our algorithm with $m=3$. For this simple example,  we verified the case with $m> 3$ for completeness. We find that for $m>3$, there do not exist any combinations that satisfy both Criterion I and II at the same time.
\end{itemize}

The second example is more plentiful, the CHY-integrand
\begin{equation}
    \mathcal{I}_{8}^{[2]}= \frac{z_{68}^2}{z_{13}^2 z_{16} z_{17} z_{24}^2 z_{28}^2 z_{35} z_{36} z_{46}^2 z_{56} z_{58}^2 z_{67} z_{78}^2}
    \label{Ex_I82}
\end{equation}
which will contain 28 cubic Feynman diagrams:
$$
\begin{aligned}
    &+\frac{1}{s_{13} s_{24} s_{78} s_{136} s_{578}}+\frac{1}{s_{13} s_{24} s_{58} s_{246} s_{578}}
    +\frac{1}{s_{13} s_{46} s_{58} s_{246} s_{578}}+\frac{1}{s_{13} s_{24} s_{78} s_{246} s_{578}}\\
    &+\frac{1}{s_{13} s_{46} s_{78} s_{246} s_{578}}+\frac{1}{s_{13} s_{28} s_{46} s_{258} s_{1346}}
    +\frac{1}{s_{13} s_{46} s_{58} s_{258} s_{1346}}+\frac{1}{s_{13} s_{28} s_{136} s_{258} s_{1346}}\\
    &+\frac{1}{s_{13} s_{58} s_{136} s_{258} s_{1346}}+\frac{1}{s_{13} s_{28} s_{46} s_{278} s_{1346}}
    +\frac{1}{s_{13} s_{46} s_{78} s_{278} s_{1346}}+\frac{1}{s_{13} s_{28} s_{136} s_{278} s_{1346}}\\
    &+\frac{1}{s_{13} s_{78} s_{136} s_{278} s_{1346}}+\frac{1}{s_{13} s_{46} s_{58} s_{578} s_{1346}}
    +\frac{1}{s_{13} s_{46} s_{78} s_{578} s_{1346}}+\frac{1}{s_{13} s_{58} s_{136} s_{578} s_{1346}}\\
    &+\frac{1}{s_{13} s_{78} s_{136} s_{578} s_{1346}}+\frac{1}{s_{13} s_{24} s_{78} s_{136} s_{1356}}
    +\frac{1}{s_{13} s_{24} s_{136} s_{248} s_{1356}}+\frac{1}{s_{13} s_{28} s_{136} s_{248} s_{1356}}\\
    &+\frac{1}{s_{13} s_{28} s_{136} s_{278} s_{1356}}+\frac{1}{s_{13} s_{78} s_{136} s_{278} s_{1356}}
    +\frac{1}{s_{13} s_{24} s_{58} s_{136} s_{1367}}+\frac{1}{s_{13} s_{24} s_{136} s_{248} s_{1367}}\\
    &+\frac{1}{s_{13} s_{28} s_{136} s_{248} s_{1367}}+\frac{1}{s_{13} s_{28} s_{136} s_{258} s_{1367}}
    +\frac{1}{s_{13} s_{58} s_{136} s_{258} s_{1367}}+\frac{1}{s_{13} s_{24} s_{58} s_{136} s_{578}}\\
\end{aligned}
$$
Let us pick the pole $s_{258}$ which appears in $6$ terms. The linking between the subset
 $\{2,5,8\}$ and its complement $\{1,3,4,6,7\}$ are containing solid lines $\{2,4\}$, $\{5,3\}$, $\{5,6\}$ and the dashed lines $\{8,6\}$.
Following the algorithm, let us construct the collection (I) with following pure primary cross-ratios
$$
\begin{aligned}
&\mathcal{P}^{24}_{35}=\frac{z_{24} z_{35}}{z_{23} z_{45}} \quad
\mathcal{P}^{24}_{56}=\frac{z_{24} z_{56}}{z_{26} z_{45}} \quad
\mathcal{P}^{24}_{78}=\frac{z_{24} z_{78}}{z_{27} z_{48}} \\
&\mathcal{P}^{35}_{78}=\frac{z_{35} z_{78}}{z_{38} z_{57}} \quad
\mathcal{P}^{56}_{78}=\frac{z_{56} z_{78}}{z_{57} z_{68}}
\end{aligned}
$$
Multiplying all of them together, we will get $\frac{z_{24}^3 z_{35}^2 z_{56}^2 z_{78}^3}{z_{23} z_{26} z_{27} z_{38} z_{45}^2 z_{48} z_{57}^2 z_{68}}$, which will produce new pole $s_{57}$, $s_{45}$ when multiplying the original $\mathcal{I}_{8}^{[2]}$. We need to remove the pure primary cross-ratio until there are non new added poles as before. We have presented all combinations that satisfying the Criterion I, but do not satisfy the Criterion II in Figure \ref{table4} and label the number of kept incompatible poles.
\begin{figure}[h]
    \centering

    \begin{tabular}[t]{m{0.66\linewidth}c}

    \hline
    \specialrule{0pt}{5pt}{5pt}
    $\mathcal{I}$ & {\rm \# of kept~incompatible~poles} \\
    \specialrule{0.1pt}{5pt}{5pt}

    \specialrule{0pt}{0pt}{5pt}
    $\mathcal{I}_{8}^{[2]}{\cal T}_{2;1}\equiv\mathcal{I}_{8}^{[2]}\mathcal{P}^{24}_{78}\mathcal{P}^{56}_{78}=\frac{z_{68}}{z_{13}^2 z_{16} z_{17} z_{24} z_{27} z_{28}^2 z_{35} z_{36} z_{46}^2 z_{48} z_{57} z_{58}^2 z_{67}}$ & $1$ \\
    \specialrule{0pt}{0pt}{5pt}
    $\mathcal{I}_{8}^{[2]}{\cal T}_{2;2}\equiv\mathcal{I}_{8}^{[2]}\mathcal{P}^{24}_{78}\mathcal{P}^{35}_{78}=\frac{z_{68}^2}{z_{13}^2 z_{16} z_{17} z_{24} z_{27} z_{28}^2 z_{36} z_{38} z_{46}^2 z_{48} z_{56} z_{57} z_{58}^2 z_{67}}$ & $1$ \\
    \specialrule{0pt}{0pt}{5pt}
    $\mathcal{I}_{8}^{[2]}{\cal T}_{2;3}\equiv\mathcal{I}_{8}^{[2]}\mathcal{P}^{24}_{56}\mathcal{P}^{56}_{78}=\frac{z_{56} z_{68}}{z_{13}^2 z_{16} z_{17} z_{24} z_{26} z_{28}^2 z_{35} z_{36} z_{45} z_{46}^2 z_{57} z_{58}^2 z_{67} z_{78}}$ & $2$ \\
    \specialrule{0pt}{0pt}{5pt}
    $\mathcal{I}_{8}^{[2]}{\cal T}_{2;4}\equiv\mathcal{I}_{8}^{[2]}\mathcal{P}^{24}_{56}\mathcal{P}^{35}_{78}=\frac{z_{68}^2}{z_{13}^2 z_{16} z_{17} z_{24} z_{26} z_{28}^2 z_{36} z_{38} z_{45} z_{46}^2 z_{57} z_{58}^2 z_{67} z_{78}}$ & $2$ \\
    \specialrule{0pt}{0pt}{5pt}
    $\mathcal{I}_{8}^{[2]}{\cal T}_{2;5}\equiv\mathcal{I}_{8}^{[2]}\mathcal{P}^{24}_{35}\mathcal{P}^{56}_{78}=\frac{z_{68}}{z_{13}^2 z_{16} z_{17} z_{23} z_{24} z_{28}^2 z_{36} z_{45} z_{46}^2 z_{57} z_{58}^2 z_{67} z_{78}}$ & $1$ \\
    \specialrule{0pt}{0pt}{5pt}
    $\mathcal{I}_{8}^{[2]}{\cal T}_{1;1}\equiv\mathcal{I}_{8}^{[2]}\mathcal{P}^{56}_{78}=\frac{z_{68}}{z_{13}^2 z_{16} z_{17} z_{24}^2 z_{28}^2 z_{35} z_{36} z_{46}^2 z_{57} z_{58}^2 z_{67} z_{78}}$ & $4$ \\
    \specialrule{0pt}{0pt}{5pt}
    $\mathcal{I}_{8}^{[2]}{\cal T}_{1;2}\equiv\mathcal{I}_{8}^{[2]}\mathcal{P}^{35}_{78}=\frac{z_{68}^2}{z_{13}^2 z_{16} z_{17} z_{24}^2 z_{28}^2 z_{36} z_{38} z_{46}^2 z_{56} z_{57} z_{58}^2 z_{67} z_{78}}$ & $4$ \\
    \specialrule{0pt}{0pt}{5pt}
    $\mathcal{I}_{8}^{[2]}{\cal T}_{1;3}\equiv\mathcal{I}_{8}^{[2]}\mathcal{P}^{24}_{78}=\frac{z_{68}^2}{z_{13}^2 z_{16} z_{17} z_{24} z_{27} z_{28}^2 z_{35} z_{36} z_{46}^2 z_{48} z_{56} z_{58}^2 z_{67} z_{78}}$ & $3$ \\
    \specialrule{0pt}{0pt}{5pt}
    $\mathcal{I}_{8}^{[2]}{\cal T}_{1;4}\equiv\mathcal{I}_{8}^{[2]}\mathcal{P}^{24}_{56}=\frac{z_{68}^2}{z_{13}^2 z_{16} z_{17} z_{24} z_{26} z_{28}^2 z_{35} z_{36} z_{45} z_{46}^2 z_{58}^2 z_{67} z_{78}^2}$ & $4$ \\
    \specialrule{0pt}{0pt}{5pt}

    \specialrule{0pt}{5pt}{15pt}
    \specialrule{0.1pt}{5pt}{5pt}
    \end{tabular}
\caption{\label{fig:i} All the combinations without add new poles of (\ref{Ex_I82})}
\label{table4}
\end{figure}
Since they all kept incompatible poles, we can not find the right pick-factor from just the collection (I). According to our algorithm we need to included the collection (III) with following five primary cross-ratio factors
\begin{equation}
\begin{aligned}
    &a_1\equiv \mathcal{P}^{35}_{68;24}=\frac{z_{26} z_{35} z_{48}}{z_{23} z_{45} z_{68}}
    & &a_2\equiv \mathcal{P}^{24}_{68;35}=\frac{z_{24} z_{38} z_{56}}{z_{23} z_{45} z_{68}}
    & &a_3\equiv \mathcal{P}^{56}_{68;24}=\frac{z_{48} z_{56}}{z_{45} z_{68}} \\
    &a_4\equiv \mathcal{P}^{78}_{68;24}=\frac{z_{26} z_{78}}{z_{27} z_{68}}
    & &a_5\equiv \mathcal{P}^{78}_{68;35}=\frac{z_{56} z_{78}}{z_{57} z_{68}}
\end{aligned}
\label{I82-mixed}
\end{equation}
There are $9$'s ${\cal T}_{i;k}$ according to \eref{rule-7b} with $i=2,1$ and $N_2=5, N_1=4$ respectively. Searching along the nested loops, we find that
\begin{itemize}

\item (1) With $m=1$, we search all $9\times 5$ combinations, five of them satisfy both Criterion I and II:

        \begin{tabular}[t]{m{0.01\linewidth}m{0.06\linewidth}l}
        \specialrule{0pt}{0pt}{5pt}
        & ${\cal T}_{2;1}$ : & $\mathcal{P}^{24}_{78} \mathcal{P}^{56}_{78} \mathcal{P}^{35}_{68;24}=\frac{z_{24} z_{26} z_{35} z_{56} z_{78}^2}{z_{23} z_{27} z_{45} z_{57} z_{68}^2}$ \  $\mathcal{P}^{24}_{78} \mathcal{P}^{56}_{78} \mathcal{P}^{56}_{68;24}=\frac{z_{24} z_{56}^2 z_{78}^2}{z_{27} z_{45} z_{57} z_{68}^2}$      \\
        \specialrule{0pt}{5pt}{5pt}
        & ${\cal T}_{2;2}$ : & $\mathcal{P}^{24}_{78} \mathcal{P}^{35}_{78} \mathcal{P}^{56}_{68;24}=\frac{z_{24} z_{35} z_{56} z_{78}^2}{z_{27} z_{38} z_{45} z_{57} z_{68}}$      \\
        \specialrule{0pt}{5pt}{5pt}
        & ${\cal T}_{2;3}$ : & $\mathcal{P}^{24}_{56} \mathcal{P}^{56}_{78} \mathcal{P}^{78}_{68;24}=\frac{z_{24} z_{56}^2 z_{78}^2}{z_{27} z_{45} z_{57} z_{68}^2}$      \\
        \specialrule{0pt}{5pt}{5pt}
        & ${\cal T}_{2;4}$ : & $\mathcal{P}^{24}_{56} \mathcal{P}^{35}_{78} \mathcal{P}^{78}_{68;24}=\frac{z_{24} z_{35} z_{56} z_{78}^2}{z_{27} z_{38} z_{45} z_{57} z_{68}}$      \\
        \specialrule{0pt}{5pt}{5pt}
        & ${\cal T}_{2;5}$ : & $\mathcal{P}^{24}_{35} \mathcal{P}^{56}_{78} \mathcal{P}^{78}_{68;24}=\frac{z_{24} z_{26} z_{35} z_{56} z_{78}^2}{z_{23} z_{27} z_{45} z_{57} z_{68}^2}$      \\
        \specialrule{0pt}{0pt}{5pt}
        \end{tabular}
\item (2) With $m=2$, we test all $9\times 10$ items that four combinations satisfy both Criterion I and II at the same time.

        \begin{tabular}[t]{m{0.01\linewidth}m{0.06\linewidth}l}
        \specialrule{0pt}{0pt}{5pt}
         & ${\cal T}_{1;1}$ : & $\mathcal{P}^{56}_{78} \mathcal{P}^{24}_{68;35} \mathcal{P}^{78}_{68;24}=\frac{z_{24} z_{26} z_{38} z_{56}^2 z_{78}^2}{z_{23} z_{27} z_{45} z_{57} z_{68}^3}$      \\
        \specialrule{0pt}{5pt}{5pt}
        & ${\cal T}_{1;2}$ : & $\mathcal{P}^{35}_{78} \mathcal{P}^{24}_{68;35} \mathcal{P}^{78}_{68;24}=\frac{z_{24} z_{26} z_{35} z_{56} z_{78}^2}{z_{23} z_{27} z_{45} z_{57} z_{68}^2}$      \\
        \specialrule{0pt}{5pt}{5pt}
        & ${\cal T}_{2;1}$ : & $\mathcal{P}^{24}_{78} \mathcal{P}^{56}_{78} \mathcal{P}^{35}_{68;24}=\frac{z_{24} z_{26} z_{35} z_{56} z_{78}^2}{z_{23} z_{27} z_{45} z_{57} z_{68}^2}$ \  $\mathcal{P}^{24}_{78} \mathcal{P}^{56}_{78} \mathcal{P}^{56}_{68;24}=\frac{z_{24} z_{56}^2 z_{78}^2}{z_{27} z_{45} z_{57} z_{68}^2}$      \\
        \specialrule{0pt}{5pt}{5pt}
        & ${\cal T}_{2;3}$ : & $\mathcal{P}^{24}_{56} \mathcal{P}^{56}_{78} \mathcal{P}^{78}_{68;24}=\frac{z_{24} z_{56}^2 z_{78}^2}{z_{27} z_{45} z_{57} z_{68}^2}$      \\
        \specialrule{0pt}{0pt}{5pt}
        \end{tabular}
\item (3) We verified that for $m>2$, there do not exist any combinations that satisfy both Criterion I and II at the same time.
\end{itemize}
All of above combinations of primary cross-ratio factors give four different pick-factors:
\begin{equation}
   (a): \frac{z_{24} z_{26} z_{35} z_{56} z_{78}^2}{z_{23} z_{27} z_{45} z_{57} z_{68}^2} \quad
   (b): \frac{z_{24} z_{35} z_{56} z_{78}^2}{z_{27} z_{38} z_{45} z_{57} z_{68}}\quad
    (c):\frac{z_{24} z_{56}^2 z_{78}^2}{z_{27} z_{45} z_{57} z_{68}^2}\quad
    (d):\frac{z_{24} z_{26} z_{38} z_{56}^2 z_{78}^2}{z_{23} z_{27} z_{45} z_{57} z_{68}^3}
\end{equation}
When multiplying each of them to the original CHY-integrand, they all produce the same Feynman diagrams containing the pole $s_{258}$.
\begin{align*}
    & \frac{1}{s_{13} s_{46} s_{58} s_{258} s_{1346}}+\frac{1}{s_{13} s_{28} s_{136} s_{258} s_{1346}}+\frac{1}{s_{13} s_{28} s_{46} s_{258} s_{1346}}    \\
    & +\frac{1}{s_{13} s_{28} s_{136} s_{258} s_{1367}}+\frac{1}{s_{13} s_{58} s_{136} s_{258} s_{1367}}+\frac{1}{s_{13} s_{58} s_{136} s_{258} s_{1346}} \\
\end{align*}

Let us give the third example that requires the collection (III)
\begin{equation}
    \mathcal{I}_{8}^{[3]}= \frac{z_{38}^2}{z_{13}^2 z_{18}^2 z_{25} z_{26} z_{28}^2 z_{34} z_{36}^2 z_{37} z_{45}^2 z_{47} z_{56} z_{78}^2}
    \label{Ex_I83}
\end{equation}
which will contain 10 cubic Feynman diagrams:
$$
\begin{aligned}
    &+\frac{1}{s_{28} s_{36} s_{45} s_{136} s_{278}}+\frac{1}{s_{13} s_{45} s_{78} s_{136} s_{278}}+\frac{1}{s_{36} s_{45} s_{78} s_{136} s_{278}}+\frac{1}{s_{13} s_{28} s_{45} s_{136} s_{278}}\\
    &+\frac{1}{s_{28} s_{36} s_{45} s_{128} s_{1278}}+\frac{1}{s_{18} s_{36} s_{45} s_{178} s_{1278}}+\frac{1}{s_{36} s_{45} s_{78} s_{178} s_{1278}}+\frac{1}{s_{18} s_{36} s_{45} s_{128} s_{1278}}\\
    &+\frac{1}{s_{36} s_{45} s_{78} s_{278} s_{1278}}+\frac{1}{s_{28} s_{36} s_{45} s_{278} s_{1278}} \\
\end{aligned}
$$
Let us pick the pole $s_{1278}$ which appears in $6$ terms. From the Figure \ref{Com-8-1}, the linking between the subset $\{1,2,7,8\}$ and its complement  $\{3,4,5,6\}$ are solid lines $\{1,3\}$, $\{2,5\}$, $\{2,6\}$, $\{7,3\}$, $\{7,4\}$ and the dashed lines $\{8,3\}$.
Following the algorithm, let us construct the collection (I) with following pure primary cross-ratios
$$
\begin{aligned}
&\mathcal{P}^{13}_{25}=\frac{z_{13} z_{25}}{z_{15} z_{23}} \quad
\mathcal{P}^{13}_{26}=\frac{z_{13} z_{26}}{z_{16} z_{23}} \quad
\mathcal{P}^{13}_{47}=\frac{z_{13} z_{47}}{z_{14} z_{37}} \quad
\mathcal{P}^{25}_{37}=\frac{z_{25} z_{37}}{z_{23} z_{57}} \\
&\mathcal{P}^{25}_{47}=\frac{z_{25} z_{47}}{z_{24} z_{57}} \quad
\mathcal{P}^{26}_{37}=\frac{z_{26} z_{37}}{z_{23} z_{67}} \quad
\mathcal{P}^{26}_{47}=\frac{z_{26} z_{47}}{z_{24} z_{67}} \quad
 \\
\end{aligned}
$$
Multiplying all of them together, we will get $\frac{z_{13}^3 z_{25}^3 z_{26}^3 z_{37} z_{47}^3}{z_{14} z_{15} z_{16} z_{23}^4 z_{24}^2 z_{57}^2 z_{67}^2}$, which will produce new pole $s_{23}$, $s_{24}$, $s_{57}$, $s_{67}$ when multiplying with the original CHY-integrand $\mathcal{I}_{8}^{[3]}$. We have presented combinations satisfying the Criterion I in Figure \ref{table5} with the number of kept incompatible poles.
\begin{figure}[h]
    \centering
    \begin{tabular}[t]{m{0.66\linewidth}c}
    \hline
    \specialrule{0pt}{5pt}{5pt}
    $\mathcal{I}$ & {\rm \# of kept~incompatible~poles} \\
    \specialrule{0.1pt}{5pt}{5pt}
    \specialrule{0pt}{0pt}{5pt}
    $\mathcal{I}_{8}^{[3]}{\cal T}_{2;1}\equiv\mathcal{I}_{8}^{[3]}\mathcal{P}^{25}_{47}\mathcal{P}^{26}_{37}=\frac{z_{38}^2}{z_{13}^2 z_{18}^2 z_{23} z_{24} z_{28}^2 z_{34} z_{36}^2 z_{45}^2 z_{56} z_{57} z_{67} z_{78}^2}$ & $2$ \\
    \specialrule{0pt}{0pt}{5pt}
    $\mathcal{I}_{8}^{[3]}{\cal T}_{2;2}\equiv\mathcal{I}_{8}^{[3]}\mathcal{P}^{25}_{37}\mathcal{P}^{26}_{47}=\frac{z_{38}^2}{z_{13}^2 z_{18}^2 z_{23} z_{24} z_{28}^2 z_{34} z_{36}^2 z_{45}^2 z_{56} z_{57} z_{67} z_{78}^2}$ & $2$ \\
    \specialrule{0pt}{0pt}{5pt}
    $\mathcal{I}_{8}^{[3]}{\cal T}_{2;3}\equiv\mathcal{I}_{8}^{[3]}\mathcal{P}^{13}_{26}\mathcal{P}^{25}_{47}=\frac{z_{38}^2}{z_{13} z_{16} z_{18}^2 z_{23} z_{24} z_{28}^2 z_{34} z_{36}^2 z_{37} z_{45}^2 z_{56} z_{57} z_{78}^2}$ & $1$ \\
    \specialrule{0pt}{0pt}{5pt}
    $\mathcal{I}_{8}^{[3]}{\cal T}_{1;1}\equiv\mathcal{I}_{8}^{[3]}\mathcal{P}^{26}_{37}=\frac{z_{38}^2}{z_{13}^2 z_{18}^2 z_{23} z_{25} z_{28}^2 z_{34} z_{36}^2 z_{45}^2 z_{47} z_{56} z_{67} z_{78}^2}$ & $2$ \\
    \specialrule{0pt}{0pt}{5pt}
    $\mathcal{I}_{8}^{[3]}{\cal T}_{1;2}\equiv\mathcal{I}_{8}^{[3]}\mathcal{P}^{25}_{47}=\frac{z_{38}^2}{z_{13}^2 z_{18}^2 z_{24} z_{26} z_{28}^2 z_{34} z_{36}^2 z_{37} z_{45}^2 z_{56} z_{57} z_{78}^2}$ & $2$ \\
    \specialrule{0pt}{0pt}{5pt}
    $\mathcal{I}_{8}^{[3]}{\cal T}_{1;3}\equiv\mathcal{I}_{8}^{[3]}\mathcal{P}^{13}_{26}=\frac{z_{38}^2}{z_{13} z_{16} z_{18}^2 z_{23} z_{25} z_{28}^2 z_{34} z_{36}^2 z_{37} z_{45}^2 z_{47} z_{56} z_{78}^2}$ & $1$ \\
    \specialrule{0pt}{0pt}{5pt}
         \specialrule{0pt}{5pt}{15pt}
    \specialrule{0.1pt}{5pt}{5pt}
    \end{tabular}
\caption{\label{fig:i} All the combinations without add new poles of (\ref{Ex_I83})}
\label{table5}
\end{figure}
According to the algorithm, we construct the collection (III) with following $9$ primary cross-ratio factors
\begin{equation}
    \begin{aligned}
        &a_1\equiv \mathcal{P}^{13}_{38;25}=\frac{z_{13} z_{58}}{z_{15} z_{38}}
        & &a_2\equiv \mathcal{P}^{13}_{38;26}=\frac{z_{13} z_{68}}{z_{16} z_{38}}
        & &a_3\equiv \mathcal{P}^{13}_{38;47}=\frac{z_{13} z_{48}}{z_{14} z_{38}} \\
        &a_4\equiv \mathcal{P}^{37}_{38;25}=\frac{z_{37} z_{58}}{z_{38} z_{57}}
        & &a_5\equiv \mathcal{P}^{25}_{38;47}=\frac{z_{25} z_{37} z_{48}}{z_{24} z_{38} z_{57}}
        & &a_6\equiv \mathcal{P}^{47}_{38;25}=\frac{z_{23} z_{47} z_{58}}{z_{24} z_{38} z_{57}} \\
        &a_7\equiv \mathcal{P}^{37}_{38;26}=\frac{z_{37} z_{68}}{z_{38} z_{67}}
        & &a_8\equiv \mathcal{P}^{26}_{38;47}=\frac{z_{26} z_{37} z_{48}}{z_{24} z_{38} z_{67}}
        & &a_9\equiv \mathcal{P}^{47}_{38;26}=\frac{z_{23} z_{47} z_{68}}{z_{24} z_{38} z_{67}}
    \end{aligned}
\label{I82-mixed}
\end{equation}
There are $6$'s ${\cal T}_{i;k}$ according to \eref{rule-7b} with $i=2,1$ and $N_2=3, N_1=3$ respectively. Searching along the nested loops, we find that
\begin{itemize}
\item (1) With $m=1$, we calculate all $6\times 9$ terms, there are twelve satisfy both Criterion I and II:

        \begin{tabular}[t]{m{0.01\linewidth}m{0.06\linewidth}l}
        \specialrule{0pt}{0pt}{5pt}
        & ${\cal T}_{2;1}$ : & $\mathcal{P}^{25}_{47} \mathcal{P}^{26}_{37} \mathcal{P}^{13}_{38;25}=\mathcal{P}^{25}_{47} \mathcal{P}^{26}_{37} \mathcal{P}^{13}_{38;47}=\frac{z_{13} z_{25} z_{26} z_{37} z_{47} z_{48}}{z_{14} z_{23} z_{24} z_{38} z_{57} z_{67}}$    \\
        \specialrule{0pt}{5pt}{5pt}
        & ${\cal T}_{2;2}$ : & $\mathcal{P}^{25}_{47} \mathcal{P}^{26}_{37} \mathcal{P}^{13}_{38;25}=\mathcal{P}^{25}_{47} \mathcal{P}^{26}_{37} \mathcal{P}^{13}_{38;47}=\frac{z_{13} z_{25} z_{26} z_{37} z_{47} z_{48}}{z_{14} z_{23} z_{24} z_{38} z_{57} z_{67}}$      \\
        \specialrule{0pt}{5pt}{5pt}
        & ${\cal T}_{2;3}$ : & $\mathcal{P}^{13}_{26} \mathcal{P}^{25}_{47} \mathcal{P}^{13}_{38;25}=\mathcal{P}^{13}_{26} \mathcal{P}^{25}_{47} \mathcal{P}^{13}_{38;47}=\frac{z_{13}^2 z_{25} z_{26} z_{47} z_{48}}{z_{14} z_{16} z_{23} z_{24} z_{38} z_{57}}$      \\
        \specialrule{0pt}{5pt}{5pt}
        & ${\cal T}_{1;1}$ : & $\mathcal{P}^{26}_{37} \mathcal{P}^{13}_{38;25}=\mathcal{P}^{26}_{37} \mathcal{P}^{13}_{38;47}=\frac{z_{13} z_{26} z_{37} z_{48}}{z_{14} z_{23} z_{38} z_{67}}$      \\
        \specialrule{0pt}{5pt}{5pt}
        & ${\cal T}_{1;2}$ : & $\mathcal{P}^{25}_{47} \mathcal{P}^{13}_{38;25}=\mathcal{P}^{25}_{47} \mathcal{P}^{13}_{38;47}=\frac{z_{13} z_{25} z_{47} z_{48}}{z_{14} z_{24} z_{38} z_{57}}$      \\
        \specialrule{0pt}{5pt}{5pt}
        & ${\cal T}_{1;3}$ : & $\mathcal{P}^{13}_{26} \mathcal{P}^{13}_{38;25}=\mathcal{P}^{13}_{26} \mathcal{P}^{13}_{38;47}=\frac{z_{13}^2 z_{26} z_{48}}{z_{14} z_{16} z_{23} z_{38}}$      \\
        \specialrule{0pt}{0pt}{5pt}
        \end{tabular}
\item (2) With $m=2$, we enumerate all $6\times 36$ that fourteen combinations satisfy both Criterion I and II at the same time.

        \begin{tabular}[t]{m{0.01\linewidth}m{0.06\linewidth}l}
        \specialrule{0pt}{0pt}{5pt}
        & ${\cal T}_{2;1}$ : & $\mathcal{P}^{25}_{47} \mathcal{P}^{26}_{37} \mathcal{P}^{13}_{38;25} \mathcal{P}^{13}_{38;26}=\mathcal{P}^{25}_{47} \mathcal{P}^{26}_{37} \mathcal{P}^{13}_{38;26} \mathcal{P}^{13}_{38;47}=\frac{z_{13}^2 z_{25} z_{26} z_{37} z_{47} z_{48} z_{68}}{z_{14} z_{16} z_{23} z_{24} z_{38}^2 z_{57} z_{67}}$    \\
        \specialrule{0pt}{5pt}{5pt}
        & ${\cal T}_{2;2}$ : & $\mathcal{P}^{25}_{47} \mathcal{P}^{26}_{37} \mathcal{P}^{13}_{38;25} \mathcal{P}^{13}_{38;26}=\mathcal{P}^{25}_{47} \mathcal{P}^{26}_{37} \mathcal{P}^{13}_{38;26} \mathcal{P}^{13}_{38;47}=\frac{z_{13}^2 z_{25} z_{26} z_{37} z_{47} z_{48} z_{68}}{z_{14} z_{16} z_{23} z_{24} z_{38}^2 z_{57} z_{67}}$      \\
        \specialrule{0pt}{5pt}{5pt}
        & ${\cal T}_{2;3}$ : & $\mathcal{P}^{13}_{26} \mathcal{P}^{25}_{47} \mathcal{P}^{13}_{38;25} \mathcal{P}^{37}_{38;26}=\mathcal{P}^{13}_{26} \mathcal{P}^{25}_{47} \mathcal{P}^{13}_{38;47} \mathcal{P}^{37}_{38;26}=\frac{z_{13}^2 z_{25} z_{26} z_{37} z_{47} z_{48} z_{68}}{z_{14} z_{16} z_{23} z_{24} z_{38}^2 z_{57} z_{67}}$      \\
        \specialrule{0pt}{5pt}{5pt}
        & ${\cal T}_{1;1}$ : & $\mathcal{P}^{26}_{37} \mathcal{P}^{13}_{38;25} \mathcal{P}^{13}_{38;26}=\mathcal{P}^{26}_{37} \mathcal{P}^{13}_{38;26} \mathcal{P}^{13}_{38;47}=\frac{z_{13}^2 z_{26} z_{37} z_{48} z_{68}}{z_{14} z_{16} z_{23} z_{38}^2 z_{67}}$      \\
        \specialrule{0pt}{5pt}{5pt}
        & ${\cal T}_{1;2}$ : & $\mathcal{P}^{25}_{47} \mathcal{P}^{13}_{38;25} \mathcal{P}^{13}_{38;26}=\mathcal{P}^{25}_{47} \mathcal{P}^{13}_{38;26} \mathcal{P}^{13}_{38;47}=\frac{z_{13}^2 z_{25} z_{47} z_{48} z_{68}}{z_{14} z_{16} z_{24} z_{38}^2 z_{57}}$      \\
        &                    & $\mathcal{P}^{25}_{47} \mathcal{P}^{13}_{38;25} \mathcal{P}^{37}_{38;26}=\mathcal{P}^{25}_{47} \mathcal{P}^{13}_{38;47} \mathcal{P}^{37}_{38;26}=\frac{z_{13} z_{25} z_{37} z_{47} z_{48} z_{68}}{z_{14} z_{24} z_{38}^2 z_{57} z_{67}}$      \\
        \specialrule{0pt}{5pt}{5pt}
        & ${\cal T}_{1;3}$ : & $\mathcal{P}^{13}_{26} \mathcal{P}^{13}_{38;25} \mathcal{P}^{37}_{38;26}=\mathcal{P}^{13}_{26} \mathcal{P}^{13}_{38;47} \mathcal{P}^{37}_{38;26}=\frac{z_{13}^2 z_{26} z_{37} z_{48} z_{68}}{z_{14} z_{16} z_{23} z_{38}^2 z_{67}}$      \\
        \specialrule{0pt}{0pt}{5pt}
        \end{tabular}
\item (3) With $m=3$, there are two combinations satisfy both Criterion I and II in $6\times 84$ terms.

        \begin{tabular}[t]{m{0.01\linewidth}m{0.06\linewidth}l}
        \specialrule{0pt}{0pt}{10pt}
        & ${\cal T}_{1;2}$ : & $\mathcal{P}^{25}_{47} \mathcal{P}^{13}_{38;25} \mathcal{P}^{13}_{38;26} \mathcal{P}^{37}_{38;26}=\mathcal{P}^{25}_{47} \mathcal{P}^{13}_{38;26} \mathcal{P}^{13}_{38;47} \mathcal{P}^{37}_{38;26}=\frac{z_{13}^2 z_{25} z_{37} z_{47} z_{48} z_{68}^2}{z_{14} z_{16} z_{24} z_{38}^3 z_{57} z_{67}}$    \\
        \specialrule{0pt}{0pt}{10pt}
        \end{tabular}
\item (4) We verified that for $m>3$, there do not exist any combinations that satisfy both Criterion I and II at the same time.
\end{itemize}
All of above eleven combinations of primary cross-ratio factors give nine different pick-factors:
\begin{align*}
    & \frac{z_{13} z_{25} z_{26} z_{37} z_{47} z_{58}}{z_{15} z_{23} z_{24} z_{38} z_{57} z_{67}}; \quad
    \frac{z_{13}^2 z_{26} z_{58}}{z_{15} z_{16} z_{23} z_{38}}; \quad
    \frac{z_{13}^2 z_{25} z_{26} z_{47} z_{58}}{z_{15} z_{16} z_{23} z_{24} z_{38} z_{57}};\quad
    \frac{z_{13} z_{26} z_{37} z_{58}}{z_{15} z_{23} z_{38} z_{67}};\quad
    \frac{z_{13}^2 z_{26} z_{37} z_{58} z_{68}}{z_{15} z_{16} z_{23} z_{38}^2 z_{67}};\\
    & \frac{z_{13} z_{25} z_{47} z_{58}}{z_{15} z_{24} z_{38} z_{57}};\quad
    \frac{z_{13}^2 z_{25} z_{47} z_{58} z_{68}}{z_{15} z_{16} z_{24} z_{38}^2 z_{57}};\quad
    \frac{z_{13}^2 z_{25} z_{37} z_{47} z_{58} z_{68}^2}{z_{15} z_{16} z_{24} z_{38}^3 z_{57} z_{67}};\quad
    \frac{z_{13}^2 z_{25} z_{26} z_{37} z_{47} z_{58} z_{68}}{z_{15} z_{16} z_{23} z_{24} z_{38}^2 z_{57} z_{67}}; \\
\end{align*}
When multiplying each of them to the original CHY-integrand, they all give the same Feynman diagrams  containing the pole $s_{1278}$ only:
\begin{align*}
    & \frac{1}{s_{28} s_{36} s_{45} s_{128} s_{1278}}+\frac{1}{s_{18} s_{36} s_{45} s_{178} s_{1278}}+\frac{1}{s_{18} s_{36} s_{45} s_{128} s_{1278}}    \\
    & +\frac{1}{s_{28} s_{36} s_{45} s_{278} s_{1278}}+\frac{1}{s_{36} s_{45} s_{78} s_{278} s_{1278}}+\frac{1}{s_{36} s_{45} s_{78} s_{178} s_{1278}} \\
\end{align*}
%

\section{Conclusion}
\label{Conclusion}

In this note, we have presented two results. In the first result, starting with the Cayley tree type of CHY-integrands, we show how to write down the corresponding effective Feynman diagrams. Base on the topological structure of the Cayley tree graph, there are two types of primary effective Feynman vertex: the colour ordered  type $V_{C}$  corresponding the line subgraph of Cayley tree  and the permutation type $V_P$  corresponding to vertex  with multiple branches. For the $V_{C}$-type effective vertex, the $|V_{C}\{(l_1,l_2,...,l_n)\}|$ represents all the cubic Feynman diagrams  respecting  the colour order of the list $\{(l_1,l_2,...,l_n)\}$. For the $P$-type vertex $V_P(k; \W K_1\shuffle \W K_2 \shuffle ... \shuffle \W K_t; n)$, it represents
the comb-like cubic Feynman graph diagrams with  ordering from the shuffle algebra. We show
how to determine two types of vertexes from the Cayley tree by decomposing the tree into
line pieces and vertexes.

Our second result is the construction of picking up factor for general CHY-integrands containing only simple poles. Our algorithm has generalized the result in  \cite{feng2016chy,huang2018permutation} for the bi-adjoint scalar theories. Unlike the situation of bi-adjoint theory where one  needs only a cross-ratio factor, for general CHY-integrand, we need to construct all possible cross-ratio factors from the linking set  ${\rm Links}[A, \O A]$,
including the denominators and numerators. When multiplying them together, we need to
introduce two criteria to select the right combination. We demonstrate our algorithm by   several non-trivial examples.

Based on results in this note, there are several interesting directions one can investigate.
First, we have constructed the picking up factor for selecting a particular pole from all Feynman diagrams. Since one can represent these diagrams using the polytope, it is natural to ask, could
we using the geometric picture, i.e., projecting from one dimension higher object to its specific
face, to understand and construct the same picking up factor.  Secondly, in \cite{feng2020one},
using the picking pole technique, the one-loop CHY-integrand for general bi-adjoint scalar theory
has been constructed by removing singular poles (i.e., tadpoles and massless bubbles). For general
theory, such as Gravity and Yang-Mills, the CHY-integrands are general. Thus one can try if it is possible to construct the one-loop integrands for these general theory by removing singular contribution using
the picking up factor developed in this note. Thirdly, it is well known the general CHY-integrand could contain higher-order poles.  The standard method to deal with them is given in  \cite{huang2016feynman} by  decomposing them as a linear combination of CHY-integrands with only simple poles using cross-ratio identity. It is interesting to ask if our picking pole factor provides another way to reduce higher poles to lower poles.

Fourthly, in this paper we considered only the effective Feynman Diagram to CHY-integrand of $m(C \mid C)$ type and have seen the connection between polytopes and effective Feynman Diagrams.
However, for the $m(C_1 \mid C_2)$, as pointed out in  \cite{gao2017labelled}, it is   the intersection of $m(C_1 \mid C_1)$ and $m(C_2 \mid C_2)$. This point can also easily seen from
the effective Feynman diagrams. In \cite{mizera2018scattering,frost2018biadjoint}, they pointed out that such pairing is essentially the same thing as the intersection pairing of two cocycles in a certain cohomology theory. Thus it is an interesting direction to investigate the symmetry of the intersection between two effective Feynman diagrams and the corresponding geometric representation.

\section*{Acknowledgments}

This work is supported by Qiu-Shi Funding and the
National Natural Science Foundation of China (NSFC) with Grant No.11935013, No.11575156.

\bibliographystyle{JHEP}

\bibliography{mybibtex}

\end{document}